\documentclass[twocolumn,prd,aps,superscriptaddress,preprintnumbers,tightenlines,showpacs,nofootinbib,eqsecnum,amsfonts,amsmath]{revtex4-1}
\usepackage{placeins}
\usepackage{color}
\usepackage{calc}
\usepackage{amsmath,amssymb,graphicx}
\usepackage{tensor}
\usepackage{bm}
\usepackage{times}
\usepackage[varg]{txfonts}
\usepackage[colorlinks, pdfborder={0 0 0}]{hyperref}
\usepackage{float}
\usepackage{dcolumn}
\usepackage[nolist,nohyperlinks]{acronym}
\usepackage{xspace}
\usepackage[abs]{overpic}
\usepackage{pict2e}
\usepackage{enumitem}
\usepackage[usenames,dvipsnames]{xcolor}
\usepackage[utf8]{inputenc}
\usepackage{acronym}
\usepackage{gensymb}
\usepackage{cleveref}
\usepackage[normalem]{ulem}
\usepackage{longtable}

\newcommand{\totalsims}{157}

\newcommand{\code}[1]{{\tt #1}}
\newcommand{\AEI}{\affiliation{Max Planck Institute for Gravitational Physics (Albert Einstein Institute), Am M\"uhlenberg 1, Potsdam 14476, Germany}}
\newcommand{\Maryland}{\affiliation{Department of Physics, University of Maryland, College Park, MD 20742, USA}}
\newcommand{\CIFAR}{\affiliation{Canadian Institute for Advanced Research, Toronto M5G 1Z8, Canada}} %
\newcommand{\CITA}{\affiliation{Canadian Institute for Theoretical Astrophysics, University of Toronto, Toronto M5S 3H8, Canada}}
\newcommand{\Cornell}{\affiliation{Cornell Center for Astrophysics and Planetary Science, Cornell University, Ithaca, New York, 14853, USA}}
\newcommand{\Caltech}{\affiliation{Theoretical Astrophysics 350-17, California Institute of Technology, Pasadena, CA 91125, USA}}
\newcommand{\Princeton}{\affiliation{Department of Physics, Princeton University, Jadwin Hall, Princeton, NJ 08544, USA}}
\newcommand{\Toronto}{\affiliation{Department of Physics, University of Toronto, Toronto M5S 3H8, Canada}}
\newcommand{\JPL}{\affiliation{Jet Propulsion Laboratory, California Institute of Technology, 4800 Oak Grove Drive, Pasadena, CA 91109, USA}}
\newcommand{\Fullerton}{\affiliation{Gravitational Wave Physics and Astronomy Center, California State University Fullerton, Fullerton, California 92834, USA}}

\def\be{\begin{equation}}
\def\ee{\end{equation}}
\def\bea{\begin{eqnarray}}
\def\eea{\end{eqnarray}}
\newcommand{\bes}{\begin{subequations}}
\newcommand{\ees}{\end{subequations}}

\newcommand{\doubleline}{\hline \hline}

\allowdisplaybreaks

\begin{document}

\title{An improved effective-one-body model of spinning, nonprecessing binary black holes\\for the era of gravitational-wave astrophysics with advanced detectors}

\author{Alejandro Boh\'e}
\email{alejandro.bohe@aei.mpg.de}
\AEI

\author{Lijing Shao}
\email{lijing.shao@aei.mpg.de}
\AEI

\author{Andrea Taracchini}
\email{andrea.taracchini@aei.mpg.de}
\AEI

\author{Alessandra Buonanno}
\AEI \Maryland

\author{Stanislav Babak}
\AEI

\author{Ian W. Harry}
\AEI

\author{Ian Hinder}
\AEI

\author{Serguei Ossokine}
\AEI

\author{Michael P\"urrer}
\AEI

\author{Vivien Raymond}
\AEI

\author{Tony Chu}
\Princeton \CITA

\author{Heather Fong}
\CITA \Toronto

\author{Prayush Kumar}
\CITA \Toronto

\author{Harald P. Pfeiffer}
\CITA \AEI \CIFAR

\author{Michael Boyle}
\Cornell

\author{Daniel A. Hemberger}
\Caltech

\author{Lawrence E. Kidder}
\Cornell

\author{Geoffrey Lovelace}
\Fullerton

\author{Mark A. Scheel}
\Caltech

\author{B\'{e}la Szil\'{a}gyi}
\Caltech \JPL

\date{\today}

\begin{abstract}
  We improve the accuracy of the effective-one-body (EOB) waveforms
  that were employed during the first observing run of Advanced
    LIGO for binaries of spinning, nonprecessing black holes by calibrating
  them to a set of 141 numerical-relativity (NR) waveforms. The NR
  simulations expand the domain of calibration towards larger mass
  ratios and spins, as compared to the previous EOBNR
    model. Merger-ringdown waveforms computed in black-hole
  perturbation theory for Kerr spins close to extremal provide
  additional inputs to the calibration.  For the inspiral-plunge
  phase, we use a Markov-chain Monte Carlo algorithm to efficiently
  explore the calibration space. For the merger-ringdown phase, we fit
  the NR signals with phenomenological formulae. After extrapolation
  of the calibrated model to arbitrary mass ratios and spins, the
  (dominant-mode) EOBNR waveforms have faithfulness --- at design
  Advanced-LIGO sensitivity --- above $99\%$ against all the NR
  waveforms, including 16 additional waveforms used for validation, when
  maximizing only on initial phase and time. This implies a negligible
  loss in event rate due to modeling for these binary
  configurations. We find that future NR simulations at mass ratios
  $\gtrsim 4$ and double spin $\gtrsim 0.8$ will be
  crucial to resolve discrepancies between different ways of
  extrapolating waveform models.  We also find that some of the NR
  simulations that already exist in such region of parameter space are
  too short to constrain the low-frequency portion of the
  models. Finally, we build a reduced-order version of the EOBNR model
  to speed up waveform generation by orders of magnitude, thus
  enabling intensive data-analysis applications during the upcoming
  observation runs of Advanced LIGO.
\end{abstract}
\pacs{04.25.D-, 04.25.dg, 04.30.-w}

\maketitle

\section{Introduction}

During its first observing run (O1), the Advanced Laser Interferometer Gravitational wave Observatory (LIGO)
detected gravitational waves (GWs) emitted by the coalescence of two  binary black holes (BBHs), GW150914 and GW151226~
\cite{Abbott:2016blz,Abbott:2016nmj}. A third candidate event, LVT151012, was also recorded~\cite{TheLIGOScientific:2016pea},
but with not high enough statistical significance to claim a detection. These discoveries opened the possibility of observing and probing the most
extreme astrophysical objects in the Universe.

GW150914 was a loud event, detected with a signal-to-noise ratio (SNR) of $\sim 24$. It was
initially identified by an (online) generic-transient search~\cite{TheLIGOScientific:2016uux},
which uses minimal assumptions about waveforms. The highest statistical confidence was obtained with
the (offline) optimal matched-filtering searches~\cite{TheLIGOScientific:2016uux} that employ waveforms predicted by
general relativity. By contrast, matched-filtering searches were essential for detecting GW151226~\cite{Abbott:2016nmj,TheLIGOScientific:2016pea},
which was an event quieter than GW150914, having a SNR of $\sim 13$ and an energy spread over
about $1 \,{\rm sec}$ ($\sim 55$ GW cycles), instead of $0.2 \,{\rm sec}$ ($\sim 10$ GW cycles).

During the O1 run, the Advanced-LIGO matched-filtering search targeted GWs from binary systems with component masses between $1\,M_\odot$ and $99\,M_\odot$,
total mass smaller than $100\,M_\odot$, and dimensionless aligned-spin magnitudes up to 0.99. For total masses larger than $4\,M_\odot$, it used a template bank~\cite{TheLIGOScientific:2016qqj,Capano:2016dsf,TheLIGOScientific:2016pea} of $\sim 200,000$ (semi-analytical) spinning, nonprecessing templates developed in Refs.~\cite{Taracchini:2013rva,Purrer:2015tud} within the effective-one-body (EOB) formalism. This analytical approach, which describes the dynamics of coalescing BBHs and the associated gravitational radiation, was originally developed in late 90s~\cite{Buonanno:1998gg,Buonanno:2000ef} and over the years
it has been further improved~\cite{Damour:2000we,Damour:2001tu,Buonanno:2005xu,Damour:2008qf,Damour:2008gu,Damour:2009kr,Barausse:2009aa,Barausse:2009xi,Pan:2010hz,Barausse:2011dq,Damour:2012ky,Pan:2011gk,Pan:2013rra,Nagar:2015xqa,Taracchini:2013rva,Bini:2016cje,Babak:2016tgq}. In particular, newly available results from perturbative approaches to the two-body problem in general relativity (GR) (post-Newtonian expansion, BH perturbation theory, and gravitational self-force formalism), as well as crucial nonperturbative information accessible only through numerical relativity (NR) have been incorporated into the framework, gradually extending the region in parameter space over which the model provides highly accurate predictions for inspiral, merger and ringdown gravitational waveforms. As a result, since the first LIGO search for BBHs in 2005~\cite{Abbott:2005kq}, EOBNR waveforms have been employed in LIGO/Virgo data analyses, and, as discussed above, have played a central role in the detection~\cite{Abbott:2016nmj}, and subsequent parameter-estimation analyses~\cite{TheLIGOScientific:2016wfe,TheLIGOScientific:2016pea,Abbott:2016izl} and GR tests~\cite{TheLIGOScientific:2016src,TheLIGOScientific:2016pea}
 of the GW observations announced earlier this year. EOB waveform models have also
been employed to build frequency-domain, phenomenological models~\cite{Pan:2007nw,Ajith:2007kx} for the inspiral, merger and ringdown stages of the
BBH coalescence. Those models~\cite{Hannam:2013oca,Khan:2015jqa} have also been used to infer the properties and carry out tests of GR with GW150914 and
GW151226~\cite{TheLIGOScientific:2016wfe,TheLIGOScientific:2016src,TheLIGOScientific:2016pea}.

In this paper, we build an improved version of \texttt{SEOBNRv2}~\cite{Taracchini:2013rva,Purrer:2015tud}, the spinning, nonprecessing EOBNR waveform model that was used for O1, and whose accuracy was recently found to degrade~\cite{Kumar:2016dhh} in some regions of the BBH parameter space, notably large aligned spins and unequal masses, where the model was extrapolating away from the NR waveforms that were available at the time of its calibration. The improvements developed in this paper include: (i) the addition of all 4PN terms to the EOB radial potential and of higher-order PN corrections to the polarization modes, (ii) a recalibration of the EOB model to a large set of recently produced NR waveforms, which expand the domain of calibration towards larger mass ratios and aligned-spin components, 
and (iii) a more robust description of the merger-ringdown signal.
The updated EOBNR model (\texttt{SEOBNRv4}) has been coded in the LIGO Algorithm Library (LAL)~\cite{LAL}, so
that it can be employed during the second observing (O2) run of Advanced LIGO, which is scheduled to start later this year, and later runs, enhancing our ability to extract physical information and interpret future GW detections.

The paper is organized as follows. In Sec.~\ref{sec:model} we discuss the EOBNR model of spinning, nonprecessing BBHs, emphasizing the new ingredients with respect to the previous model~\cite{Taracchini:2013rva}. In Sec.~\ref{sec:NR} we review the catalog of NR simulations and BH perturbation-theory waveforms that we use to calibrate the model. In Sec.~\ref{sec:cali} we describe how to tune the inspiral-plunge calibration parameters to NR simulations using a newly developed Markov-chain Monte Carlo code. We discuss the performance of the model after interpolating and extrapolating it to the entire parameter space, and in Sec.~\ref{sec:comparisons} we compare it to previously developed spinning, nonprecessing waveform models. In Sec.~\ref{sec:length} we highlight how short NR simulations cannot constrain the low-frequency portion of waveform models. Section~\ref{EOBNR:ROM} describes the reduced-order version of the EOBNR model, which is used to speed up the waveform generation, allowing its application in Advanced LIGO and Virgo data analyses. In Sec.~\ref{sec:concl} we provide our concluding remarks. In Appendix~\ref{app:IV} we summarize information on the NR (dominant-mode) waveforms around the time of merger, which are also included in the EOBNR model. Finally, Appendix~\ref{app:newRD} provides fitting formulae for constructing the merger-ringdown signal of the EOBNR model.

Henceforth, we adopt geometric units: $G=1=c$.

\section{Motivation and overview of the effective-one-body formalism}
\label{sec:model}

The problem of describing the GW signal generated by a pair of BHs that (quasi-circularly)
orbit each other and eventually merge into a single BH is challenging because of the
different dynamical regimes that this process spans. When the binary is wide --- say, as compared to the BH
horizons --- the component objects move at orbital speeds (in the center-of-mass frame) that are small with
respect to the speed of light. During this phase of the coalescence, the post-Newtonian (PN) (i.e.,
slow-motion and weak-field) approximation  to general relativity can be employed to model the orbital
dynamics and the associated GW emission (see., e.g., Ref.~\cite{Blanchet:2013haa} for an extensive review of the
current status of PN theory applied to the two-body problem). As the BHs spiral in, plunge and eventually merge,
their orbital motion becomes more relativistic and the GW energy flux is stronger. NR
techniques are required to obtain highly-accurate waveforms during this stage of the process. State-of-the-art codes can now accurately evolve BBHs for several tens of orbits ($\sim 40\mbox{--}60$) in large regions of the parameter space~\cite{Mroue:2013xna,Chu:2015kft,Lovelace:2010ne,Scheel:2014ina, Husa:2015iqa}: (i) at large mass ratios ($\sim 8$), but for moderate BH (dimensionless) spin magnitudes
($\sim 0.6$), and (ii) at higher BH spin magnitudes, but for comparable masses ($\sim 1\mbox{--}3$). Shorter simulations
($\sim 10$) can also be produced for mass ratios $\sim 20$ and spin magnitudes $\sim 0.8$. The longest NR run to date covers 175 orbits of a nonspinning BBH with mass ratio 7~\cite{Szilagyi:2015rwa}. Finally, soon after the merger, a distorted remnant BH is born. This relaxes into a stationary Kerr spacetime by radiating GWs that are well described
by BH perturbation theory, as well as by NR.

In spite of tremendous progress, a purely NR approach to simulating BBHs for any
possible configuration down to the lower edge of the sensitive frequency band of
current ground-based GW detectors is not feasible yet. This motivated the need to develop more sophisticated semi-analytical waveform models~\cite{Taracchini:2013rva,Purrer:2015tud,Hannam:2013oca,Khan:2015jqa,Babak:2016tgq,Nagar2016}
that, while being cheap to compute for data analysis, are very good approximations to general relativity.

In this Section we review the main features of the EOB approach
for spinning, nonprecessing binary BHs and describe the improvements that we introduce with respect to the
previous EOBNR model~\cite{Taracchini:2012ig,Taracchini:2013rva}, which was 
employed during the O1 data analyses.

\subsection{Conservative dynamics}

The EOB formalism aims at combining all available
results that describe the general-relativistic two-body problem --- both analytical
and numerical --- into a unified description. In the case of a binary composed
of BHs, let $m_{1,2}$
and $\boldsymbol{S}_{1,2}$ be the masses and spins of the two component
objects as used in the PN description of the \emph{real
  problem}. Let $q \equiv m_1/m_2 \geq 1$ be the mass ratio of the
binary. The key ingredient of the EOB model is a resummation of the
conservative PN dynamics of a generic BBH in terms of the conservative
dynamics of a test particle with mass $\mu$ and spin
$\boldsymbol{S}_*$ in a deformed Kerr metric with mass $M$ and
spin $\boldsymbol{S}_{\textrm{Kerr}}$ (\emph{effective problem}),
the deformation parameter being $\mu/M$. In
analogy with the Newtonian treatment of a self-gravitating binary,
here $\mu$ is the reduced mass of the BBH, while $M$ is its total
mass. As to the spins $\boldsymbol{S}_{*}$ and
$\boldsymbol{S}_{\textrm{Kerr}}$, these are given as functions of
$m_{1,2}$, $\boldsymbol{S}_{1,2}$ and the dynamical
variables. These relationships between mass
and spin parameters of the real and effective problem are obtained 
imposing (i) a precise energy mapping
between the two systems, and (ii) requiring that the Hamiltonian describing the
effective problem reduces to that of the real problem in the
slow-motion, weak-field limit. In particular, the \emph{energy
  mapping} prescribes that~\cite{Buonanno:1998gg}
\begin{equation}
H_{\textrm{EOB}} = M \sqrt{1+2\nu\left(\frac{H_{\textrm{eff}}}{\mu} - 1 \right)} - M\,,
\end{equation}
where $\nu \equiv \mu / M$ is the symmetric mass ratio of the real
system, $H_{\textrm{EOB}}$ is the EOB resummed Hamiltonian for the real problem and $H_{\textrm{eff}}$ is the
Hamiltonian for the effective problem. The explicit form of
$H_{\textrm{EOB}}$ that we shall adopt in this paper
was derived in Refs.~\cite{Barausse:2009xi,Barausse:2011ys}, based on the
linear-in-spin Hamiltonian for spinning test particles of
Ref.~\cite{Barausse:2009aa}. When PN-expanded, the EOB Hamiltonian that
we employ in this paper reproduces: (i) spin-spin couplings at leading order for any mass
ratio, (ii) spin-orbit couplings up to next-to-next-to-leading order
for any mass ratio, and (iii) all spin-orbit couplings in the
test-particle limit.

We describe the EOB orbital dynamics in terms of the following quantities: the (dimensionless) radial separation $r$ (in
units of $M$), the orbital phase $\phi$, and their conjugate (dimensionless) momenta, $p_r$ and $p_{\phi}$ (in units of $\mu$). Of course, since we consider only spins that are aligned or antialigned with the orbital angular momentum, their projections on $\boldsymbol{\hat{L}}_N$ (with $\boldsymbol{\hat{L}}_N$ being the direction of the Newtonian angular momentum) are constant; we denote them via $\chi_{1,2} \equiv (\boldsymbol{S}_{1,2}\cdot\boldsymbol{\hat{L}}_N)/m_{1,2}^2$. Note that $-1 \leq \chi_{1,2} \leq 1$.

The effective Hamiltonian depends on a radial potential that enters the 00-component of the effective deformed metric, it
reads~\cite{Taracchini:2012ig,Taracchini:2013rva}
\begin{align}
\Delta_u &= \chi_{\textrm{Kerr}}^2 \left(u - \frac{1}{r_+^{\textrm{EOB}}}\right)\left(u - \frac{1}{r_-^{\textrm{EOB}}}\right)\nonumber\\
&\times\left[1+\nu\,\Delta_0 + \log{\left(1 + \sum_{i=1}^5\frac{\Delta_i}{r^i}\right)}\right]\,,\label{Apotential}
\end{align}
where $\chi_{\textrm{Kerr}} \equiv (\boldsymbol{S}_{\textrm{Kerr}} \cdot \boldsymbol{\hat{L}}_N)/M^2$, $u$ is the inverse of the EOB radial coordinate, $r_{\pm}^{\textrm{EOB}} \equiv (1 - \nu K)\left[1 \pm (1 - \chi_{\textrm{Kerr}}^2)^{1/2}\right]$ (with $K$ a calibration parameter). We observe that $\Delta_u$ vanishes at the EOB horizon $r=r_+^{\textrm{EOB}}$. The $\Delta_i$'s used in the previous version of the spinning, nonprecessing EOBNR model~\cite{Taracchini:2012ig,Taracchini:2013rva} can be found in Appendix~A of Ref.~\cite{Steinhoff:2016rfi}. Here, we augment the 4PN coefficient $\Delta_5$ by the quadratic-in-$\nu$ corrections that were computed in Ref.~\cite{Bini:2013zaa}
\be
\Delta_5 \supset (1 - \nu K)^2 \left(\frac{41\,\pi^2}{32} -  \frac{221}{6}\right) \nu\,.
\ee
We adopt the same mapping between the spin variables in the real and effective descriptions of Refs.~\cite{Taracchini:2012ig,Taracchini:2013rva}, and include the same spin-orbit and spin-spin calibration parameters, $d_{\textrm{SO}}$ and $d_{\textrm{SS}}$.

\subsection{Inspiral-plunge waveforms and dissipative dynamics}
\label{sec:inspplungewf}

The conservative dynamics described above is complemented by analytic formulae for the inspiral-plunge GW multipolar modes $h_{\ell m}^{\textrm{insp-plunge}}$. These expressions are a recasting of PN results~\cite{Arun:2008kb,Buonanno:2012rv} into a factorized form~\cite{Damour:2007xr,Damour:2008gu,Pan:2010hz} that is meant to capture strong-field features that are observed when numerically computing the gravitational perturbations of isolated Kerr BHs by test particles on circular, equatorial orbits down to the innermost stable circular orbit (ISCO). The factorized formulae are functions of the EOB orbital dynamics. Here, we adopt the same expressions that were used in Ref.~\cite{Taracchini:2013rva} with the addition of new amplitude corrections to the $(2,2)$ mode that enter the $\rho_{22}$ factor (see Ref.~\cite{Pan:2010hz}) at 2.5PN order, that is
\be
\left[\left(-\frac{34}{21}+\frac{49\,\nu}{18}+\frac{209\,\nu^2}{126}\right)\chi_{\textrm{S}}
-\left(\frac{34}{21}+\frac{19\,\nu}{42}\right)\chi_{\textrm{A}}\delta \right] v_{\Omega}^5\,,\label{rho25PN}
\ee
and at 3.5PN order, that is
\begin{align}
&\left[\left(\frac{18733}{15876} + \frac{74749\,\nu}{5292} - \frac{245717\,\nu^2}{63504} + \frac{50803\,\nu^3}{63504}\right)\right.\chi_{\textrm{S}}\nonumber\\
& \left.+
\left(\frac{18733}{15876} + \frac{50140\,\nu}{3969} + \frac{97865\,\nu^2}{63504}\right)\chi_{\textrm{A}} \delta \right]v_{\Omega}^7\,,\label{rho35PN}
\end{align}
as well as new phase corrections that enter the $\delta_{22}$ factor (again, see Ref.~\cite{Pan:2010hz}) at 3PN order, that is
\be
-\frac{4}{3}\left[ \chi_{\textrm{S}}(1-2\nu) +  \chi_{\textrm{A}}\delta \right](\Omega H_{\textrm{EOB}})^2\,.
\label{eq:delta22corr}
\ee
Here $\chi_{\textrm{S,A}} \equiv (\chi_1 \pm \chi_2)/2$, $\delta \equiv (m_1 - m_2)/M$, and $v_{\Omega} \equiv (M\Omega)^{1/3}$, where $\Omega \equiv \textrm{d} \phi/\textrm{d}t$ is the orbital frequency. Note that the amplitude corrections in Eqs.~(\ref{rho25PN})-(\ref{rho35PN}) replace the spinning test-particle-limit terms that were used in Ref.~\cite{Taracchini:2013rva} at those PN orders. The 2.5PN correction \eqref{rho25PN} to the factorized waveform was derived in Ref.~\cite{Damour:2014sva}. We derived the ones in the phase at 3PN order and in the amplitude at 3.5PN order (Eqs.~\eqref{eq:delta22corr} and \eqref{rho35PN}, respectively) for this paper starting from the Taylor expanded PN results of Refs.~\cite{Blanchet:2011zv} and \cite{Bohe:2013cla}, respectively.

Like in previous EOBNR models, we apply nonquasicircular (NQC) corrections to the $(2,2)$ mode with the aim of describing the NR
merger signals at and around the peak of GW emission in the most accurate way. In particular, we multiply the factorized formula for the $(2,2)$ mode by~\cite{Taracchini:2012ig,Taracchini:2013rva}
\begin{align}
N_{22} &= \left[1 + \left(\frac{p_{r^*}}{r M\Omega}\right)^2\left(a_1^{h_{22}} + \frac{a_2^{h_{22}}}{r} + \frac{a_3^{h_{22}}}{r^{3/2}}\right)\right]\nonumber\\
&\times \exp{\left[ \frac{ip_{r^*}}{r M\Omega}\left(b_1^{h_{22}} + b_2^{h_{22}} p_{r^*}^2\right)\right]}\,,
\end{align}
where $p_{r^*}$ is the conjugate momentum to the tortoise coordinate $r^*$ (see
Ref.~\cite{Pan:2009wj}), and the coefficients $a_i^{h_{22}}$ ($i=1,2,3$) and
$b_i^{h_{22}}$ ($i=1,2$) are fixed by imposing that amplitude, curvature of the
amplitude, GW frequency, and slope of the GW frequency match fits of such
quantities (often referred to as ``input values'') to NR data. This amounts to
solving 2 linear systems of equations, one for the $a_i^{h_{22}}$'s ($i=1,2,3$)
and one for the $b_i^{h_{22}}$'s. In the model, the input values are enforced at
a time $t_{\textrm{peak}}^{22} \equiv t_{\textrm{peak}}^{\Omega} + \Delta
t^{22}_{\textrm{peak}}$, where $t_{\textrm{peak}}^{\Omega}$ is the time when the peak of $\Omega$ occurs and $\Delta t^{22}_{\textrm{peak}}$ is a calibration parameter. The introduction of the time lag $\Delta t^{22}_{\textrm{peak}}$ between peak of orbital frequency and peak of radiation mimics what one observes in the test-particle limit using BH perturbation theory~\cite{Barausse:2011kb,Taracchini:2014zpa,Price:2016ywk}.
The input values are fits to NR that depend on $\nu$ and the variable $\chi$~\cite{Taracchini:2012ig,Taracchini:2013rva},
\be
\chi \equiv \frac{\chi_{\textrm{Kerr}}}{1-2\nu} = \chi_{\textrm{S}} +\frac{ \chi_{\textrm{A}}}{1-2\nu}\,\delta\,.
\label{eq:defchi}
\ee
Explicit formulae for the input values can be found in Appendix~\ref{app:IV}.

The EOB GW radiation-reaction force $\bm{\mathcal{F}}$ is modeled adding up the amplitudes of the factorized modes themselves~\cite{Buonanno:2005xu,Damour:2007xr,Damour:2008gu,Pan:2010hz}
\be
\bm{\mathcal{F}} \equiv \frac{\Omega}{16\pi} \frac{\boldsymbol{p}}{\left|\boldsymbol{L}\right|}\sum_{\ell=2}^{8}\sum_{m=-\ell}^{\ell} m^2\left\vert D_{\textrm{L}} h_{\ell m}^{\textrm{insp-plunge}}\right\vert^2\,,\label{RR}
\ee
where $D_{\textrm{L}}$ is the luminosity distance of the BBH to the observer. Whereas in previous versions of the EOBNR model the NQC factor $N_{22}$ was included in the computation of the radiation-reaction force $\bm{\mathcal{F}}$, here we only apply it to the $(2,2)$ waveform after the orbital dynamics has been computed without this factor. This has the advantage of speeding up the generation of waveforms, since it avoids the costly procedure of determining the NQC coefficients in an iterative manner, as outlined in Ref.~\cite{Taracchini:2012ig}, while still providing the correct merger signal.

Reference~\cite{Buonanno:2005xu} provided an algorithm to set up inspiraling quasicircular initial conditions for a generic BBH within the EOB approach. One can then numerically evolve such initial conditions by solving Hamilton's equations for $H_{\textrm{EOB}}$, supplemented by the nonconservative force in Eq.~(\ref{RR}). The evolution is carried out until the light-ring (or photon-orbit) crossing.

\subsection{Merger-ringdown waveforms}
\label{subsec:RDmodel}
The description of the ringdown (RD) signal differs significantly from that of
Refs.~\cite{Taracchini:2012ig,Taracchini:2013rva}. Here, instead of employing a
linear combination of quasinormal modes (QNMs) of the remnant BH that forms
after merger, we use a simple analytic ansatz in the spirit of
Refs.~\cite{Baker:2008mj,Damour:2014yha,Nagar:2016iwa}. We use the full catalog
of NR waveforms presented in Sec.~\ref{sec:NR} (including the Teukolsky
waveforms described in Sec.~\ref{subsec:Teukolskywaveforms}) to determine the
free coefficients in the model. A detailed study of the accuracy of the
phenomenological model that we present below as well as comparison with the
model of Ref.~\cite{Nagar:2016iwa} is presented in Appendix~\ref{app:newRD}.
Here, we simply summarize the main conclusions, namely that our model allows us to faithfully reproduce the ringdown signal of NR waveforms across the NR catalog: replacing the NR ringdown by our model and computing the mismatch against the original NR waveform at a total mass such that the peak of the waveform is at 50\,Hz (so that the ringdown falls in the most sensitive frequency band of the detector), we obtain values below $0.001$ across the NR catalog with typical values around $10^{-4}$ (see Fig.~\ref{fig:rdcomparisonshist} below). By comparison, the distribution obtained with the model presented in Ref.~\cite{Nagar:2016iwa} (which was only calibrated to a subset of the NR catalog used here) peaks close to $0.001$ and features a tail extending above $1\%$.

The RD waveform is attached to the inspiral-plunge waveform at its amplitude peak, that is at time $t_{\textrm{match}}^{22} \equiv t_{\textrm{peak}}^{22}$, where the NQC correction guarantees agreement with the NR input values. For $t \geq t_{\textrm{match}}^{22}$, we define
\be
\label{eq:RDansatz}
h^{\textrm{merger-RD}}_{22}(t) = \nu\, \tilde{A}_{22}(t)\,e^{i \tilde{\phi}_{22}(t)}\,e^{-i \sigma_{220} (t-t_{\textrm{match}}^{22})}\,,
\ee
where $\sigma_{220}$ is the least-damped QNM of the BH that forms after merger. We denote $\sigma^R \equiv \textrm{Im}\, \sigma_{220} < 0$ and $\sigma^I \equiv -\textrm{Re}\, \sigma_{220} < 0$. The value of $\sigma_{220}$ is computed from Ref.~\cite{Berti:2005ys} using the final mass and spin of the remnant. These, in turn, are computed using fitting formulae that connect the masses and spins of the initial BBH to the properties of the final object. In particular, we employ the same final mass formula of Ref.~\cite{Taracchini:2013rva}, which is based on Refs.~\cite{Barausse:2012qz,Hemberger:2013hsa}. We use the final spin formula of Ref.~\cite{Hofmann:2016yih}, which collected 619 NR simulations available in the literature.

The amplitude term $\tilde{A}_{22}$ and the phase term $\tilde{\phi}_{22}$ are simple analytic ans\"{a}tze described below with free coefficients fitted to our catalog of NR simulations.

Finally, the complete inspiral-merger-ringdown signal is given by
\begin{align}
h_{22}(t) &= h_{22}^{\textrm{insp-plunge}}(t)\theta{(t_{\textrm{match}}^{22}-t)}\nonumber\\
&+ h^{\textrm{merger-RD}}_{22}(t)\theta{(t-t_{\textrm{match}}^{22})}\,.
\end{align}

\begin{figure*}[!ht]
\includegraphics[width=.85\columnwidth]{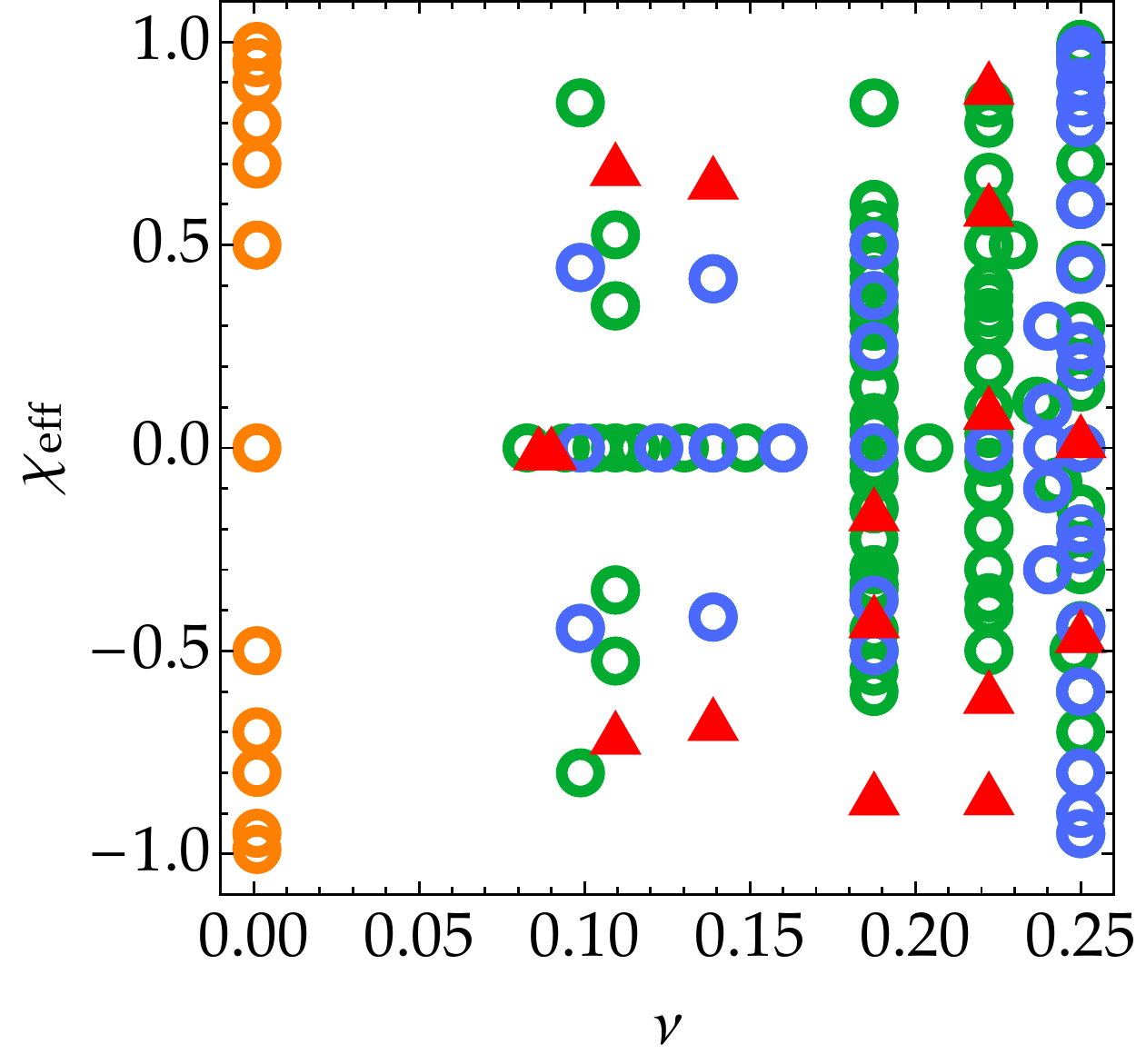}
\includegraphics[width=.85\columnwidth]{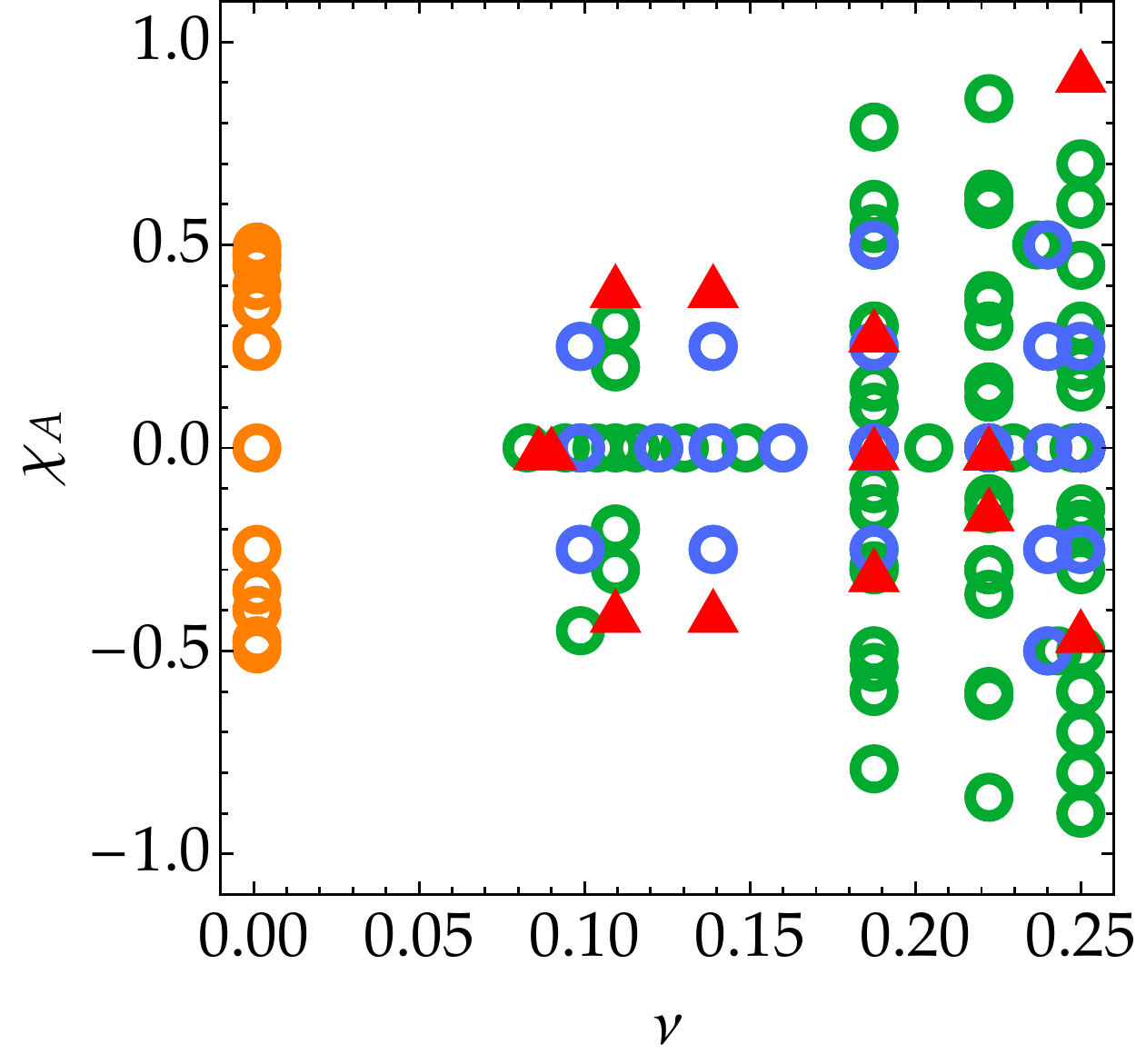}
\raisebox{1.2\height}{\includegraphics[width=.35\columnwidth]{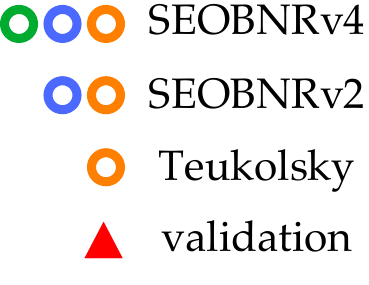}}
\caption{\label{fig:calibspacechieff} Numerical-relativity~\cite{Mroue:2013xna,Chu:2015kft,Kumar:2015tha,Lovelace:2010ne,Scheel:2014ina,Husa:2015iqa} and Teukolsky~\cite{Taracchini:2014zpa} waveforms used for calibration and validation of the EOBNR waveform model. We project the 3D parameter space of spinning, nonprecessing waveforms using the symmetric mass ratio $\nu$ and two BH spin combinations: $\chi_{\textrm{eff}}\equiv  (m_1 \chi_1+m_2 \chi_2)/M$ and $\chi_{\textrm{A}} \equiv (\chi_1 - \chi_2)/2$. We note the better coverage of the large, positive $\chi_{\textrm{eff}}$ region and of the $\chi_{\textrm{A}}$ dimension in the NR catalog used to calibrate the EOBNR model of this paper (\texttt{SEOBNRv4}) with respect to the one used in the EOBNR model of Ref.~\cite{Taracchini:2013rva} (\texttt{SEOBNRv2}). Red triangles indicate NR waveforms used for validation of the calibration.}
\end{figure*}

\subsubsection{Merger-ringdown amplitude}

To model the RD amplitude, we use the same ansatz of Ref.~\cite{Damour:2014yha}
\be
\label{eq:ampansatz}
\tilde{A}_{22}(t) \equiv c_1^c \tanh\left[c_1^f (t - t_{\textrm{match}}^{22}) + c_2^f\right] + c_2^c\,.
\ee
The superscripts $c$ stand for ``constrained'' as these coefficients are fixed by imposing that the amplitude is of class $C^1$ at the attachment point, while the superscripts $f$ stand for ``free'' and correspond to coefficients that will be fitted to NR simulations. Requiring that $|h_{22}|$ is of class $C^1$ at $t=t^{22}_{\textrm{match}}$ allows us to express $c_1^c$ and $c_2^c$ as functions of $c_1^f$, $c_2^f$, $\sigma^R$, $|h_{22}^{\textrm{insp-plunge}}(t^{22}_{\textrm{match}})|$, and $\partial_t |h_{22}^{\textrm{insp-plunge}}|(t^{22}_{\textrm{match}})$. Note that the last two values match the NR input values thanks to the NQC corrections to the merger waveform. In particular, $\partial_t |h_{22}^{\textrm{insp-plunge}}|(t^{22}_{\textrm{match}}) = 0$ since the attachment is done at the amplitude peak of the inspiral-plunge waveform.

After plugging these constraints in Eq.~\eqref{eq:ampansatz}, we are left with a function of the two free parameters $c_{1,2}^f$, which can be determined for each point in parameter space $(\nu,\chi_1,\chi_2)$ where we have a NR waveform by performing a least-squares fit. The resulting values are well represented using simple polynomial fits
\bea
\label{eq:ampcoeffs}
c_1^f =&-&0.0893454\, \nu ^2 +0.0612892\,\nu +0.00146142\,\nu  \chi \nonumber\\
 &-&0.0136459\, \chi ^2-0.0196758\, \chi +0.0830664\,,\\
c_2^f =&-&1.82173 \,\nu ^2-5.25339\, \nu ^2 \chi +2.40203 \,\nu  \chi \nonumber\\
&+&1.39777 \,\nu -0.371365\, \chi -0.623953.
\eea
Note that the amplitude is only smooth up to its first derivative at the attachment point.

\subsubsection{Merger-ringdown phase}

To model the RD phase, we find that a sligthly simplified version of the ansatz proposed in Ref.~\cite{Damour:2014yha}, namely
\be
\tilde{\phi}_{22}(t) = \phi_0 - d_1^c \log\left[\frac{1+d_2^f e^{-d_1^f (t-t_{\textrm{match}}^{22})}}{1+d_2^f}\right]\,,
\ee
is sufficient to accurately reproduce the NR waveforms. We impose that the phase is $C^1$ at the attachment point (i.e., the GW frequency is $C^0$). This allows us to express $d_1^c$ in terms of $d_1^f$, $d_2^f$, $\sigma^I$, and the GW frequency at the attachment point $\omega_{22}^{\textrm{insp-plunge}}(t_{\textrm{match}}^{22})$. Here, $\phi_0$ trivially corresponds to the phase of the inspiral-plunge waveform at $t_{\textrm{match}}^{22}$. Note that $\omega_{22}^{\textrm{insp-plunge}}(t_{\textrm{match}}^{22})$ is set equal to the corresponding NR input value by the NQC corrections. Values for $d_{1,2}^f$ are then obtained for each NR waveform in the catalog using a least-squares fit. The resulting values are again well represented by simple polynomials
\bea
d_1^f =&-&0.808987\, \nu ^2 +0.263456\, \nu -0.120853\, \nu  \chi\nonumber \\
&-&0.0244358\, \chi ^2+0.00779176\, \chi +0.147584\,,\\
d_2^f =&+&17.5646\, \nu ^2 -6.99396\, \nu -9.61861\, \nu  \chi\nonumber\\
&+&0.581626\, \chi ^2+3.13067\, \chi +2.46654\,.
\eea
Note that these expressions guarantee a monotonic evolution of the frequency after the attachment point.

In Appendix~\ref{app:newRD} we discuss the accuracy of this phenomenological RD model and compare it with the RD
model of Ref.~\cite{Nagar:2016iwa}.

\section{Numerical-relativity and black-hole-perturbation-theory waveforms}
\label{sec:NR}

To calibrate the inspiral-plunge part of EOB waveforms we use 140 NR waveforms~\cite{Chu:2015kft} generated by the (pseudo) Spectral Einstein code (\code{SpEC}) of the Simulating eXtreme Spacetime (SXS) project, and 1 NR waveform~\cite{Husa:2015iqa} produced by the \code{BAM} code. We also incorporate information from merger waveforms computed in BH perturbation theory~\cite{Barausse:2011kb,Taracchini:2014zpa}. After calibration, 
we further assess the accuracy of the EOBNR waveforms by comparing them to 4 waveforms produced by the \code{SpEC} code
and 2 by the \code{Einstein Toolkit} code. Those waveforms were generated for this paper. 

In Fig.~\ref{fig:calibspacechieff} we show, in the intrinsic BBH parameter space, the location 
of the NR and BH-perturbation-theory waveforms employed to build the new EOBNR model (\texttt{SEOBNRv4}) 
and the previous EOBNR model (\texttt{SEOBNRv2}), as well.  

\subsection{Numerical-relativity methods}

SpEC~\cite{SpECwebsite,Scheel:2006gg,Szilagyi:2009qz,Buchman:2012dw}
is a pseudospectral code capable of efficiently solving many types of
elliptic and hyperbolic differential equations, with the primary goal
of modeling compact-object binaries.  For smooth problems, spectral
methods are exponentially convergent and high accuracy can be achieved
even for long simulations.  \code{SpEC} evolves the first order
formulation~\cite{Lindblom:2006} of the generalized harmonic
formulation of Einstein's
equations~\cite{Friedrich1985,Pretorius2005a}. The damped harmonic
gauge~\cite{Lindblom:2009tu} is used to provide stable coordinate
conditions. Singularities inside BHs are dynamically excised
from the computational domain using feedback control systems
~\cite{Hemberger:2012jz,Scheel:2014ina}. \code{SpEC} uses h-p adaptivity 
to dynamically control numerical truncation error and to increase
computational efficiency~\cite{Szilagyi:2014fna}. Waveforms are
extracted using the Reggie-Wheeler-Zerilli formalism
on a series of coordinate spherical shells and extrapolated to null
infinity using polynomial expansions in powers of the areal
radius~\cite{Boyle:2009vi}.

\begin{table*}
\begin{ruledtabular}
\begin{tabular}{ccccccc}
ID & $q$ & $\chi_1$ & $\chi_2$ & $e$ & $M\omega_{22}$ & $N_\mathrm{orb}$\\
\hline
SXS:BBH:0610 & $1.2$ & $-0.50$ & $-0.50$ & $7.4\times 10^{-5}$ & $0.01872$ & $12.1$ \\
SXS:BBH:0611 & $1.4$ & $-0.50$ & $+0.50$ & $6.0\times 10^{-4}$ & $0.02033$ & $12.5$ \\
SXS:BBH:0612 & $1.6$ & $+0.50$ & $-0.50$ & $3.7\times 10^{-4}$ & $0.02156$ & $12.8$ \\
SXS:BBH:0613 & $1.8$ & $+0.50$ & $+0.50$ & $1.8\times 10^{-4}$ & $0.02383$ & $13.1$ \\
SXS:BBH:0614 & $2.0$ & $+0.75$ & $-0.50$ & $6.7\times 10^{-4}$ & $0.02355$ & $13.1$ \\
SXS:BBH:0615 & $2.0$ & $+0.75$ & $+0.00$ & $7.0\times 10^{-4}$ & $0.02401$ & $13.3$ \\
SXS:BBH:0616 & $2.0$ & $+0.75$ & $+0.50$ & $8.0\times 10^{-4}$ & $0.02475$ & $13.3$ \\
SXS:BBH:0617 & $2.0$ & $+0.50$ & $+0.75$ & $7.8\times 10^{-4}$ & $0.02342$ & $13.1$ \\
SXS:BBH:0618 & $2.0$ & $+0.80$ & $+0.80$ & $5.9\times 10^{-4}$ & $0.02578$ & $13.4$ \\
SXS:BBH:0622 & $8.0$ & $-0.90$ & $+0.00$ & $1.1\times 10^{-3}$ & $0.01559$ & $28.0$ \\
\hline
SXS:BBH:0620 & $5.0$ & $-0.80$ & $+0.00$ & $3.4\times 10^{-3}$ & $0.02527$ & $8.2$ \\
SXS:BBH:0621 & $7.0$ & $-0.80$ & $+0.00$ & $3.2\times 10^{-3}$ & $0.02784$ & $7.1$ \\
SXS:BBH:0619 & $2.0$ & $+0.90$ & $+0.90$ & $2.9\times 10^{-4}$ & $0.02520$ & $13.5$ \\
ET:AEI:0001 & $5.0$ & $+0.80$ & $+0.00$ & $9.2\times 10^{-4}$ & $0.03077$ & $10.5$ \\
ET:AEI:0002 & $7.0$ & $+0.80$ & $+0.00$ & $6.1\times 10^{-4}$ & $0.03503$ & $10.4$ \\
ET:AEI:0004 & $8.0$ & $+0.85$ & $+0.85$ & $3.0\times 10^{-3}$ & $0.04368$ & $7.4$ \\
\end{tabular}

\end{ruledtabular}
\caption{We display the binary configurations of the new NR simulations produced for
  this paper.  Those used for calibration of the new model are shown first, followed by those used for validation, separated by a horizontal line.
  Shown are the mass ratio $q=m_1/m_2$, the dimensionless spins
$\chi_{1,2}=(\boldsymbol{S}_{1,2}\cdot\boldsymbol{\hat{L}}_N)/m_{1,2}^2$, the eccentricity
$e$, the initial frequency of the $(\ell,m)=(2,2)$ mode of the waveform strain $\omega_{22}$, and the
number of orbits (up to the waveform peak) $N_{\textrm{orb}}$.  All quantities are measured at an
early time after the effects of junk radiation are no longer important.}
\label{table:nrsims}
\end{table*}

The \code{Einstein Toolkit}~\cite{Loffler:2011ay} is a collection of open source NR 
components built around the \code{Cactus} framework~\cite{Goodale:2002a}. The initial
data is computed in the Bowen-York
formalism~\cite{Bowen:1980yu,Brandt:1997tf} using \code{TwoPunctures}~\cite{Ansorg:2004ds}, with low eccentricity parameters determined
through our implementation of \cite{Pfeiffer:2007yz} for the Einstein
Toolkit. The time evolution is performed in the
BSSN~\cite{Nakamura:1987zz, Shibata:1995we, Baumgarte:1998te}
formulation of the Einstein equations using
\code{McLachlan}~\cite{Brown:2008sb}, and the BHs are evolved with
the coordinate conditions of the moving-puncture
method~\cite{Baker:2005vv,Campanelli:2005dd} using 8th order accurate
finite differencing.  Adaptive mesh refinement, in which regions of
high resolution follow the BHs, is provided by
\code{Carpet}~\cite{Schnetter:2003rb}.  The near zone is computed using Cartesian grids, and the wave zone is computed on
spherical grids using the \code{Llama} multipatch
infrastructure~\cite{Pollney:2009yz}, enabling the efficient
computation of high-accuracy waveforms at large distances from the
source.  Apparent horizons are computed using
\code{AHFinderDirect}~\cite{Thornburg:2003sf} and spins are computed in the
dynamical horizon formalism using
\code{QuasiLocalMeasures}~\cite{Dreyer:2002mx}.  Gravitational waves are
computed using \code{WeylScal4}, and the GW strain $h$ is computed from the Newman-Penrose curvature component
$\Psi_4$ at finite radius $r \in [100\,M,500\,M]$ using fixed-frequency integration~\cite{Reisswig:2010di} with a cutoff frequency
equal to 3/4 the initial waveform frequency, and extrapolated
to $\mathcal{J}^+$ using second and first order extrapolation for the amplitude and phase respectively.
\code{WeylScal4} and \code{McLachlan} are both generated using the
\code{Kranc}~\cite{Husa:2004ip} automated-code-generation package.
Simulations are managed using the \code{Simulation
Factory}~\cite{Thomas:2010aa}, and the BBH evolution
parameters are based on the open source \code{Einstein Toolkit} GW150914
example~\cite{wardell_barry_2016_155394}.  Analysis and postprocessing
is performed using the open-source
\code{SimulationTools}~\cite{SimulationToolsWeb} for Mathematica.

The \code{BAM} code~\cite{Bruegmann:2006at,Husa:2007hp} uses broadly the same
methods as the \code{Einstein Toolkit} but, with the exception of the
\code{TwoPunctures} initial data solver, was developed independently. The
spatial derivatives are discretized using 6th order accurate finite
differencing, and the wave zone is computed on Cartesian grids.

\subsection{Numerical-relativity waveforms}

We use a total of \totalsims{} NR waveforms: 141 for calibrating the model and 16 for validation. The full list can be found in Appendix \ref{app:NRsims}, which contains separate tables for the different data sets that constitute our catalog. Here, we give a brief description of each set, denoting 
with $\{C,V \} = \{141,16\}$ the number of simulations used for calibration and validation in each of them.

A first set of $\{38,0\}$ waveforms belongs to the original SXS public catalog 
\cite{SpECwebsite,Mroue:2013xna}, which was also used to calibrate the previous EOBNR
model~\cite{Taracchini:2013rva}. Additional sets of $\{6,0\}$ long waveforms (between $36$ and $88$ orbits) with mass ratios $5$ and $7$ and spins on the largest BH of $\chi_1=\pm0.4$ or $\pm0.6$ (and no spin on the companion)~\cite{Kumar:2015tha} and of $\{2,1\}$ near equal-mass and near extremal spins~\cite{Lovelace:2010ne,Scheel:2014ina} were subsequently added to our catalog. Another set of $\{84,10\}$ SXS
waveforms were produced in the last few years and are described in
Ref.~\cite{Chu:2015kft}. These waveforms extend the
coverage in the region of parameter space with $1 \le q\le 3$ to systems where both spin magnitudes go up to $0.85$, including many antisymmetric spin configurations. The length of the waveforms ranges between 20 and 40 orbits.

Furthermore, $\{9, 4\}$ new SXS waveforms were produced for this paper and are
summarized in Table \ref{table:nrsims}. These waveforms can be broadly
divided into 2 categories: those filling in the gaps of coverage in
the aligned-spin catalogue~\cite{Chu:2015kft} ($1<q\le2,\ 0.5\leq|\chi_{1,2}|\leq0.9$), and
those extending coverage to even higher mass ratios $(5\leq q\leq7)$ for single spin binaries where the largest BH has a large anti-aligned spin ($-0.9\leq\chi_{1}\leq -0.8$). We also use
for calibration one waveform with physical parameters
$(q,\chi_1,\chi_2)=(8,0.85,0.85)$ that was previously produced using
\code{BAM}~\cite{Bruegmann:2006at,Husa:2007hp} and it was employed to calibrate the 
\texttt{IMRPhenomD} model~\cite{Khan:2015jqa}. However, this waveform was available at
only one resolution, was not extrapolated to infinity, and had a relativity high
eccentricity of $1.2 \times 10^{-2}$.  In order to check that these effects were not dominant, we
also produced a waveform for this configuration, listed as ET:AEI:0004 in Table \ref{table:nrsims}, with the \code{Einstein Toolkit} including multiple resolutions, extrapolation, and with a lower eccentricity $3.0 \times 10^{-3}$. We have found excellent agreement between the waveforms 
produced with the two codes, and negligible
effects due to resolution or extrapolation errors when comparing with the EOBNR waveform.  We have also employed the \code{Einstein Toolkit} to
produce two $(q,\chi_1,\chi_2)=(5,0.8,0)$ and $(7,0.8,0)$ waveforms
listed as ET:AEI:0001 and
ET:AEI:0002 in Table \ref{table:nrsims}.  The new waveforms are
between 7 and 28 orbits in length and used for validation.

\subsection{Merger-ringdown waveforms from black-hole perturbation theory}
\label{subsec:Teukolskywaveforms}

Although NR is currently capable of accurately simulating the full coalescence of BBHs at moderately large mass ratios and spins (up to $\sim 20$ and $0.8$~\cite{Husa:2015iqa}), evolutions at even higher mass ratios and larger spin magnitudes are still not tractable within this framework. Nonetheless, for such systems it is possible to extract valuable information from BH perturbation theory. In particular, the  merger-RD GW emission of test particles inspiraling and merging into a BH can be modeled by numerical solutions of the Regge-Wheeler-Zerilli~\cite{Regge:1957td,Zerilli} (Teukolsky~\cite{Teukolsky:1973ha}) equation, describing metric (curvature) perturbations to Schwarzschild (Kerr) spacetimes. Several studies in recent years~\cite{Bernuzzi:2010xj,Barausse:2011kb,Taracchini:2014zpa} have employed time-domain Teukolsky codes sourced by particles on plunging, equatorial trajectories to compute the dominant and leading subdominant multipolar modes of the merger-RD waveforms.

In this paper, we employ the Teukolsky waveforms of Ref.~\cite{Taracchini:2014zpa} to build the test-particle-limit fits to the input values (see Appendix~\ref{app:IV}). This means that we extract amplitude, curvature of the amplitude, GW frequency, and slope of the GW frequency of the $(2,2)$ mode at its peak. Teukolsky waveforms are also used in the construction of the phenomenological RD model described in Sec. \ref{subsec:RDmodel}. Note that we cannot exploit the inspiral portion of Teukolsky waveforms because of the approximations that are involved in the modeling of the perturbing trajectory. In fact, Ref.~\cite{Taracchini:2014zpa} only used the conservative dynamics of geodesics in Kerr spacetime and the dissipation provided by leading-order BH perturbation theory.

\section{Calibration of inspiral-plunge parameters to numerical-relativity waveforms}
\label{sec:cali}

Given a point in BBH parameter space $(m_1, m_2, \chi_1, \chi_2)$, the EOB model described in Sec.~\ref{sec:model} provides an inspiral-merger-ringdown $(2,2)$-mode waveform that depends on four inspiral-plunge calibration parameters. These parameters are: (i) a parameter $K$ that determines the position of the EOB horizon $r_+^{\textrm{EOB}}$ and the shape of the radial potential $\Delta_u$ (see Eq.~(\ref{Apotential}))
in the strong-field region below and around the ISCO, (ii) a 4.5PN spin-orbit parameter $d_{\textrm{SO}}$ that enters the EOB spin mapping between 
the real and effective descriptions, (iii) a 3PN spin-spin parameter $d_{\textrm{SS}}$ that enters the EOB spin mapping, and (iv) a parameter $\Delta t_{\textrm{peak}}^{22}$ that determines the time delay between the peak of orbital frequency and the peak of radiation. In this Section, we describe how these four parameters are tuned to NR waveforms.

\subsection{Calibration requirements}
\label{subsec:calireq}

The goal of the calibration is to obtain EOB waveforms that can be employed in the analysis of current ground-based interferometric data with negligible impact from mismodeling errors, at least in the region of parameter space where BBH configurations have been simulated in full NR. Two are the main applications of waveform approximants, namely template-based detection pipelines and parameter estimation of GW sources --- including parametrized tests of general relativity. Template banks~\cite{TheLIGOScientific:2016qqj,Capano:2016dsf,TheLIGOScientific:2016pea} that are employed by LIGO-Virgo in matched-filtering searches of binary coalescences are built requiring that the loss in signal-to-noise ratio due to the discrete nature of the bank is smaller than 3\%~\cite{TheLIGOScientific:2016qqj}, which translates into a loss in detections smaller than 10\%. On the other hand, in the context of parameter estimation, the correct assessment of biases due to waveform modeling inaccuracies requires a full-fledged Bayesian inference.
In fact, no simple waveform accuracy requirements can be formulated and the criterion of indistinguishability proposed in 
Ref.~\cite{Lindblom:2009ux} is a sufficient, but not necessary criterion, and it has been shown to be highly
restrictive~\cite{Lindblom:2010mh,Littenberg:2012uj}. Here, we do not aim at addressing the question of whether our EOBNR waveform
model is completely free of biases. As done in previous studies~\cite{Pan:2011gk,Pan:2013rra,Taracchini:2013rva,Babak:2016tgq}, 
we adopt the simplified criterion of requiring that the EOBNR waveforms
have matches with NR waveforms --- in the sense specified below --- above 99\% when the optimization is done only on a global phase
and time shift.

Given two waveforms $h_1(t)$ and $h_2(t)$, their noise-weighted overlap or match is defined as~\cite{Finn:1992wt}
\be
( h_1 | h_2 ) \equiv 4\, \textrm{Re} \int_{f_l}^{f_h}
\frac{\tilde h_1(f) \tilde h^*_2(f)}{ S_n(f) } {\textrm{d}} f \,,
\label{eq:overlap}
\ee
with $\tilde h_{1,2}(f)$ indicating the Fourier transforms of the waveforms and $S_n(f)$ the one-sided
power spectral density (PSD) of the detector noise. The \emph{faithfulness} is then defined as the overlap between the normalized waveforms maximized over relative time and phase shifts
\be
\langle h_1 | h_2 \rangle = \max_{\phi_c, t_c} \frac{
    \left(h_1 (\phi_c, t_c)~|~h_2\right)}{\sqrt{ (h_1|h_1) (h_2|h_2)}} \,.
  \label{eq:match}
\ee
Another useful notion is that of \emph{effectualness}, defined as the maximum faithfulness of a waveform against a template bank. This amounts to maximizing the faithfulness over a discrete set of intrinsic physical parameters. For the calibration of the model to NR, we use the design zero-detuned high-power noise PSD of Advanced LIGO~\cite{Shoemaker:2010}. We choose $f_l$ as the starting GW frequency after the junk radiation has settled in the NR
simulation,\footnote{If the starting frequency is lower than 10\,Hz, we use
$f_l=10$\,Hz instead.} and  $f_h = 3\,{\rm kHz}$.  We taper the waveforms in the time domain (before transforming to the frequency domain) using a hyperbolic-tangent window function to reduce spectral leakage~\cite{McKechan:2010kp}. Let
\be
\bm{\theta} \equiv \left\{ K,d_{\textrm{SO}}, d_{\textrm{SS}}, \Delta
t^{22}_{\textrm{peak}} \right\}
\ee
denote the set of inspiral-plunge calibration parameters, and
\be
\boldsymbol{\lambda} \equiv \left \{m_1, m_2, \chi_1, \chi_2 \right\}
\ee
the set of intrinsic BBH parameters. In practice, the intrinsic parameter space is only 3-dimensional $(q, \chi_1, \chi_2)$, because BBH waveforms scale trivially with the total mass $M$. Since we work with dominant-mode nonprecessing waveforms, we perform all calculations omitting extrinsic BBH parameters (such as inclination, sky location, polarization, etc.). We denote the faithfulness of $h_{\textrm{EOB}}$ to $h_{\textrm{NR}}$,
at given values of the calibration parameters $\bm{\theta}$, as
\be
{\cal M}(\bm{\theta}) = \langle
    h_{\textrm{EOB}} (\bm{\lambda}; \bm{\theta}) | h_{\textrm{NR}}
    (\bm{\lambda}) \rangle.
\ee
Note that the comparison is done between waveforms with the same intrinsic parameters $\bm{\lambda}$. The \emph{unfaithfulness} is defined as $\bar{\cal M}(\bm{\theta}) \equiv 1 - {\cal M}(\bm{\theta})$.

To guide the waveform calibration, we design a figure of merit which, for each NR waveform in the catalog, is a function of (i) the faithfulness with the corresponding EOBNR waveform, and (ii) the difference $\delta t_{\textrm{peak}}^{22}(\bm{\theta})$ of the merger time (measured after low-frequency phase alignment between EOB and NR waveforms). We use the time when the amplitude peaks as a proxy for the merger time. For each configuration in the NR catalog, our goal is to find values of $\boldsymbol{\theta}$ such that $\mathcal{M} \geq 99 \%$ and $|\delta t_{\textrm{peak}}^{22}| \leq 5\,M$. Note that the requirement on $\delta t_{\textrm{peak}}^{22}$ aims at taming time-domain dephasings at merger, something to which $\mathcal{M}$ is not very sensitive.

\subsection{Markov-chain Monte Carlo analysis}
\label{subsec:MCMC}

Given the dimensionality of our calibration parameter space, a naive approach aiming at covering it with a regular grid is highly inefficient and not feasible in practice. One alternative, used in previous calibrations of the model \cite{Pan:2011gk,Taracchini:2013rva}, is to resort to local optimization algorithms (such as the numerical simplex method) which can efficiently converge to minima of our figure of merit (and to minima with values of the figure of merit satisfying our calibration requirements provided that good initial conditions are chosen). This however only provides us with best-fit values for each numerical configuration but with no notion of how much we can deviate from those values without degrading the figure of merit below some threshold. In the present work, we aim at using more information on the structure of our calibration space, and in particular on the correlations between our calibration parameters.

In general, Markov-chain Monte Carlo (MCMC) methods are well suited to exploring high-dimensional parameter spaces with limited computational costs~\cite{MacKay:2003}. Here, we employ the \texttt{emcee}\footnote{\url{http://dan.iel.fm/emcee}}~\cite{ForemanMackey:2012ig} package, which is a \texttt{Python} implementation of an affine-invariant MCMC ensemble
sampler~\cite{Goodman:2010}.  This algorithm has better performance over
traditional MCMC sampling methods (e.g., the traditional Metropolis-Hasting
method), as measured by the smaller autocorrelation time and fewer hand-tuning
input parameters.  It transforms the sampling of the parameter space by an
affine transformation such that the internal algorithm samples an isotropic
density, so it is insensitive to covariances among parameters. This is achieved
by the ``stretch move'', that simultaneously evolves an ensemble of walkers,
and determines a walker's next proposal distribution (i.e., the next possible
move) by current positions of the other walkers in the complementary ensemble (for more details,
see Refs.~\cite{Goodman:2010,ForemanMackey:2012ig}).

For each NR simulation, we want to obtain a posterior distribution in $\bm{\theta}$-space whose mean and
variance (and mutual correlations between the $\theta^j$'s) relate to the calibration requirements described in Sec.~\ref{subsec:calireq}. In the MCMC sampler, we need to assign the probability to accept a possible
move of the $k$-th walker from an old position $\bm{\theta}_k^{\textrm{(old)}}$ to a new position, $\bm{\theta}_k^{\textrm{(new)}} = \bm{\theta}_j +
Z\left[\bm{\theta}_k^{\textrm{(old)}} - \bm{\theta}_j\right]$, with the $j$-th walker randomly drawn from the remaining walkers, and $Z$ a random variable drawn from a distribution $g(z)$ (whose expression is given in Eq.~(10) of Ref.~\cite{ForemanMackey:2012ig}).  To satisfy the detailed balance condition, the probability for the move is $\min \left[ 1, Z^{N-1} P\left(\bm{\theta}_k^{\textrm{(new)}}\right)/P\left(\bm{\theta}_k^{\textrm{(old)}}\right) \right]$, with $N$ the dimension of the parameter space ($N=4$ for our $\bm{\theta}$-space), and $P(\bm{\theta})$ the
likelihood function. For each NR run, we choose the likelihood function to be,
\begin{equation}
  P(\bm{\theta}) \propto \exp \left[ -\frac{1}{2}\left( \frac{\bar{\cal M}_{\rm
  max}(\bm{\theta})}{\sigma_{\cal M}}\right)^2 - \frac{1}{2}\left(
    \frac{\delta t^{22}_{\textrm{peak}}(\bm{\theta})}{\sigma_t}
  \right)^2 \right] \,,
  \label{eq:like}
\end{equation}
where $\bar{\cal M}_{\rm max}(\bm{\theta})$ is, for a given $\bm{\theta}$, the
maximum unfaithfulness of EOB to NR over the total mass range
$10\,M_\odot \leq M \leq  200\,M_\odot$, $\sigma_{\cal M}$ is chosen to be $1\%$, and
$\sigma_t$ is chosen to be $5\,M$, consistently with our calibration requirements.

We carry out the calibration employing the 140 SXS NR waveforms plus 1 \code{BAM} NR waveform, as presented in Sec.~\ref{sec:NR}.
Furthermore, after calibration we use 6 NR waveforms with parameters $(q,\chi_1,\chi_2) =$ $(1.3, 0.96, -0.9)$, $(2, 0.9,0.9)$, $(5, -0.8,
0)$, $(5, 0.8, 0)$, $(7, -0.8, 0)$, $(7, 0.8, 0)$ to test and validate the EOBNR waveform model.
Initial values of $\boldsymbol{\theta}$ were constructed with the
help of a coarse grid for each NR waveform.  For each NR simulation, we used 44 walkers and accumulated
$\sim40,000$ points, a number large enough to result in well sampled posteriors while keeping the computational cost manageable.

For each chain in $\boldsymbol{\theta}$-space, we discard the first half of the
points, as the burn-in phase of the MCMC run~\cite{Brooks:2011}. We also
discard points with $\bar{\cal M}_{\rm max} > 1\%$ and $|\delta
t^{22}_{\textrm{peak}}| > 5\,M$.  The 2D projections of these 4D distributions
were then examined by eye. Some cases featured a secondary mode, notably
in the parameter $K$. To simplify the analysis presented in Sec.~\ref{sec:inter:extra}  where a multidimensional Gaussian distribution is
assumed, some modes are pruned away by hand. For example,
we consistently keep the points in the chains that correspond to smaller $K$ if a secondary
(higher) mode exists.  From the remaining points, we extract the vector of the means $\langle
\bm{\theta} \rangle_{(n)}$ and the covariance matrices ${\cal C}_{(n)}$, where $n$
labels the NR simulation.
To check that simply taking the mean of each 1D posterior provides a good calibration point for each configuration, we compute the faithfulness
between the EOBNR and NR waveforms using $\boldsymbol{\theta}=\langle \bm{\theta} \rangle_{(n)}$ and
find a worst value over the catalog $\gtrsim 99.5\%$. Thus, in the next Section we fit the means $\langle
\bm{\theta} \rangle_{(n)}$ --- using error bars constructed from the covariances ${\cal C}_{(n)}$ --- to obtain expressions for our calibration parameters as functions of the physical parameters.

\subsection{Interpolation and extrapolation to the entire BBH parameter
space}\label{sec:inter:extra}

We now discuss how we interpolate between and extrapolate away from the 141 BBH configurations spanned by the NR catalog.  We want $K$, $d_{\textrm{SO}}$, $d_{\textrm{SS}}$, and $\Delta t_{\textrm{peak}}^{22}$ to be prescribed functions of $(q,\chi_1,\chi_2)$ that best fit the means $\langle \bm{\theta} \rangle_{(n)}$. For simplicity, we only consider polynomial fitting functions that depend on $\nu$ and the spin combination $\chi$ (defined in Eq.~\eqref{eq:defchi}), similarly to what was done for fits of the input values and of the ringdown waveforms.

First, we want to fix the nonspinning limit of the calibration parameters. However, this is only possible for $K$ and $\Delta t^{22}_{\textrm{peak}}$, since for nonspinning BBH configurations the EOB waveforms do not depend on $d_{\textrm{SO}}$ and $d_{\textrm{SS}}$. Let ${\cal S}_{\textrm{ns}}$ be the set of 17 nonspinning NR runs that are present in the catalog. Let $\bm{\theta}_{\textrm{ns}}\equiv\left\{K,\Delta t^{22}_{\textrm{peak}}\right\}$ and ${\cal C}_{\textrm{ns}}$ be the $2\times2$ covariance matrix restricted to the $\bm{\theta}_{\textrm{ns}}$-space. We parametrize $K$ and $\Delta t^{22}_{\textrm{peak}}$ with polynomials that are at most cubic in $\nu$. We determine the coefficients of these polynomials by minimizing the following quantity:
\begin{align}
  \chi^2_{\textrm{ns}}&\equiv \sum_{n \in \mathcal{S}_{\textrm{ns}}} \frac{1}{2} \left(
  \bm{\theta}_{\textrm{ns}} - \langle \bm{\theta}_{\textrm{ns}} \rangle_{(n)} \right) ({\cal C}^{-1}_{\textrm{ns}})_{(n)}
  \left( \bm{\theta}_{\textrm{ns}} - \langle \bm{\theta}_{\textrm{ns}} \rangle_{(n)} \right)^{\textrm{T}}\nonumber\\
  &+ \chi^2_{\textrm{TPL}}\,,
  \label{eq:nonspin:chi2}
\end{align}
where the last term enforces that $K$ and $\Delta t^{22}_{\textrm{peak}}$
approach their test-particle limits~\cite{Taracchini:2013rva,Taracchini:2014zpa}, $1.712$ and $-2.5\,M$,
respectively. We obtain that the nonspinning fits of $K$ and $\Delta t^{22}_{\textrm{peak}}$ read
\begin{align}
  \left.K\right|_{\chi=0} = &+267.788247\,  \nu^3 -126.686734\,  \nu^2\nonumber\\
  &+ 10.257281\,\nu  + 1.733598 \,,
  \label{eq:K:nonspin}
\end{align}
and
\begin{align}
\left.\frac{\Delta t^{22}_{\textrm{peak}}}{M}\right|_{\chi=0} =&+716.044155\,  \nu^3 -13.087878\,  \nu^2\nonumber\\
 &-45.883834\,  \nu -2.504992 \,.
  \label{eq:dNQC:nonspin}
\end{align}

\begin{figure*}[tbp]
  \includegraphics[width=\columnwidth]{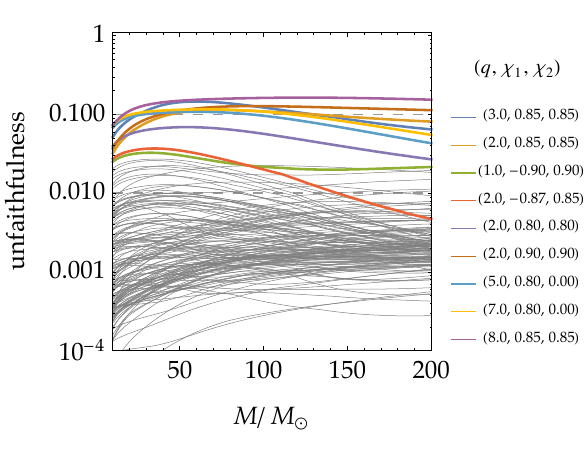}
  \includegraphics[width=\columnwidth]{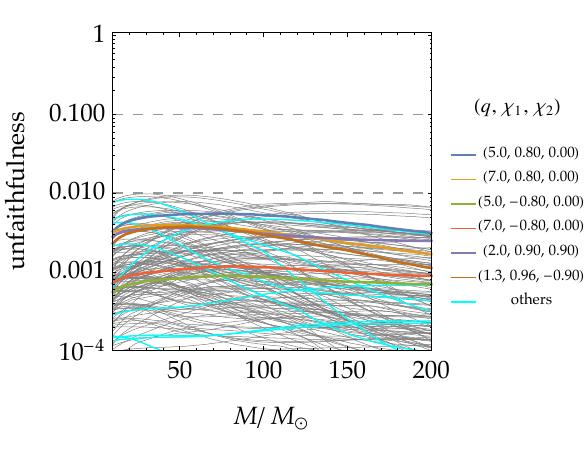}
  \caption{\label{fig:unfaithfulness:v2v4} Unfaithfulness of the
    EOBNR model of Ref.~\cite{Taracchini:2013rva} (\texttt{SEOBNRv2})
    (\emph{left panel}) and the EOBNR model of this paper
      (\texttt{SEOBNRv4}) (\emph{right panel}) against the NR catalog
      for total masses $10\,M_\odot \leq M \leq 200\,M_\odot$, using the Advanced
      LIGO design zero-detuned high-power noise PSD and a
      low-frequency cutoff equal to the initial geometric frequency of each NR
      run. In the left panel, the cases where the maximum unfaithfulness is
      $>3\%$ are highlighted in color and labeled by $( q,\chi_1,\chi_2)$. In
      the right panel, the cases that were not used in the calibration are
      highlighted with colors and 6 cases whose parameters lie close to the boundary of our calibration domain are singled out in the legend. We note that the new EOBNR model
      (\texttt{SEOBNRv4}) has unfaithfulness below 1\% against the \emph{whole}
    NR catalog. }
\end{figure*}

Let us now consider the problem of fitting the means of the calibration parameters over ${\cal S}_{\textrm{spin}}$, the set of spinning NR simulations. We now work in $\bm{\theta}$-space, and parametrize each $\theta^j$ with a polynomial that is at most cubic in $\nu$ and $\chi$, making sure that the nonspinning limits in Eqs.~(\ref{eq:K:nonspin}) and (\ref{eq:dNQC:nonspin}) are satisfied. To determine the coefficients of the fitting polynomials, we devise a quantity to be minimized, $\chi^2_{\textrm{spin}}$, containing three terms: (i) a term that restricts the domain of the four calibration parameters, (ii) a term that penalizes deviations from the test-particle limit of $\Delta t_{\textrm{peak}}^{22}$ (see the $(2,2)$-mode curve in Fig.~13 of Ref.~\cite{Taracchini:2014zpa}), and (iii) a term that depends on the MCMC means and covariances, of the form
\be
    \sum_{n \in {\cal S}_{\textrm{s}}} \frac{w}{2} \left( \bm{\theta} - \langle \bm{\theta}
    \rangle_{(n)} \right) {\cal C}^{-1}_{(n)} \left( \bm{\theta} - \langle
    \bm{\theta} \rangle_{(n)} \right)^{\rm T} \,,
\ee
where $w$ is a weighting function that reads
\be
    w \equiv \chi_1^2 + \chi^2_2 + \frac{|\chi|}{2\nu} \,.
\ee
The introduction of the weighting function $w$ is necessary to
empirically account for the inhomogeneous distribution of NR simulations in the
BBH parameter space, their different length and to improve the faithfulness of the model against NR
waveforms with large aligned-spin components. The minimization of
$\chi^2_{\textrm{spin}}$ was performed with the Nelder-Mead downhill simplex
algorithm~\cite{Nelder:1965zz}, giving
\begin{eqnarray}
  K &=& - 59.165806\,\chi^3\nu^3
      - 0.426958\,\chi^3\nu
      + 1.436589\,\chi^3  \nonumber \\
    &&  + 31.17459\,\chi^2\nu^3
      + 6.164663\,\chi^2\nu^2
      - 1.380863\,\chi^2 \nonumber \\
    &&  - 27.520106\,\chi \nu^3
      + 17.373601\,\chi\nu^2
      + 2.268313\,\chi\nu \nonumber \\
    &&  - 1.62045\,\chi +\left.K\right|_{\chi=0}\,,
  \label{eq:K:spin}
\end{eqnarray}
\begin{eqnarray}
  d_{\textrm{SO}} &=& + 147.481449\,\chi^3\nu^2 - 568.651115\chi^3\nu\nonumber \\
                && + 66.198703\,\chi^3  - 343.313058\,\chi^2\nu \nonumber \\
                 && + 2495.293427\,\chi\nu^2 - 44.532373\,,
  \label{eq:dSO:spin}
\end{eqnarray}
\begin{eqnarray}
  d_{\textrm{SS}} &=& +528.511252\,\chi^3\nu^2 - 41.000256\,\chi^3\nu  \nonumber \\
                 && + 1161.780126\,\chi^2\nu^3  - 326.324859\,\chi^2\nu^2  \nonumber \\
                 && + 37.196389\,\chi\nu + 706.958312\,\nu^3 \nonumber \\
                 && - 36.027203\,\nu + 6.068071\,,
  \label{eq:dSS:spin}
\end{eqnarray}
\begin{eqnarray}
  \frac{\Delta t^{22}_{\textrm{peak}}}{M} &=& - 0.192775\,\chi^3\nu^2  + 19.053803\,\chi^3\nu\nonumber \\
                             && - 11.543497\,\chi^2  + 40.318332\,\chi\nu\nonumber \\
                             &&  - 13.006363\,\chi  + \left. \frac{\Delta
                             t^{22}_{\textrm{peak}}}{M}\right|_{\chi=0} \,.
  \label{eq:dNQC:spin}
\end{eqnarray}
In the above expressions, we notice that not all powers of $\nu$ and $\chi$ up to cubic
order are present. Indeed, we were able to set to zero some of the terms without degrading the
performance of the fit, demonstrating that not all powers are necessary to represent the data. For
example, the expression of $d_{\rm SO}$ in Eq.~(\ref{eq:dSO:spin}) only contains six terms
instead of the sixteen allowed when simply restricting to cubic order polynomials in $\nu$ and $\chi$.

\begin{figure}[h!t]
  \includegraphics[width=\columnwidth]{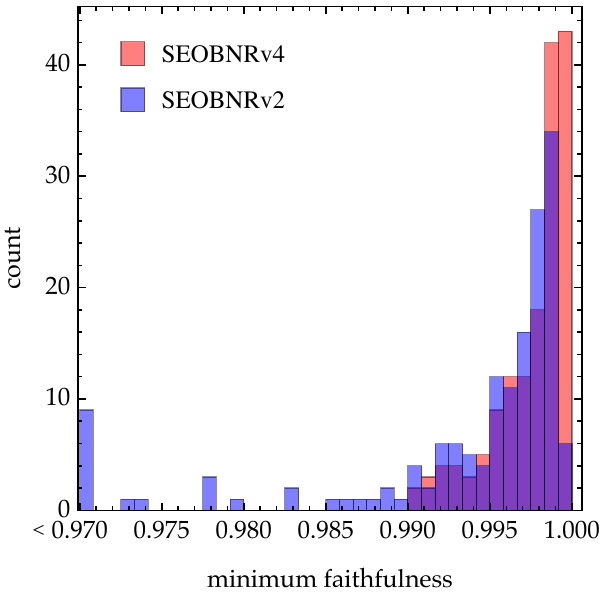}
  \caption{\label{fig:histogram:v2}Distribution of minimum faithfulness of the old EOBNR model (\texttt{SEOBNRv2})~\cite{Taracchini:2013rva} and new EOBNR model (\texttt{SEOBNRv4}) against the NR catalog. The total mass range considered is $10\,M_\odot \leq M \leq 200\,M_\odot$. The calculations are done with the Advanced LIGO design zero-detuned high-power noise PSD and a low-frequency cutoff corresponding to the initial geometric frequency of each NR simulation.}
\end{figure}

\begin{figure*}[htbp]
  \includegraphics[width=2\columnwidth]{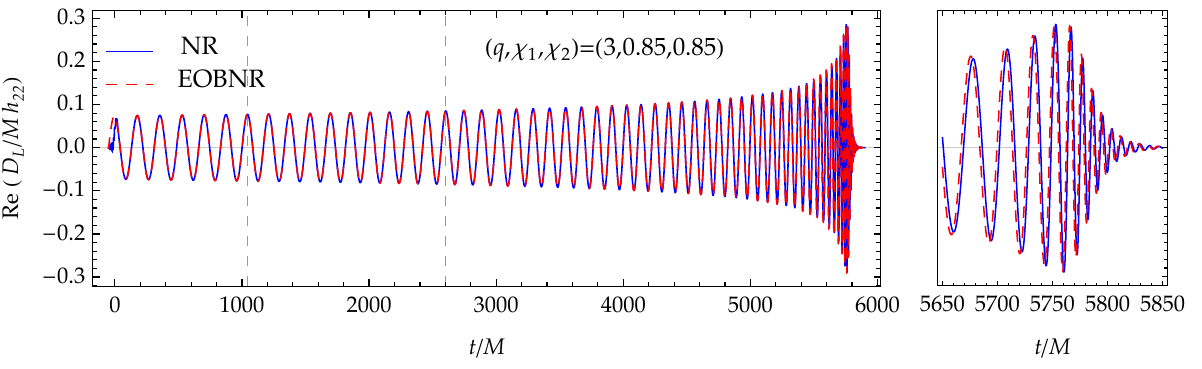}
  \caption{\label{fig:waveforms} Waveform comparison in the time domain between the (dominant mode) EOBNR waveform of this paper (\texttt{SEOBNRv4}) (dashed red) and the NR waveform (solid blue) for a BBH with
$(q,\chi_1,\chi_2) = (3,0.85, 0.85)$. The waveforms are phase aligned and time shifted at low frequency (the alignment window is indicated by the vertical dashed lines). The phase evolution throughout late inspiral, merger and ringdown is well reproduced, as well as the time to merger. }
  \end{figure*}

\subsection{Performance against the numerical-relativity catalog}
\label{subsec:performanceagainstNRcatalog}
Having completed the calibration procedure, we now investigate the performance of our final EOBNR model by computing its faithfulness against the NR catalog, including the 16 test cases that were not used in the calibration. Note that 6 of these cases lies close to the boundary of the calibration domain. Matches are computed using the setup described above Eq.~\eqref{eq:match}. In particular, the design zero-detuned high-power noise PSD of Advanced LIGO~\cite{Shoemaker:2010} is used with a lower frequency cutoff corresponding to the initial geometric frequency of each NR waveform.

In order to put results into context, in the left panel of Fig.~\ref{fig:unfaithfulness:v2v4} we first show the comparison between NR and the previous instance of the spinning, nonprecessing EOBNR model~\cite{Taracchini:2013rva} (\texttt{SEOBNRv2}), which was calibrated in 2013 to 38 NR waveforms (see Fig.~\ref{fig:calibspacechieff}). As was already pointed out in Ref.~\cite{Kumar:2016dhh}, this model performs very well against most of the NR simulations that became available after its calibration, but the faithfulness degrades noticeably for binaries with unequal masses ($\sim 2\mbox{--}3$) and large positive aligned-spin components
($\sim 0.8$), with some cases reaching an unfaithfulness of more than $10\%$. By contrast, our calibrated EOBNR model (\texttt{SEOBNRv4}) (displayed in the right panel of Fig.~\ref{fig:unfaithfulness:v2v4}) agrees to better than $1\%$ with all NR simulations. These results are summarized in Fig.~\ref{fig:histogram:v2}, where, for both models, the distribution of the maximum unfaithfulness across the mass range $10\,M_{\odot} \leq M \leq 200\,M_{\odot}$ for each NR runs is represented as a histogram.

While the faithfulness is the quantity of interest for parameter-estimation applications, it can sometimes hide inaccuracies in the waveform (that can be reabsorbed by time and phase shifts).
As a further illustration of the excellent agreement between our new EOBNR model and NR, in Fig.~\ref{fig:waveforms} we overlay both waveforms for the $(q,\chi_1,\chi_2) = (3,0.85, 0.85)$ configuration after phase aligning at low frequency. We see that the new EOBNR model accurately reproduces the full phase evolution through merger and ringdown, and that quantities such as the time to merger are also well predicted. This is due to the inclusion of the second term in Eq.~\eqref{eq:like}.

\section{Comparison with waveform models used in the first observing run of Advanced LIGO}

\label{sec:comparisons}

\begin{figure*}[tb]
\hspace{-0.2cm}
\includegraphics[width=.9\columnwidth]{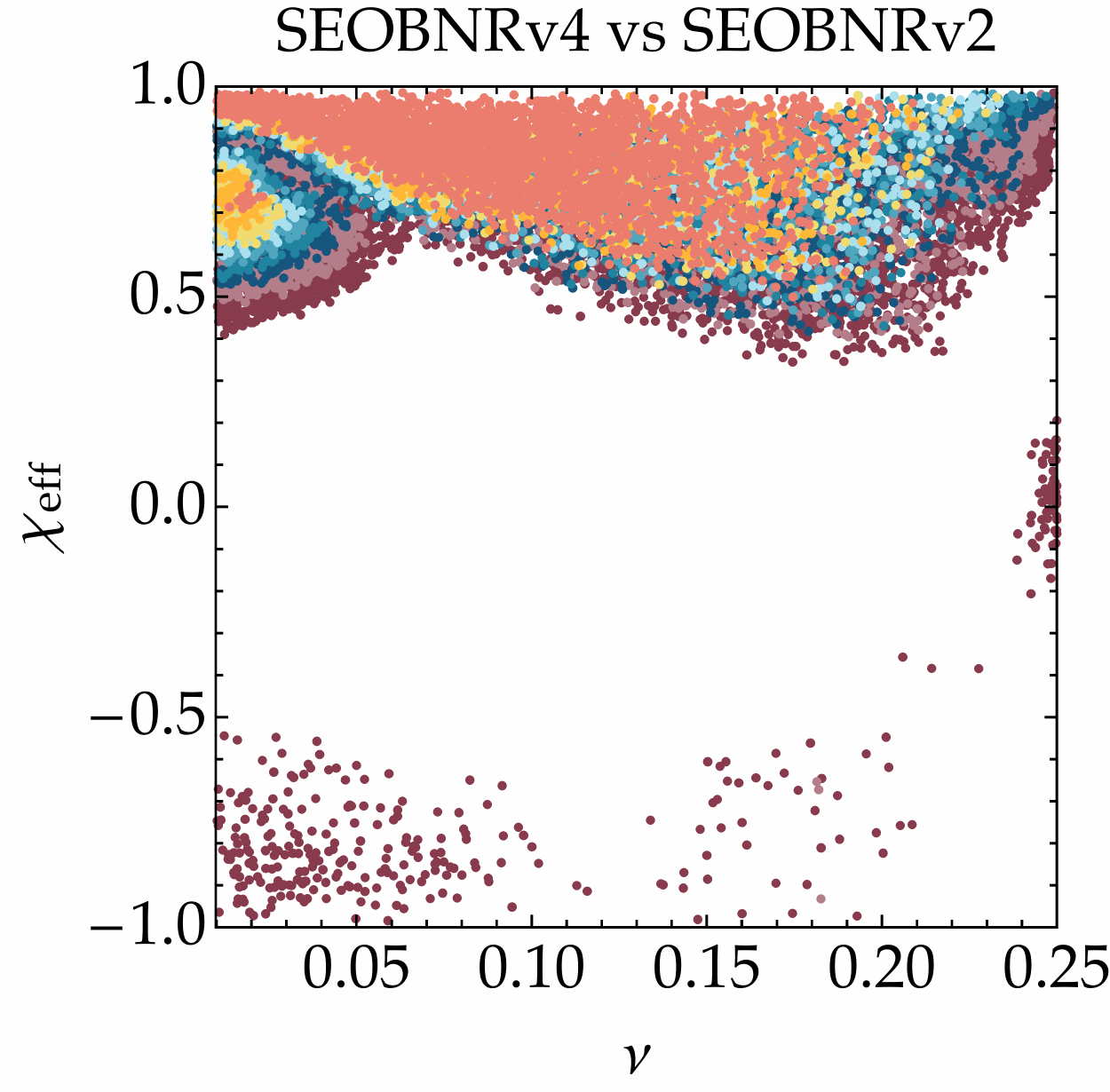}
\includegraphics[width=.9\columnwidth]{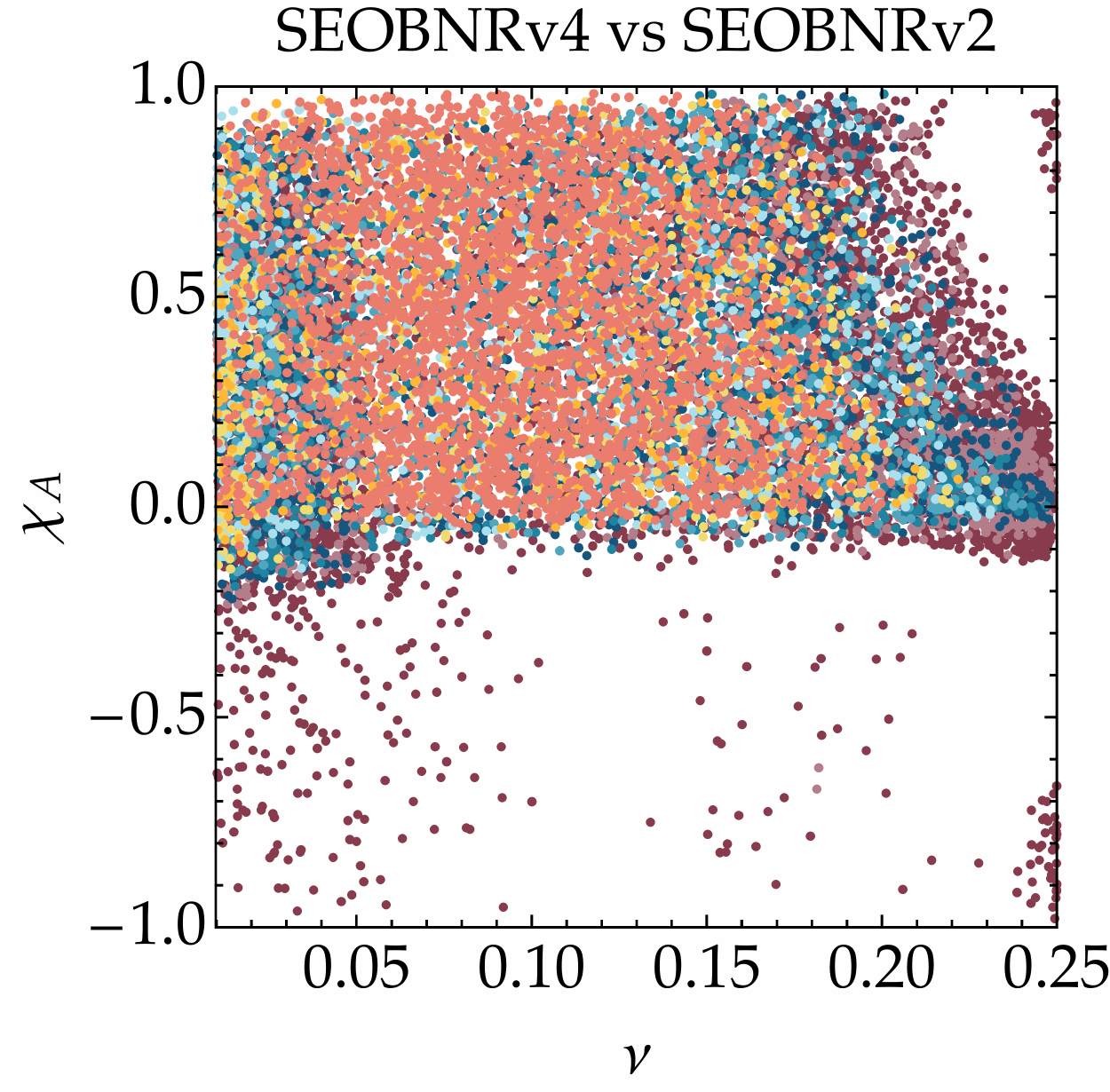}
\raisebox{0.05\height}{\includegraphics[width=0.25\columnwidth]{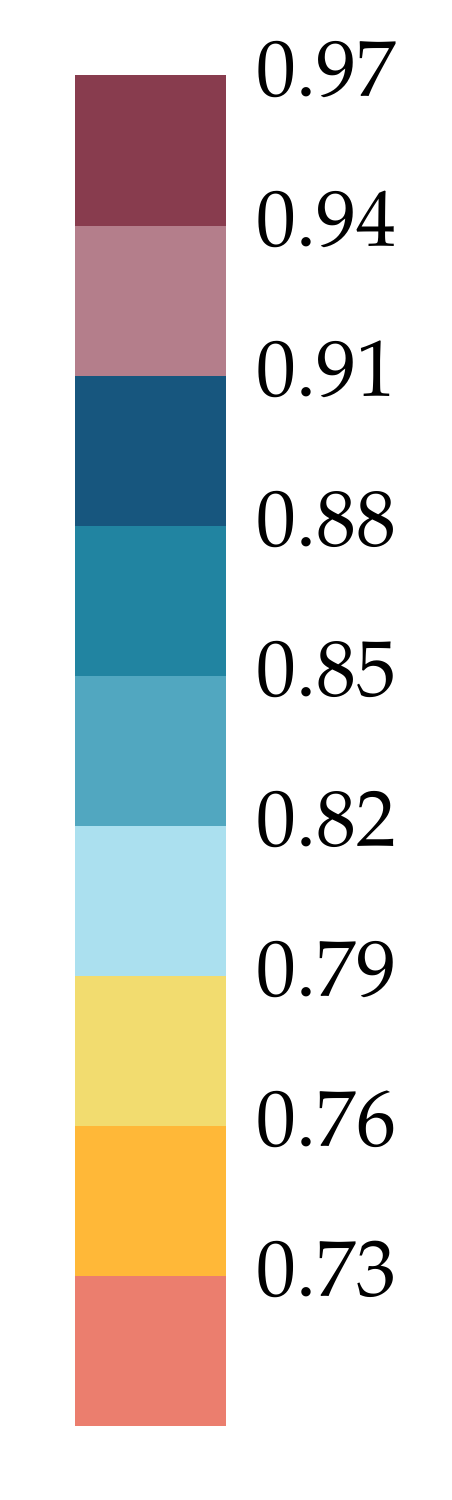}}

\includegraphics[width=.9\columnwidth]{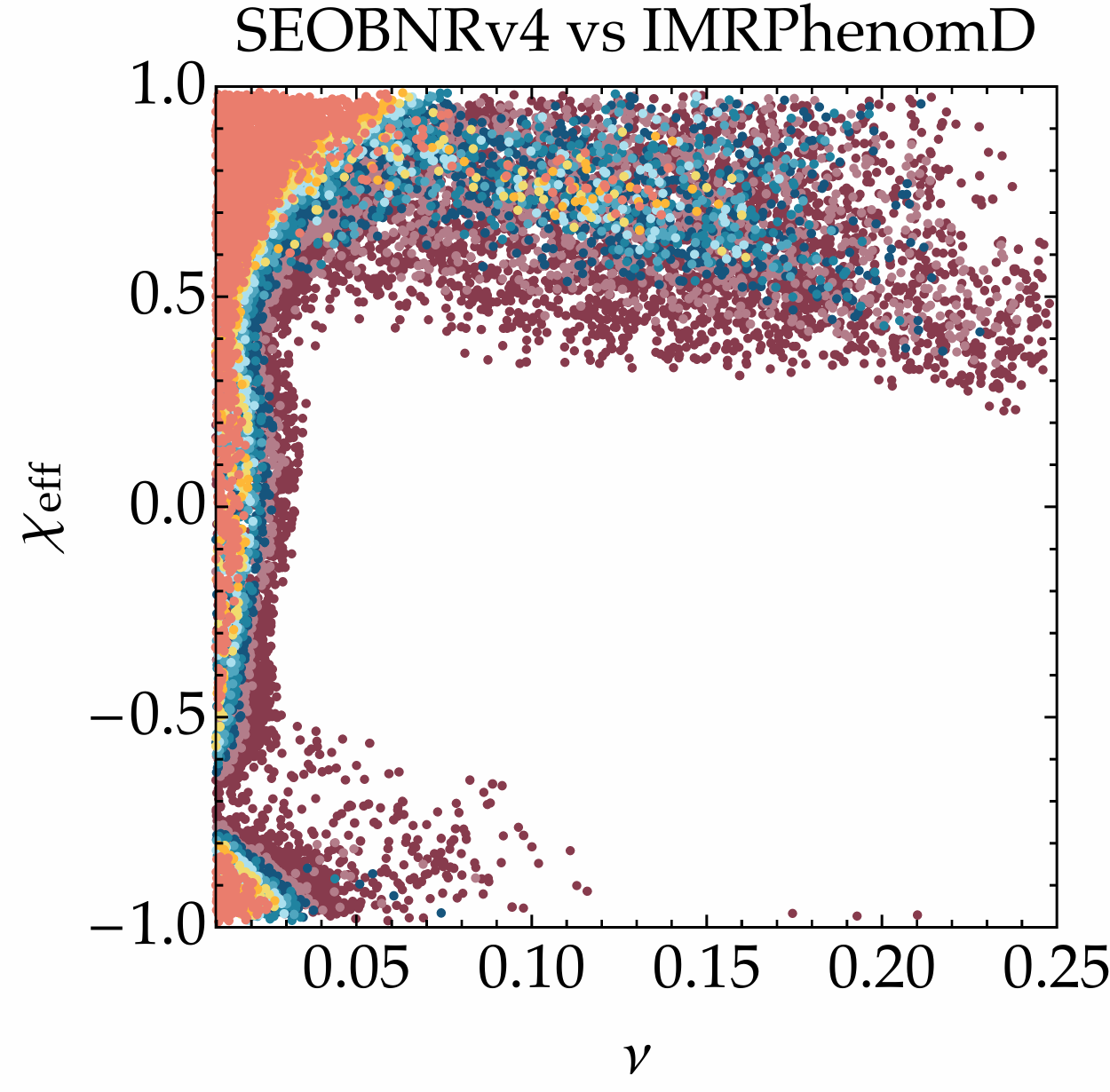}
\includegraphics[width=.9\columnwidth]{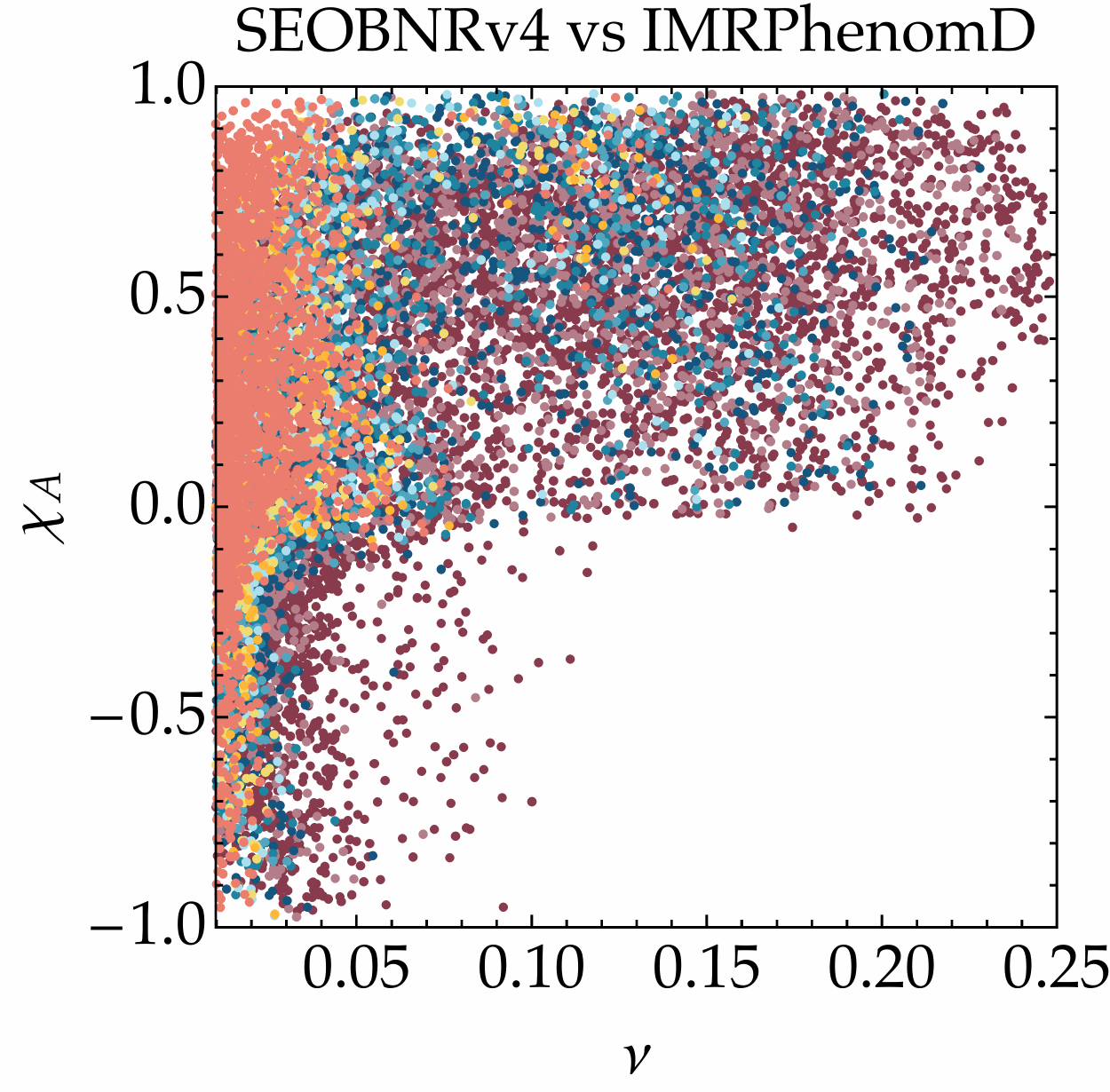}
\raisebox{0.05\height}{\includegraphics[width=0.25\columnwidth]{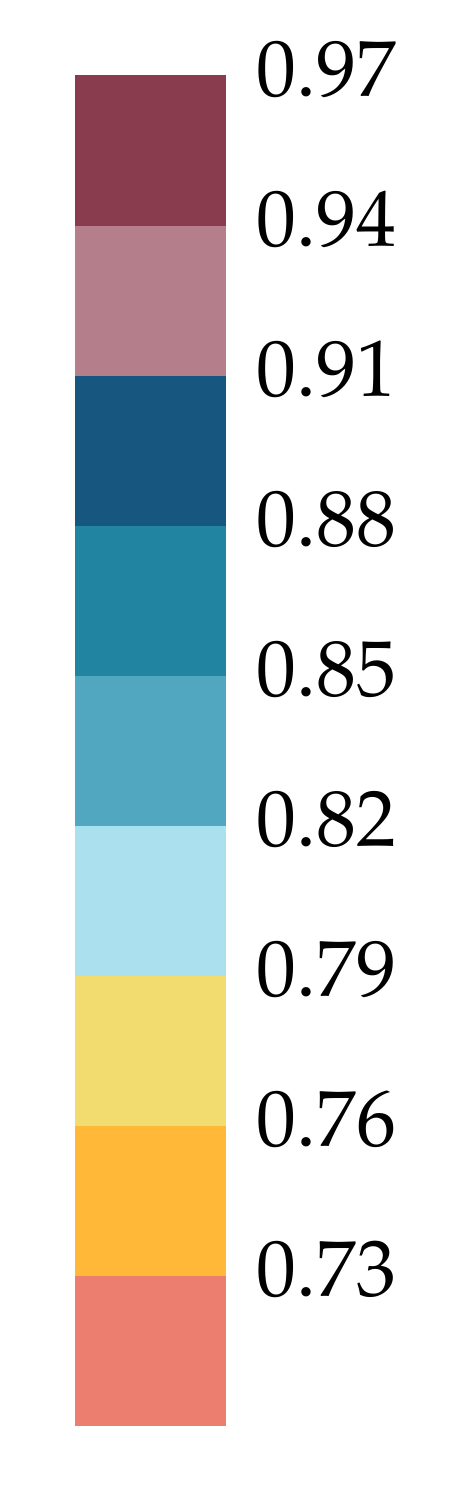}}
 \caption{\label{fig:faithsimv2v4} Faithfulness of the EOBNR model of this paper (\texttt{SEOBNRv4}) against the previous EOBNR model (\texttt{SEOBNRv2})~\cite{Taracchini:2013rva} (\emph{top row}) and the phenomenological inspiral-merger-ringdown model (\texttt{IMRPhenomD})~\cite{Khan:2015jqa} (\emph{bottom row}) for $2\times 10^5$ random spinning, nonprecessing BBHs with $4\,M_{\odot} \leq M \leq 200\, M_{\odot}$ using the Advanced LIGO O1 noise PSD and a low-frequency cutoff of 25\,Hz. Here $\chi_{\textrm{eff}}\equiv  (m_1 \chi_1+m_2 \chi_2)/M$ and $\chi_{\textrm{A}} \equiv (\chi_1 - \chi_2)/2$. Points with faithfulness above 97\% are not shown. Points with faithfulness $\leq 73\%$ are in red. We note that the biggest changes introduced by the new calibration occur for large, positive $\chi_{\textrm{eff}}$ and positive $\chi_{\textrm{A}}$. The new EOBNR model is most different from the phenomenological model in the large-$q$, large-$\chi_{\textrm{eff}}$ region, where both models are extrapolated away from the available NR simulations.}
\end{figure*}

\begin{figure*}[tb]
\hspace{-0.2cm}
\includegraphics[width=.9\columnwidth]{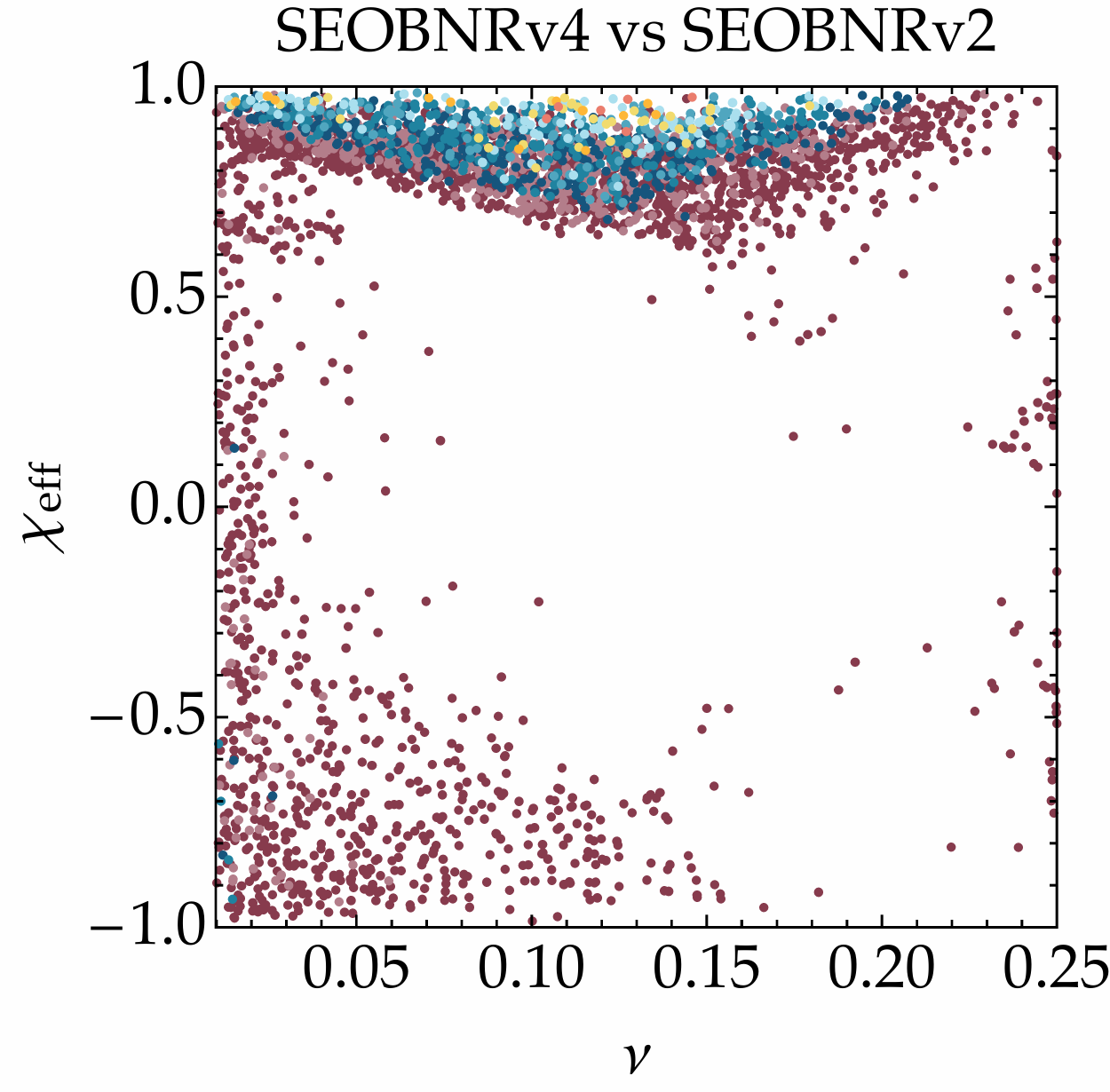}
\includegraphics[width=.9\columnwidth]{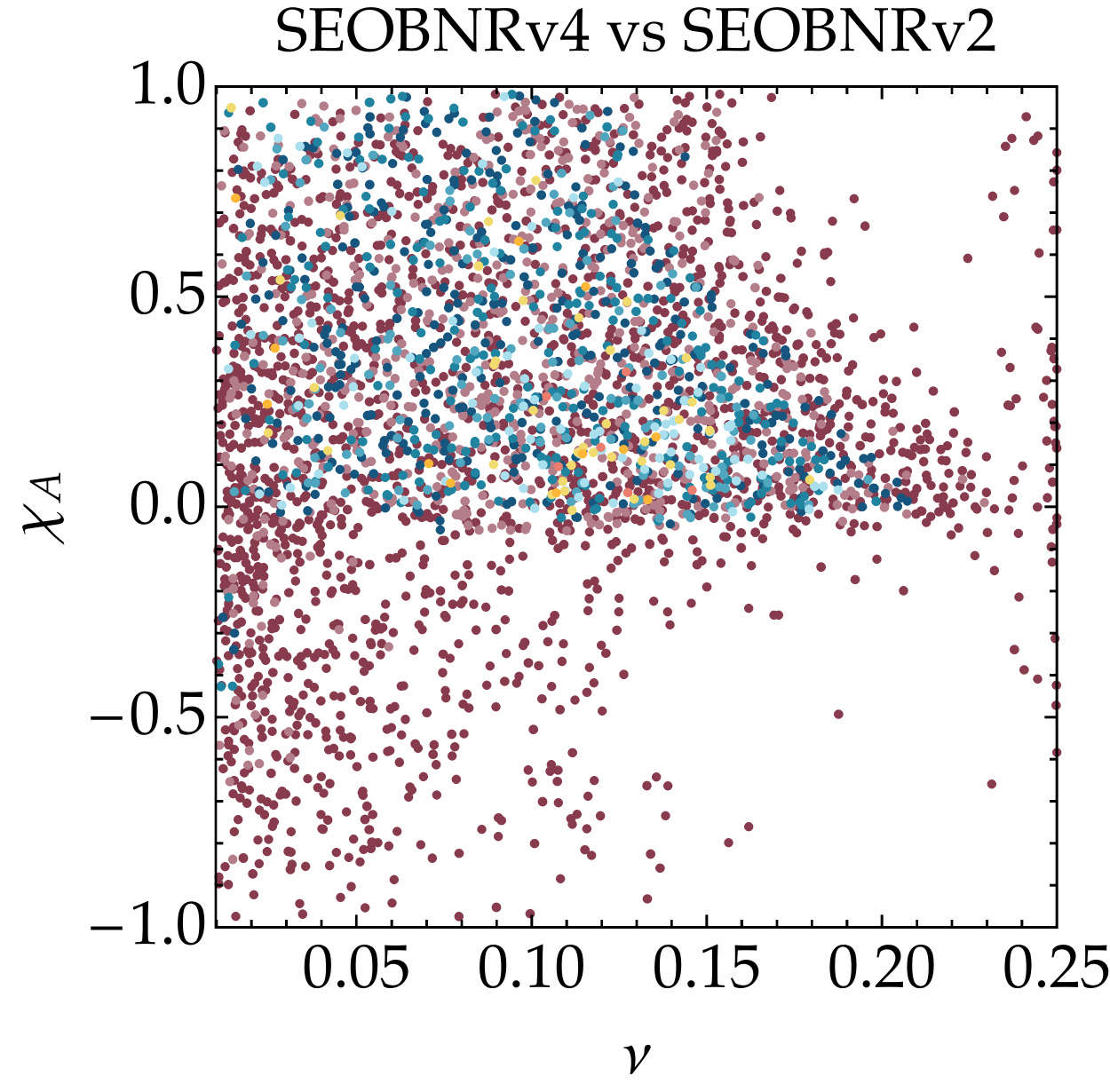}
\raisebox{0.05\height}{\includegraphics[width=0.25\columnwidth]{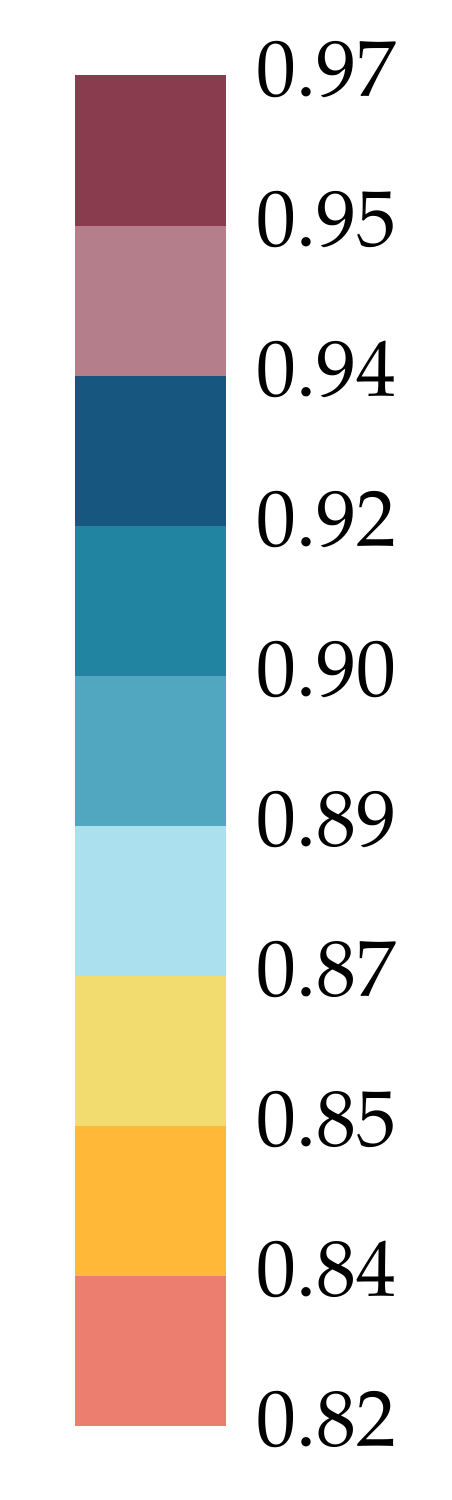}}

\includegraphics[width=.9\columnwidth]{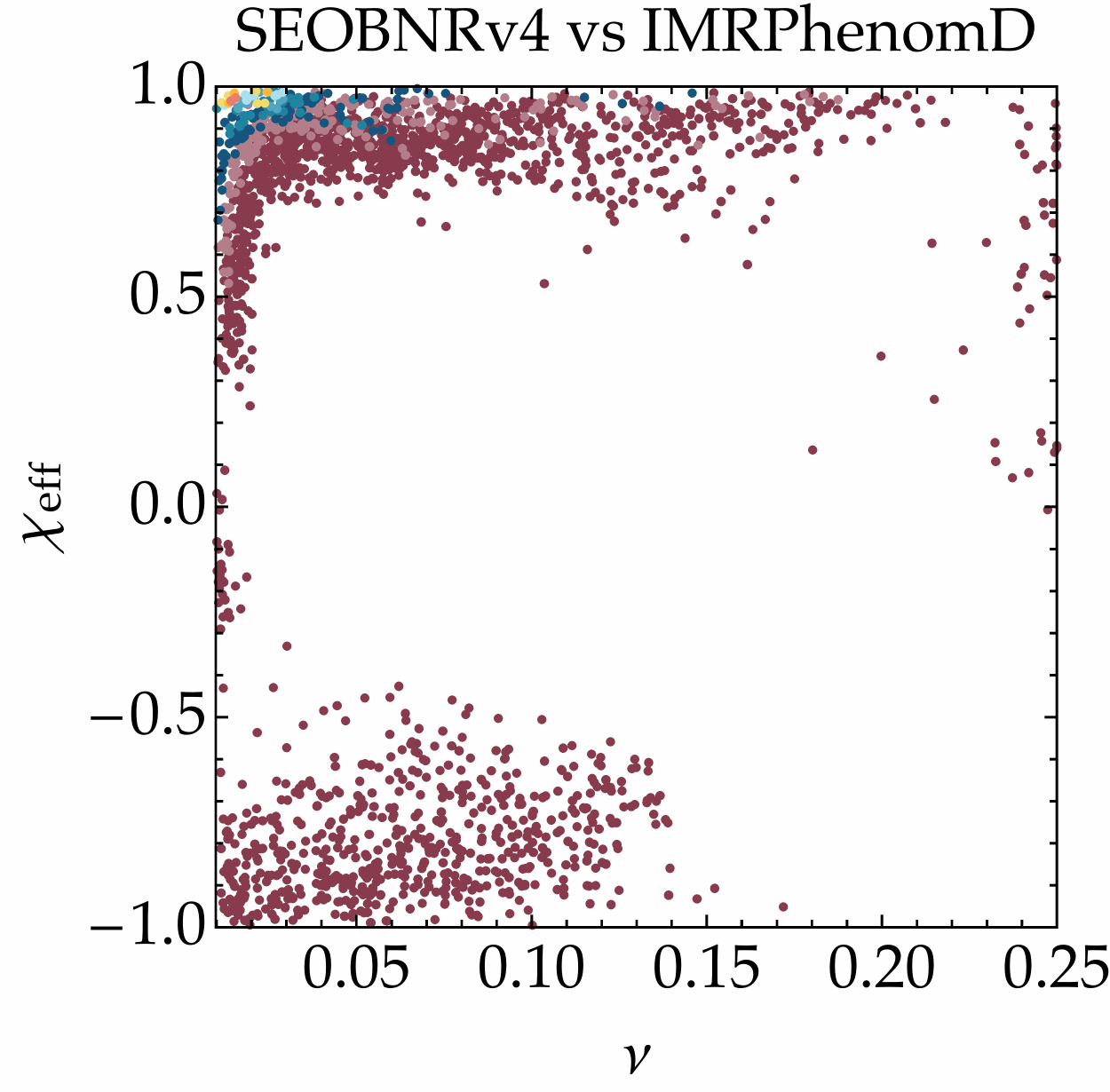}
\includegraphics[width=.9\columnwidth]{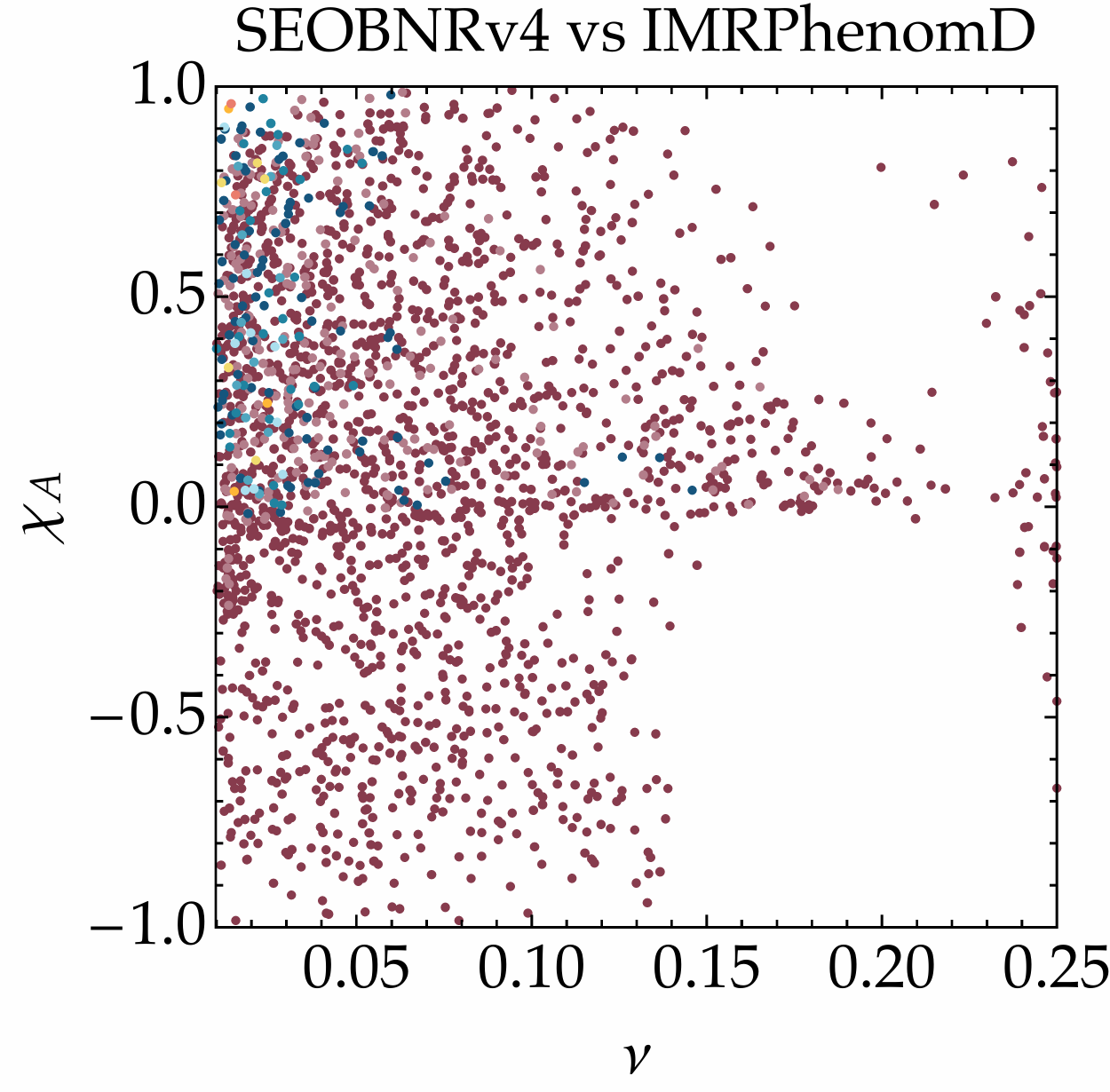}
\raisebox{0.05\height}{\includegraphics[width=0.25\columnwidth]{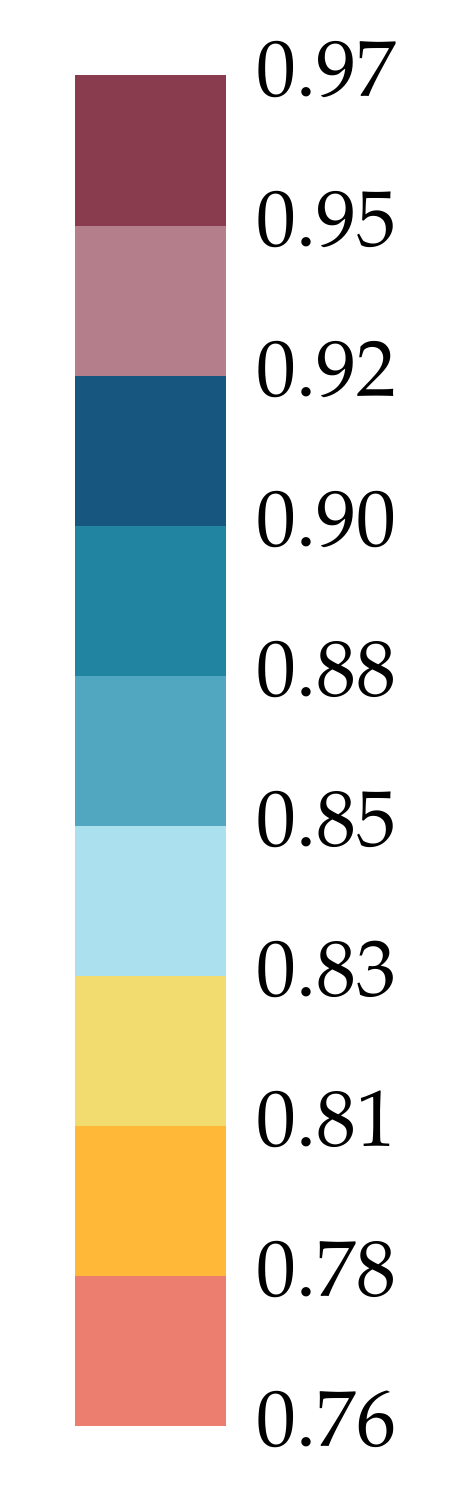}}
\caption{\label{fig:banksimv2v4}Effectualness of the EOBNR model of this paper (\texttt{SEOBNRv4}) against the previous EOBNR model (\texttt{SEOBNRv2})~\cite{Taracchini:2013rva} (\emph{top row}) and the phenomenological inspiral-merger-ringdown model (\texttt{IMRPhenomD})~\cite{Khan:2015jqa} (\emph{bottom row}): we use $10^5$ random
spinning, nonprecessing \texttt{SEOBNRv4} injections $4\,M_{\odot} \leq M \leq 100\, M_{\odot}$ that we recover using either a \texttt{SEOBNRv2} (\emph{top row}) or an \texttt{IMRPhenomD} (\emph{bottom row}) template bank~\cite{TheLIGOScientific:2016qqj}. Calculations are performed with the Advanced LIGO O1 noise PSD and a low-frequency cutoff of 25\,Hz. Here $\chi_{\textrm{eff}}\equiv  (m_1 \chi_1+m_2 \chi_2)/M$ and $\chi_{\textrm{A}} \equiv (\chi_1 - \chi_2)/2$. Points with effectualness above 97\% are not shown. The difference between the models is only present at very large values of $\chi_{\textrm{eff}}$.}
\end{figure*}

In this Section, we compare our EOBNR model (\texttt{SEOBNRv4}) across parameter space (i.e., not restricting to masses and spins for which NR waveforms are
available) with the spinning, nonprecessing models that were used for data analyses during the O1 run~\cite{TheLIGOScientific:2016qqj,TheLIGOScientific:2016wfe},
namely the previous EOBNR model (\texttt{SEOBNRv2})~\cite{Taracchini:2013rva} and the phenomenological
inspiral-merger-ringdown model (\texttt{IMRPhenomD})~\cite{Khan:2015jqa}. The goal here is very
different from that of Sec.~\ref{subsec:performanceagainstNRcatalog}, where we aimed at assessing the accuracy of the model to NR simulations. Now, we want to identify regions of parameter space where different models agree --- which gives some indication that systematic errors
due to mismodeling are small (at least for values of $M$ such that the signals are in band) --- or disagree --- thus suggesting that there the waveform models need further development.

We carry out two types of comparisons: faithfulness --- where models are compared using the same physical parameters --- and effectualness --- where additional maximization over the physical parameters is performed. While the former informs us on intrinsic differences between the models, the latter is a useful quantity in the context of GW searches, where the incoming data are compared to many templates that (discretely) cover the entire parameter space. Since our goal is to guide data-analysis applications in the forthcoming runs of Advanced LIGO, all overlap computations in this Section are performed using the noise PSD of the O1 run with a lower frequency cutoff  $f_l$ of 25\,Hz~\cite{Martynov:2016fzi}.

Figure~\ref{fig:faithsimv2v4} summarizes
the results of the faithfulness comparison of \texttt{SEOBNRv4}
against \texttt{SEOBNRv2} and \texttt{IMRPhenomD}. In
each case, $2\times 10^5$ configurations are randomly drawn with component
masses uniformly distributed in $1\, M_\odot \leq m_{1,2} \leq 200\,
M_\odot$ (with the restriction that the total mass is $4\,M_\odot \leq M
\leq 200\,M_\odot$) and component spins uniformly distributed in $-1
\leq \chi_{1,2} \leq 1$. The parameter space to explore is 4D
$(M,\nu,\chi_1,\chi_2)$. For the purpose of condensing the results into a small number of plots, while still
capturing the main features of the comparison, we resort to
projections on 2D subspaces and choose two such projections:
$(\nu,\chi_{\textrm{eff}}\equiv (m_1 \chi_1+m_2 \chi_2)/M)$ and
$(\nu,\chi_{\textrm{A}}\equiv(\chi_1-\chi_2)/2)$ (these are the same $y$-axes as in
Fig.~\ref{fig:calibspacechieff}). In order to unclutter the plots, we
remove all the points with faithfulness $> 97\%$. Therefore, the white
areas correspond to regions of parameter space where the models agree
to better than $3\%$. The remaining points are colored according to
the faithfulness. Note that, whenever points overlap with each other, those
with the lowest faithfulness are brought to the front of the plot.

Focusing first on the comparison of \texttt{SEOBNRv4} with \texttt{SEOBNRv2} (see the top row of Fig.~\ref{fig:faithsimv2v4}), we see that the model has mostly changed near the equal-mass line for very unequal spins, and for unequal masses in the region where $\chi_{\textrm{eff}}$ is positive and large, which corresponds to positive $\chi_1$, and also to positive $\chi_{\textrm{A}}$. This is the region where the \texttt{SEOBNRv2} model was extrapolated and where its performance had been known to degrade (see the left panel of Fig.~\ref{fig:unfaithfulness:v2v4}), which has now been fixed by calibrating it to NR waveforms in this region. The fraction of points with faithfulness below 97\% is only about 8\% of the total.

The \texttt{IMRPhenomD} model was calibrated to a set of \code{BAM} and SXS NR waveforms listed Table I in Ref.~\cite{Khan:2015jqa}. 
The comparison of \texttt{SEOBNRv4} against \texttt{IMRPhenomD} (see the bottom row of Fig.~\ref{fig:faithsimv2v4}) shows a wide region where the two models agree. This is not surprising since in those regions of parameter space both models were calibrated to similar NR waveforms. The fraction of points with faithfulness below 97\% is only about 7\% of the total. The largest disagreement lies in two regions: one with $\chi_{\textrm{eff}} \gtrsim 0.4$ and $\chi_{\textrm{A}}>0$ and one with $\nu \lesssim 0.03$ and $\chi_{\textrm{A}}\lesssim0$. For $\nu \gtrsim 0.18$, some mild disagreement --- a few percent --- arises for unequal-spin systems, where the \texttt{IMRPhenomD} model is known to lose accuracy (see Fig.~5 of Ref.~\cite{Kumar:2016dhh}). The region of parameter space where current spinning, nonprecessing waveform models disagree the most --- with faithfulness up to several tens of percents --- corresponds to BBHs with very unequal masses and for which the most massive BH has a large positive aligned-spin component. This is expected for at least two reasons. First, the number of cycles to merger from any given frequency increases with the mass ratio and the spin, so systems in this region spend many cycles in band, and are therefore intrinsically more difficult to model. Second, NR simulations in this region are more challenging since the large mass ratio implies the presence of very different scales to be resolved, while the large spin makes the geometry around the BH more difficult to track. As a consequence, very few NR waveforms are available in this region to calibrate the models, and those that have been produced so far~\cite{Husa:2015iqa} are shorter than simulations in less challenging regions of parameter space. As we argue below, more NR waveforms, and most crucially \emph{longer} ones, will be needed in the future to reduce the discrepancy between models in this region and control the systematic error introduced by mismodeling in GW data analysis. 

For a more detailed discussion of the faithfulness of \texttt{SEOBNRv4} against \texttt{IMRPhenomD}, see Appendix~\ref{app:faith}.

Finally, the top row of Fig.~\ref{fig:banksimv2v4} shows the effectualness of the \texttt{SEOBNRv4} model against an \texttt{SEOBNRv2} template bank~\cite{TheLIGOScientific:2016qqj} covering total masses $4\, M_{\odot} \leq M \leq 100\, M_\odot$ --- as in the O1 run of Advanced LIGO --- with a lower frequency cutoff at 25~Hz. \texttt{SEOBNRv4} injections are drawn with the same distribution as for the faithfulness comparison described above. The effectualness is below 97\% only for
3.5\% of the overall injection set, implying that the improvements in the \texttt{SEOBNRv4} model are not crucial for detecting BBH signals, although they are certainly important to extract the correct BBH parameters upon detection. Consistently with the faithfulness study, the effectualness is smaller in the region of unequal-mass BBHs with very large, positive $\chi_{\textrm{eff}}$ ($\chi_{\textrm{eff}} \gtrsim 0.8$) and positive $\chi_{\textrm{A}}$. The bottom row of Fig.~\ref{fig:banksimv2v4} shows the same effectualness computation, but now against an \texttt{IMRPhenomD} template bank. Here the fraction of points with effectualness below 97\% is 2.1\%. Points with effectualness below 90\% are only 0.06\% and are concentrated in the upper left corner of the $(\nu,\chi_{\textrm{eff}})$ plane, that is in the domain of extrapolation for both models, away from the bulk of available NR simulations. The remarkable agreement throughout the vast majority of parameter space is a welcome result. At mass ratios $q \lesssim 8$ --- at least for a noise configuration similar to that of O1 --- differences between the two models are comparable to or smaller than the tolerance of template banks construction. We repeated the effectualness computation using the design zero-detuned high-power PSD~\cite{Shoemaker:2010} with a lower frequency cutoff of 15\,Hz, finding similar results. Thus, template banks built with either \texttt{SEOBNRv4} or \texttt{IMRPhenomD} will not be significantly affected by further improvements in either model in this region. By contrast, reducing the large ineffectualness seen at large mass ratios and large aligned-spin components will require future NR simulations of sufficient length, as we argue in the next section.

\section{Length requirements on numerical-relativity simulations for calibration purposes}
\label{sec:length}

In Sec.~\ref{subsec:performanceagainstNRcatalog}, we have discussed the performance of our model against 141 NR waveforms that were used for its calibration and 5 additional waveforms used for validation, and found agreement to better than $1\%$ in unfaithfulness. It is however important to keep in mind that such a comparison only informs us on the accuracy of the model at frequencies larger than the initial frequency of each NR simulation.\footnote{We remind the reader that in Fig.~\ref{fig:unfaithfulness:v2v4}, the match integral is computed using a lower frequency cutoff corresponding to the fixed initial geometric frequency of the NR waveform.} In this Section, we investigate to what extent our calibration procedure is sufficient to constrain the entire waveform, including the low-frequency portion not covered by NR simulations.  Since no direct comparison with NR (or any other surrogate to general relativity whose error is under control) can be performed there, we have to resort to internal consistency checks to identify regions where the calibration procedure becomes unreliable.

\begin{figure}[tbp]
  \includegraphics[width=\columnwidth]{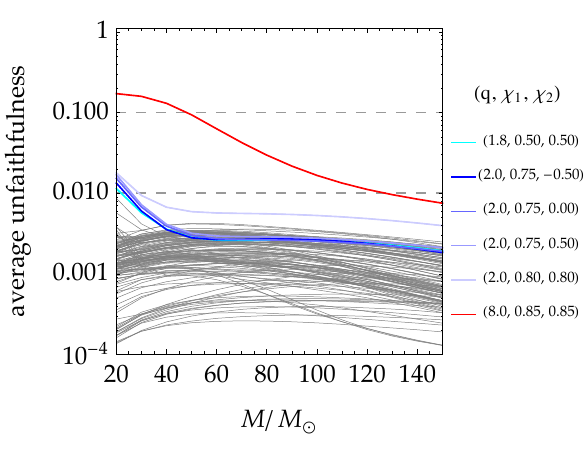}
  \caption{\label{fig:lowfreqstab} Convergence of low-frequency EOBNR waveforms upon calibration to NR. For each BBH configuration used in the calibration, we compute the average unfaithfulness between $N=1000$ EOBNR waveforms generated for values of the calibration parameters $\bm{\theta}$ that belong to regions of the MCMC posteriors where our calibration requirements (see Sec.~\ref{subsec:calireq}) are met and the fiducial EOBNR waveform with calibration parameters $\langle\bm{\theta}\rangle_{(n)}$. The matches employ a low-frequency cutoff of 25\,Hz and the Advanced LIGO design zero-detuned high-power noise PSD. The vast majority of BBHs have average unfaithfulness below 1\%, indicating that the current length of NR simulations of those configurations is sufficient to constrain the low-frequency portion of the EOBNR model. On the other hand, for the few cases listed in the legend, the calibration to NR has not lead to convergence of the model at low frequencies, and longer NR simulations are necessary.}
\end{figure}

In particular, we focus on the following question: can different sets of calibration parameters $\bm{\theta}$ allow to faithfully reproduce a given NR waveform, but lead to very different low-frequency behavior? The MCMC infrastructure developed for the calibration makes it easy to address this question, as it provides us with a whole distribution of $\bm{\theta}$'s for which the EOBNR model closely reproduces NR, on a waveform-per-waveform basis. For definiteness, for each NR configuration we restrict ourselves to those $\bm{\theta}$'s in the chain for which the unfaithfulness with NR is smaller than $1\%$ across the whole mass range \emph{and} the difference in time of merger (after low-frequency phase alignment) is smaller than $5\,M$, and randomly draw $N=1000$ points from that set. In order to understand how these $N$ different EOBNR waveforms differ at low frequency (without having to perform $\mathcal{O}(N^2)$ faithfulness computations), we compare them to a reference waveform corresponding to the set of calibration parameters $\langle\bm{\theta}\rangle$ defined in the last paragraph of Sec.~\ref{subsec:MCMC}. We use here a lower frequency cutoff of 25\,Hz and the Advanced LIGO design zero-detuned high-power noise PSD curve~\cite{Shoemaker:2010}. In Fig.~\ref{fig:lowfreqstab}, for each BBH configuration $\bm{\lambda}$ for which we had a NR run for calibration, we compute the average (over our set of $N$ points) unfaithfulness $1/N \sum_{i=1}^{N} (1-\langle h_{\textrm{EOB}}(\bm{\lambda};\langle\bm{\theta}\rangle)|h_{\textrm{EOB}}(\bm{\lambda};\bm{\theta}_{i})\rangle)$ as a function of the total mass.

The variability in the low-frequency behavior across the $\bm{\theta}_{i}$'s obviously depends on the length of each NR waveform, as well as on the physical parameters $\bm{\lambda}$. For almost all cases, we find an average faithfulness well below the $1\%$ level, with a worst value very close to this threshold. We therefore conclude that those NR waveforms are long enough to constrain the low-frequency content as well: all sets of calibration parameters that allow to reproduce the NR portion give very similar inspirals from 25\,Hz. For some short SXS runs with $q=1.8$ or $2$ (indicated in the legend of Fig.~\ref{fig:lowfreqstab}, of approximate length $N_{\rm GW\, cycles} = 26,\, 26,\,27,\,27,\,27$ from top to bottom), the average mismatch grows above the $1\%$ level as the total mass decreases, indicating that those waveforms alone would be too short to calibrate the model with this accuracy at low masses at those BBH configurations. However, this region of parameter space is covered by many other longer NR runs that do not suffer from the same issue, thus we argue that the final model is not affected by this. By contrast, with the more isolated $(q,\chi_1,\chi_2)=(8,0.85,0.85)$ NR waveform (see Fig.~\ref{fig:calibspacechieff}) which contains approximately 15 GW cycles before merger, the average mismatch exceeds $10\%$ at low masses and can be as high as $50\%$ in certain cases. Here, the NR waveform is too short to fully constrain the calibration parameters and our procedure can only ensure that the high-frequency part of the model is correct.

\begin{figure}[tbp]
  \includegraphics[width=\columnwidth]{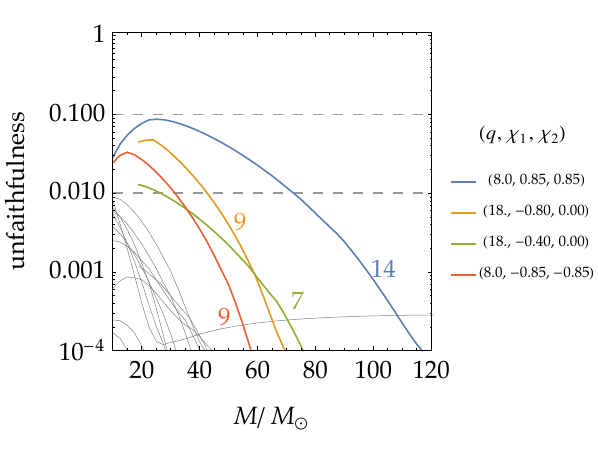}
  \caption{\label{fig:lowfreqstabhybrid} Impact of NR length on hybrid waveforms used for \texttt{IMRPhenomD} model~\cite{Khan:2015jqa} calibration. For the BBH configurations that were employed in the calibration of \texttt{IMRPhenomD} model, we hybridize high-frequency \texttt{SEOBNRv4} waveforms with low-frequency uncalibrated EOB waveforms at the frequency reported in Table~I of Ref.~\cite{Khan:2015jqa}. We then compare these hybrid waveforms to purely \texttt{SEOBNRv4} waveforms by computing faithfulness from 25\,Hz with the Advanced LIGO design zero-detuned high-power noise PSD. The number of GW cycles between the hybridization frequency and the merger is shown next to the 4 curves highlighted in the legend. }
\end{figure}

Naturally, the limitations due to the finite length of NR waveforms are not specific to the calibration of  the \texttt{SEOBNRv4} model, and will affect the construction of any inspiral-merger-ringdown model. As an illustration, we discuss the case of the \texttt{IMRPhenomD} model, which calibrates its phenomenological ansatz to NR waveforms hybridized with an \emph{uncalibrated} version\footnote{An important motivation behind this choice is to avoid calibrating models against each other, an independence which is crucial in order to estimate systematic errors introduced by modeling by comparing two models as done in Ref.~\cite{TheLIGOScientific:2016wfe}).} of the EOBNR model. While at sufficiently low frequencies both the calibrated and the uncalibrated model should agree (between themselves and with any PN approximant), they become different in the late inspiral (as soon as the EOB calibration starts to play a role). The use of uncalibrated EOB as an inspiral approximant is justified as long as  this difference kicks in after the hybridization frequency. In order to investigate whether this is the case, we build hybrids between the calibrated model (for the high-frequency part, as a surrogate for the actual NR waveforms used in the \texttt{IMRPhenomD} construction, which are not public) and the uncalibrated one (at low frequencies), and compare them to the calibrated model. We focus on the configurations actually used in the \texttt{IMRPhenomD} construction, which are listed in Table~I of Ref.~\cite{Khan:2015jqa}, together with the respective hybridization frequencies.\footnote{Note that the frequency reported for the $(q, \chi_1,\chi_2)=(8,0.85,0.85)$ waveform is incorrect and should read $Mf_{\rm hyb} = 0.0175$. 
}
The results are reported in Fig.~\ref{fig:lowfreqstabhybrid}. For most cases, using the uncalibrated EOB model below the hybridization frequency instead of the calibrated one only introduces a mismatch smaller than $1\%$, even at low masses. In a few cases however, the difference is well above that threshold, indicating that the hybridization frequency is too high --- or, equivalently, that the NR waveform actually used in the construction is too short --- for the uncalibrated EOB model to still be a good approximant. This is for instance the case for the $(q, \chi_1,\chi_2)=(8,0.85,0.85)$ \code{BAM} waveform (the same used in this paper). The number of cycles (predicetd by \texttt{SEOBNRv4}) between the hybridization frequency and the merger is approximately $N_{\rm GW\, cycles} = 14,\, 9,\,7,\,9$ from top to bottom for the cases highlighted in the legend.

\section{Construction of the reduced-order model}
\label{EOBNR:ROM}

\begin{figure}[htb]
\hspace{-1cm}\includegraphics[width=0.8\columnwidth]{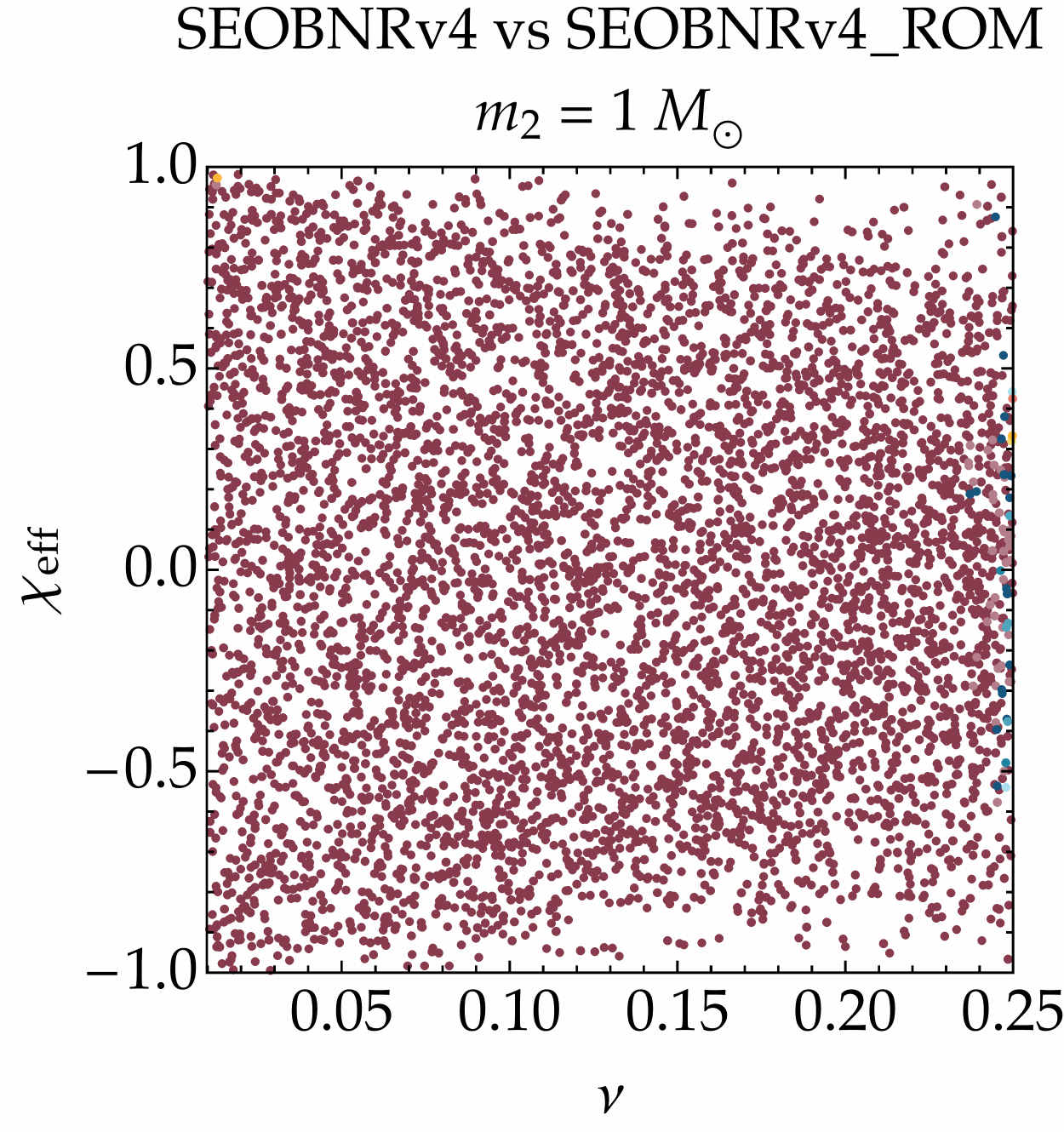} \raisebox{0.05\height}{\includegraphics[width=0.25\columnwidth]{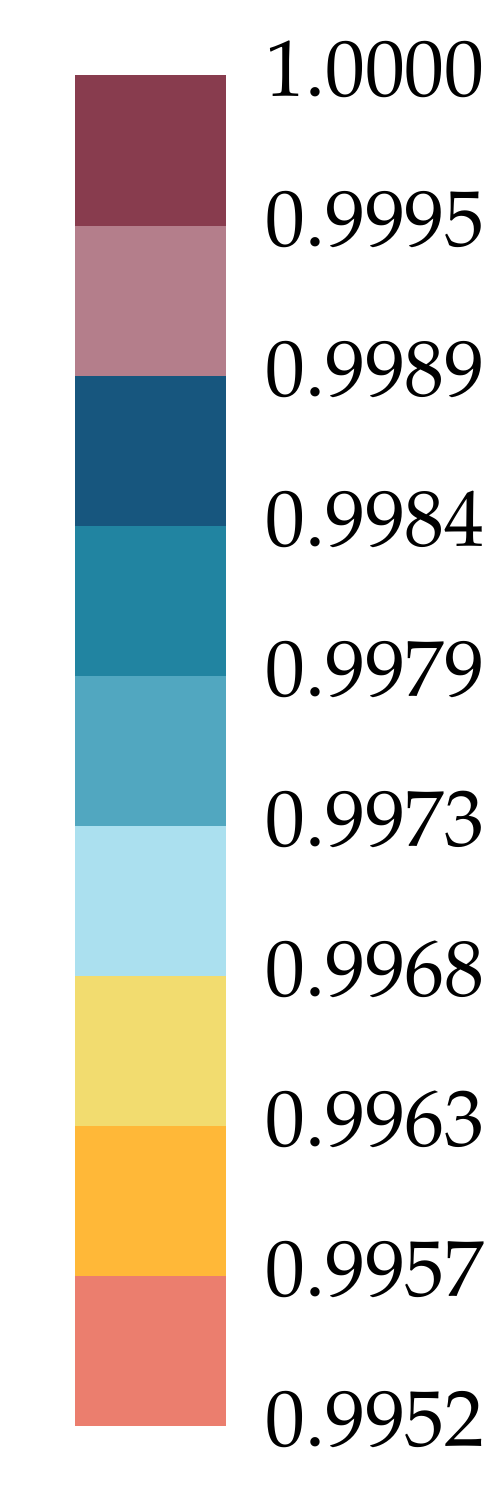}}

\hspace{-1cm}\includegraphics[width=0.8\columnwidth]{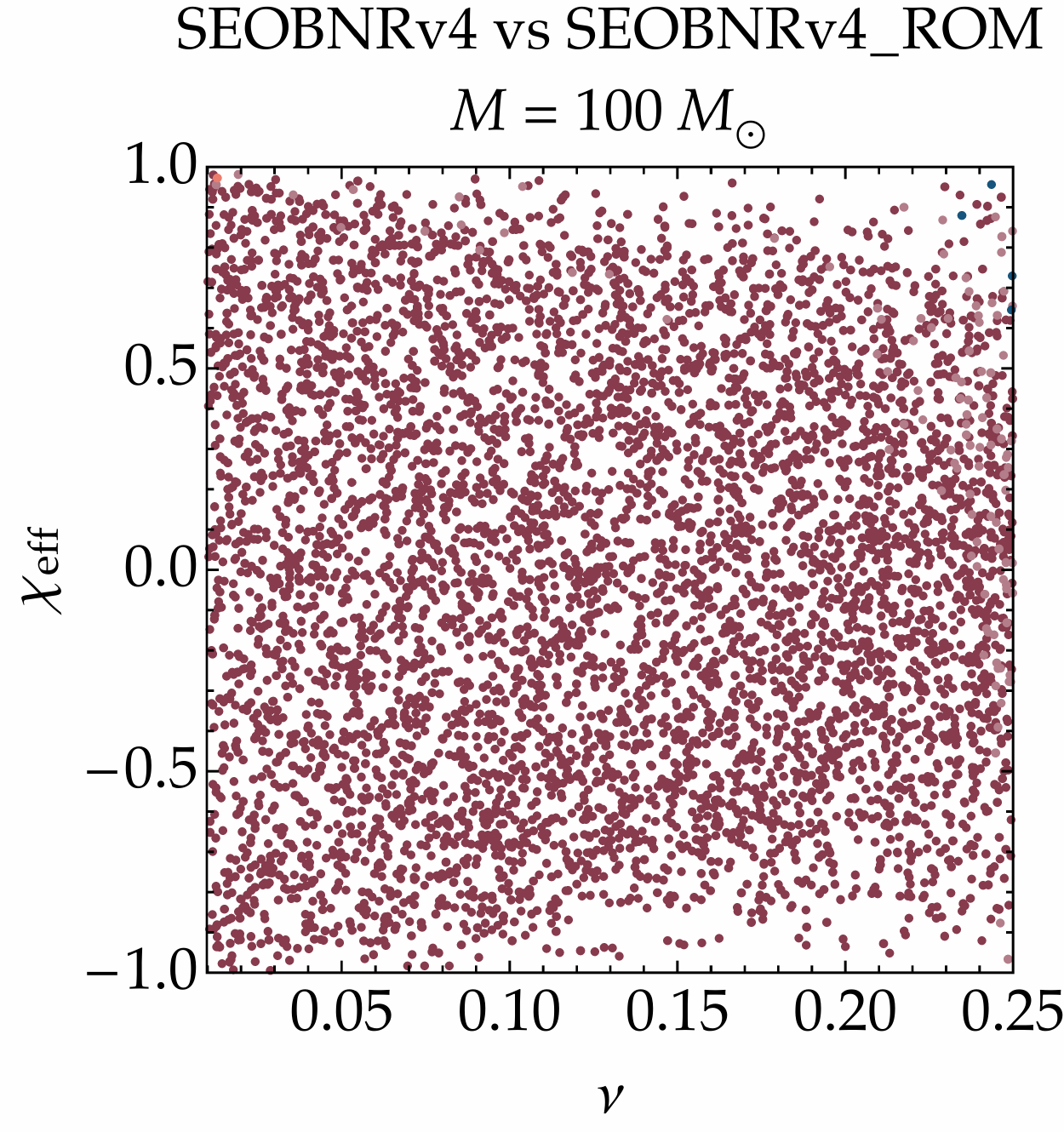} \raisebox{0.05\height}{\includegraphics[width=0.25\columnwidth]{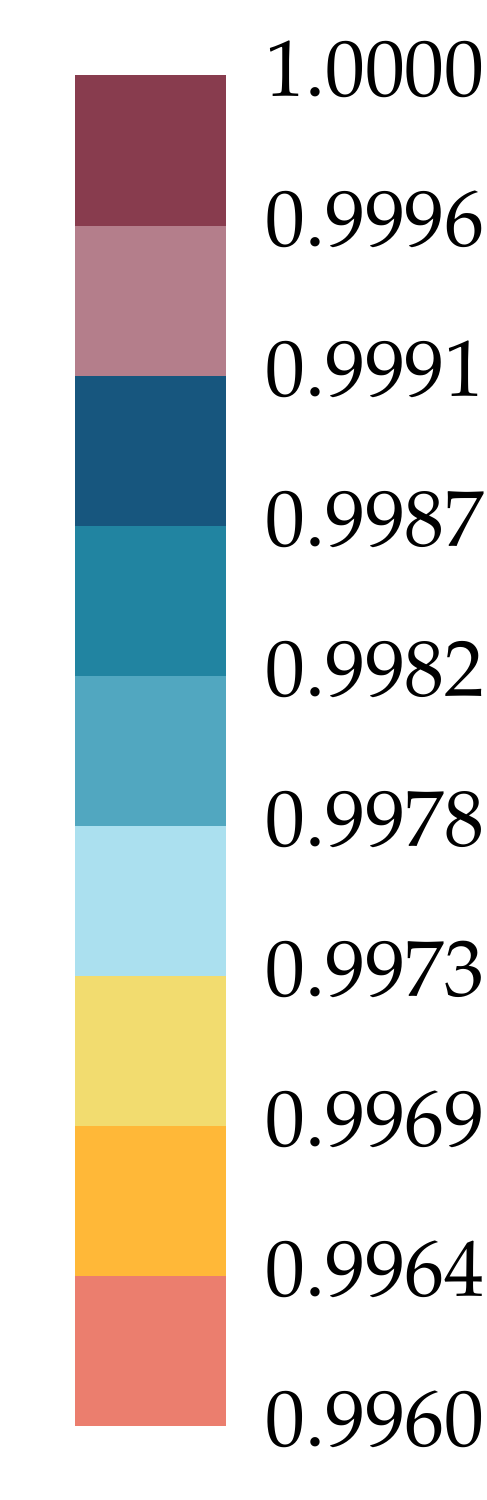}}
\caption{\label{fig:SEOBNRv4ROM_mismatch} Faithfulness of the \texttt{SEOBNRv4\_ROM} model against \texttt{SEOBNRv4} model as a function of the symmetric mass ratio $\nu$ and the effective spin combination $\chi_{\textrm{eff}}$. The aLIGO O1 PSD~\cite{Martynov:2016fzi} is used with a low frequency cutoff of 20\,Hz. The top panel shows unfaithfulness for BBHs where the smaller body is fixed at $1\,M_\odot$, while in the bottom panel the total mass is fixed at $100\,M_\odot$.}
\end{figure}

\begin{figure*}[tb]
  \includegraphics[width=.95\columnwidth]{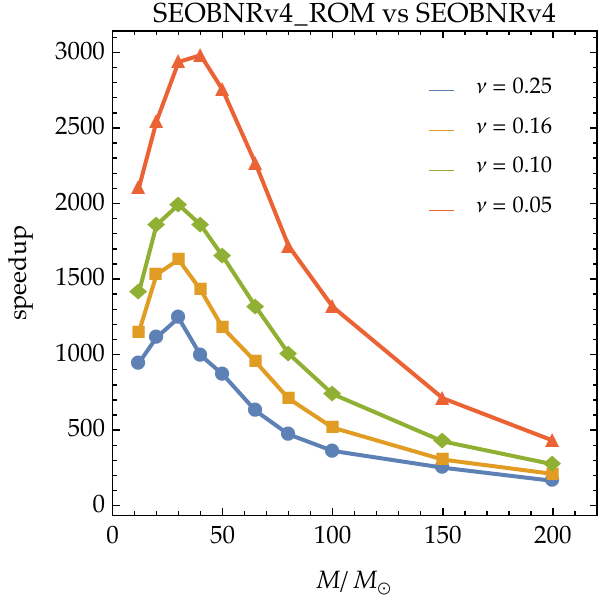}
  \hspace{1cm}
  \includegraphics[width=.95\columnwidth]{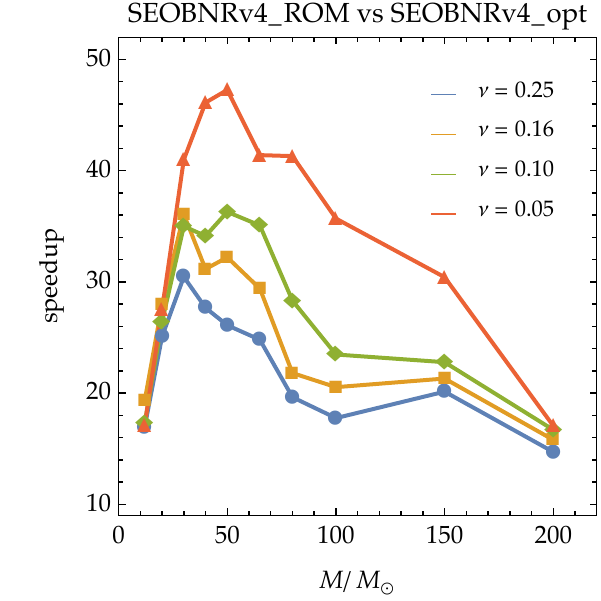}
  \caption{\label{fig:SEOBNRv4ROM_speedup}
 Speedup of the \texttt{SEOBNRv4\_ROM} model compared to \texttt{SEOBNRv4} (\emph{left panel}) and \texttt{SEOBNRv4\_opt} (\emph{right panel}) models as a
function of the total mass for several mass ratios.}
\end{figure*}

The generation of stochastic template banks and Bayesian parameter-estimation simulations
require on the order of $10^6 \mbox{--} 10^8$ waveform evaluations. Since the numerical integration
of the EOB orbital dynamics through Hamilton's equations and the generation of gravitational waveforms
can take from seconds to hours, depending on the binary's parameters,
it can be quite slow to produce EOBNR waveforms for data-analysis applications.
Reduced order modeling (ROM) allows for the construction of fast and accurate surrogates of
waveform models.  ROMs combine methods for building reduced bases of
waveforms (or their amplitudes and phasings) and interpolation techniques
of expansion coefficients over the binary's parameter space. Established methods
use singular-value decomposition (SVD) with tensor product spline
interpolation~\cite{Purrer:2015tud,Purrer:2014fza,Smith:2012du,Cannon:2012gq,Cannon:2011rj}
or the greedy-basis algorithm and empirical interpolation
method~\cite{Field:2013cfa,Field:2011mf,Blackman:2014maa,Blackman:2015pia}.  Previous work on
ROMs for EOBNR models~\cite{Purrer:2015tud,Purrer:2014fza} has
demonstrated speed-ups to up to several thousands with ROM errors
smaller than the EOBNR's calibration errors, that is $\sim 1\%$.

In this study, we follow the ROM construction described in
Refs.~\cite{Purrer:2015tud} and~\cite{Purrer:2014fza}. The Fourier-domain 
amplitude and phase of \texttt{SEOBNRv4} input waveforms are interpolated
onto a sparse, geometric frequency grid spanning $[9.85 \times
10^{-5},0.3]$. The ringdown is extended by fitting the amplitude of
all input waveforms to an exponentially damped Lorentzian function. 
We use patching in the frequency domain and over the parameter space as introduced in Ref.~\cite{Purrer:2015tud}.
It is efficient to split the construction into low and high frequency ROMs. At low frequencies the ROM needs to capture the early inspiral part of the waveform, which varies smoothly and only requires moderate resolution over the parameter space, while finer structure from the merger and ringdown must be resolved at high frequencies with a finer grid of input waveforms.
In addition, resolution requirements are not uniform over the parameter space. Since tensor product interpolation does not allow for local refinement regions we switch between multiple overlapping patches which provide a covering of the mass-ratio and aligned spin space.

A single low-frequency patch is joined with two high-frequency patches at a frequency $Mf_\textrm{m} = 0.01$. The low-frequency patch uses a grid of $70 \times 12 \times 12$ waveforms in $\{\nu, \chi_1, \chi_2\}$ with $259$ sparse frequency points. It spans the domain of the
Cartesian product of intervals $0.01 \leq \nu \leq 0.25$ and $-1\leq \chi_{1,2} \leq 1$.
The first high-frequency patch spans this same domain on a grid of $57 \times 33 \times 21$ waveforms with $149$ frequency points. This is complemented by a second high-frequency patch covering the domain with $ 0.01\leq \nu \leq 0.025$, $0.995 \leq \chi_1 \leq 1$, and $-1 \leq \chi_2 \leq 1$ on a grid of $13 \times 11 \times 21$ waveforms and $325$ frequency points to provide higher resolution for the merger-ringdown part of the waveforms. Outside the domain of the second high-frequency patch, the first high-frequency patch is used. 

In Fig.~\ref{fig:SEOBNRv4ROM_mismatch} we show the faithfulness of the \texttt{SEOBNRv4\_ROM} model
against the \texttt{SEOBNRv4} model for the aLIGO O1 PSD~\cite{Martynov:2016fzi} with a low-frequency cutoff of 20\,Hz.
We find that, overall, the \texttt{SEOBNRv4\_ROM} model is accurate to better than $1\%$ in unfaithfulness for BBHs with a total mass of $2\, M_\odot$ or higher.
For low masses, the unfaithfulness in the bulk is lower than $0.1\%$, except for some configurations near equal mass where it can rise to $0.4\%$. For total masses of $M = 100\,M_\odot$ the unfaithfulness is again lower than $0.1\%$ and slightly above that in the very high mass-ratio - spin corner of the parameter space. The unfaithfulness rises to $0.4\%$ for $M= 300\,M_\odot$. For total masses higher than about $500\,M_\odot$ differences between the ringdown description in the ROM and SEOBNRv4 can result in unfaithfulness above $1\%$ for some configurations.

In Fig.~\ref{fig:SEOBNRv4ROM_speedup} we show the speed-up of the \texttt{SEOBNRv4\_ROM} model against the \texttt{SEOBNRv4} model and the \texttt{SEOBNRv4\_opt} model, which is a version
of the \texttt{SEOBNRv4} code with significant optimizations~\cite{Devine:2016ovp}. We see that the \texttt{SEOBNRv4\_ROM} model
is several thousand times faster than the \texttt{SEOBNRv4} model and a factor $20 \mbox{--} 50$ faster than the \texttt{SEOBNRv4\_opt} model.

\section{Conclusions}
\label{sec:concl}

We built a new, calibrated EOBNR waveform model for spinning, nonprecessing BBHs (\texttt{SEOBNRv4}) using  12 BH-perturbation-theory waveforms~\cite{Taracchini:2014zpa}
and 141 NR simulations~\cite{Mroue:2013xna,Chu:2015kft,Kumar:2015tha,Lovelace:2010ne,Scheel:2014ina,Husa:2015iqa}, which extend to larger positive aligned-spin
  components and more spin-asymmetric configurations as compared to the NR waveforms employed in the previous version of the EOBNR model~\cite{Taracchini:2013rva}. After calibrating the model, interpolating and extrapolating it to arbitrary mass ratios and spins, we found that the model can reproduce the NR waveforms with a faithfulness larger than 99\% using the Advanced LIGO design zero-detuned high-power
  noise PSD~\cite{Shoemaker:2010} for total masses $10\,M_\odot \leq M \leq 200\, M_\odot$. This also holds true for 6 new NR simulations that were produced for this paper but were not used in the calibration and 10 NR waveforms that were employed for validation. To achieve
this level of accuracy, we employed MCMC techniques to explore the 4D space of inspiral-plunge calibration parameters and we employed phenomenological fitting formulae for the ringdown signal.

We compared the improved EOBNR model to spinning, nonprecessing
  waveform models that were used in the data analysis of the O1 run
of Advanced LIGO, namely \texttt{SEOBNRv2}~\cite{Taracchini:2013rva}
and \texttt{IMRPhenomD}~\cite{Khan:2015jqa}. We carried out
faithfulness comparisons from 25\,Hz with the O1 noise
PSD~\cite{Martynov:2016fzi}. We found that \texttt{SEOBNRv4} has
faithfulness:~\footnote{We notice that the regions of low faithfulness are 
far from the parameter space where GW150914 and GW151226 were observed.} 
(i) as low as 43\% against \texttt{SEOBNRv2} in the
region of large, positive aligned-spin components and spin-asymmetric
BBHs, irrespective of the mass ratio, where new NR simulations became
available for calibration; (ii) as low as 35\% against
\texttt{IMRPhenomD} in the region of large, positive aligned-spin
components and large mass ratios, where both models are extrapolating
away from the respective calibration domain and the number of GW
cycles in band is larger than in any other part of parameter space. We note that the fraction
of points with faithfulness below 97\% is only about 8\% (7\%) of the total
when comparing \texttt{SEOBNRv4} to \texttt{SEOBNRv2} (\texttt{IMRPhenomD}).
The faithfulness results against \texttt{IMRPhenomD} waveforms 
at mass ratios $\gtrsim 4$ and aligned-spin components $\gtrsim 0.8$ 
strongly suggest the importance of producing new NR simulations in this 
region of the parameter space, so that discrepancies between 
different ways of extrapolating waveform models can be resolved.
By contrast, the high effectualness between \texttt{SEOBNRv4} and 
\texttt{IMRPhenomD} waveform models in almost all parameter space (both for O1 and design noise curves), 
suggests that for Advanced LIGO detection purposes the dominant-mode 
models do not need to be further improved. However, the inclusion 
of higher modes is likely to be important to increase our chance of 
detecting binary coalescences in some regions of the parameter space~\cite{Capano:2013raa}, 
notably large mass ratios.

Several studies were carried out in the past to try to understand 
how to build semi-analytic waveform models tuned and/or hybridized to 
NR waveforms, so that they could be trusted outside the 
frequency region (or mass range) in which the NR information is employed 
(e.g., see Refs.~\cite{MacDonald:2011ne,Ohme:2011zm,Pan:2013tva} and references 
therein). Here, we assessed how much the finite length of available NR simulations
affects the calibration of inspiral-merger-ringdown models. We showed
that, for a handful of BBH configurations in the NR catalog at our disposal,
the calibration cannot yet constrain the low-frequency portion of the
model due to the small length of the NR runs. We restricted the scope
of this study to flagging points at which longer NR waveforms are
required. A more ambitious study would consist in trying to predict
the initial NR frequency necessary to satisfactorily constrain the low
frequencies in the model \emph{at points in parameter space where we
  do not have long enough waveforms yet}. As a first step, one would
need to determine for each NR waveform available the largest frequency
for which the low frequencies are well constrained (say for which the
average faithfulness is smaller than $1\%$ for all
masses). This requires running several MCMC chains where the NR
waveform is artificially cut at increasingly large initial
frequencies. Once this minimal frequency requirement has been
identified for each available waveform, one could then try to
extrapolate to other values of $(q,\chi_1,\chi_2)$. 
Given its expensive character, such a study is however beyond the scope of this
paper and we reserve it for future work. We also showed that the
construction of hybrids using uncalibrated EOB waveforms in the
low-frequency regime (as done in \texttt{IMRPhenomD}) can be
problematic when the hybridization frequency is too high.

We have shown the importance of new, long NR waveforms with both high
  mass ratio and high spin, but this region of the parameter space was
  challenging for \code{SpEC}.  As a result, we have made use of an
  existing (short) waveform from \code{BAM} for calibration, and new (short) 
  waveforms that we produced using the \code{Einstein Toolkit} for validation.
  Each code has different strengths, and combining results from all
  three allows the best possible science to be performed.

We plan to trade the spinning, nonprecessing dynamics and waveforms 
of the precessing EOBNR model of Refs.~\cite{Pan:2013rra,Babak:2016tgq} 
with the improved version developed in this paper, so that the 15-dimensional 
fully precessing EOBNR model can infer more accurately the 
properties~\cite{TheLIGOScientific:2016wfe,Abbott:2016izl} 
of future detections of coalescing binaries with Advanced LIGO.  

Finally, we built a reduced-order model of \texttt{SEOBNRv4} that can
be orders of magnitude faster than its time-domain implementation in
generating waveforms, while still being faithful to it for
data-analysis applications in Advanced LIGO.

The model described in this paper, as well as its ROM version, have already been implemented and reviewed in
LAL~\cite{LAL}, and are publicly available under the name of \texttt{SEOBNRv4} and \texttt{SEOBNRv4\_ROM}, 
respectively. 

\section*{Acknowledgments}
We would like to thank Mark Hannam and Sascha Husa for kindly providing us with the nonpublic \code{BAM} $(q,\chi_1,\chi_2)=(8,0.85,0.85)$ waveform, which was used
in Sec.~\ref{sec:cali} to calibrate the EOB model.
This work was supported in part at Caltech by the Sherman Fairchild Foundation and NSF Grants No. PHY-1404569, at Cornell by NSF Grants No. PHY-1606654 and No. AST- 1333129 and the Sherman Fairchild Foundation and at Cal State Fullerton by NSF grants PHY-1307489 and PHY-1606522.
We gratefully acknowledge support for this research at CITA from NSERC of Canada, the Ontario Early Researcher Awards Program, the Canada Research Chairs Program, and the Canadian Institute for Advanced Research. Calculations were performed at the GPC
supercomputer at the SciNet HPC Consortium; SciNet is
funded by: the Canada Foundation for Innovation (CFI) under the
auspices of Compute Canada; the Government of Ontario; Ontario
Research Fund (ORF) -- Research Excellence; and the University of
Toronto. Further calculations were performed on the Briar\'ee cluster
at Sherbrooke University, managed by Calcul Qu\'ebec and Compute
Canada and with operation funded by the Canada Foundation for
Innovation (CFI), Minist\'ere de l'\'Economie, de l'Innovation et des
Exportations du Quebec (MEIE), RMGA and the Fonds de recherche du
Qu\'ebec - Nature et Technologies (FRQ-NT). Some of the calculations were performed on the ORCA cluster at Cal State Fullerton, which is supported by the Research Corporation for Science Advancement, PHY-1429873, and Cal State Fullerton. New NR simulations were
performed on the AEI {\tt Datura} and {\tt Minerva} clusters. The Markov-chain Monte Carlo
runs were performed on the AEI {\tt Vulcan} cluster.

\appendix

\section{Input values for the dominant-mode merger waveform}
\label{app:IV}

In this Appendix we provide fitting formulae for the values of amplitude, curvature of the amplitude, GW frequency, and slope of the GW frequency of the $(2,2)$
mode at the peak of radiation.

Let $f(\nu,\chi)$ denote any such fit. First, we extract the input values from both the NR and the test-particle Teukolsky-code waveforms used in this paper (see Sec.~\ref{sec:NR}). Let us now focus on how to build fits for peak amplitude, curvature of the amplitude, and slope of the GW frequency; we will discuss the case of the peak frequency later. The fits for these three quantities are built as follows: (i) We fit the Teukolsky data as a function of $\chi$ at fixed $\nu = 10^{-3}$ (test-particle limit) with a suitable function $f_{\rm TPL}(\chi)$; (ii) We fit the NR data as a function of $\chi$ at fixed $\nu = 1/4$ (equal-mass limit) with a suitable function $f_{\rm EQ}(\chi)$; (iii) We assume a polynomial dependence in $\nu$ and impose that $f(\nu = 10^{-3},\chi)=f_{\rm TPL}(\chi)$ and $f(\nu = 1/4,\chi)=f_{\rm EQ}(\chi)$: this fixes two coefficients in the polynomial expansion; (iv) The rest of the coefficients are determined through a global fit over the whole $(\nu,\chi)$ parameter space where we have NR data. For the peak GW frequency we build test-particle-limit and equal-mass-limit fits as in (i) and (ii), but then we prescribe a linear dependence on $\nu$ for the global fit.

\subsection{Amplitude at the peak}
The test-particle-limit and equal mass (Z=TPL  or EQ) fit read
\be
f_{\rm Z}(\chi)/\nu = \sum_{i=0}^3 p^{(\rm Z)}_i \chi^i\,,
\ee
with
\begin{center}
\begin{tabular}{lcl}
$p^{(\rm TPL)}_0 =1.452857,$ & &$p^{(\rm EQ)}_0 =1.577458,$\\
$p^{(\rm TPL)}_1 = 0.166134,$ & &$p^{(\rm EQ)}_1 = - 0.007695,$\\
$p^{(\rm TPL)}_2 = 0.027356,$ & &$p^{(\rm EQ)}_2 = 0.021887,$\\
$p^{(\rm TPL)}_3 = -0.020073,$ & &$p^{(\rm EQ)}_3 = 0.023268.$
\end{tabular}
\end{center}
The global fit reads
\be
 f(\nu,\chi)/\nu= \sum_{i=0}^2 A_i \nu^i\,,
 \ee
where $A_0$ and $A_2$ are fixed by requiring that the test-particle-limit and equal-mass-limit fits are recovered exactly when $\nu = 10^{-3}$ and $\nu = 1/4$, respectively, and $A_1 = \sum_{k=0}^3 e_k\chi^k$, with
\begin{center}
\begin{tabular}{lcl}
$e_0 =  -0.034424,$&&$e_1 = -1.218066,$\\
$e_2 = -0.568373,$ && $e_3 = 0.401114.$
\end{tabular}
\end{center}

\subsection{Curvature of amplitude at the peak}
The test-particle-limit and equal mass fit read
\bea
f_{\rm TPL}(\chi)/\nu &=& \sum_{i=1}^3 p^{(\rm TPL)}_i (\chi-1)^i\,,\\
f_{\rm EQ}(\chi)/\nu &=& \sum_{i=0}^1 p^{(\rm EQ)}_i \chi^i\,,
\eea
with
\begin{center}
\begin{tabular}{lcl}
$p^{(\rm TPL)}_1 =0.00239561,$ && $p^{(\rm EQ)}_0 =-0.00412651,$\\
$p^{(\rm TPL)}_2 = -0.00019274,$ && $p^{(\rm EQ)}_1 = 0.00222400.$\\
$p^{(\rm TPL)}_3 = -0.00029666.$ &&
\end{tabular}
\end{center}
The global fit reads
\be
 f(\nu,\chi)/\nu= \sum_{i=0}^2 A_i \nu^i\,,
 \ee
where $A_0$ and $A_2$ are fixed by requiring that the test-particle-limit and equal-mass-limit fits are recovered exactly when $\nu = 10^{-3}$ and $\nu = 1/4$, respectively, and $A_1 = \sum_{k=0}^1 e_k\chi^k$, with
\begin{center}
\begin{tabular}{lcl}
$e_0 =  -0.00577654,$ && $e_1 = 0.00103086.$
\end{tabular}
\end{center}

\subsection{GW frequency at the peak}
The test-particle-limit fit reads
\begin{align}
f_{\rm TPL}(\chi) &= p^{(\rm TPL)}_0 + (p^{(\rm TPL)}_1 + p^{(\rm TPL)}_2 \chi) \nonumber\\
&\times\log{(p^{(\rm TPL)}_3 - p^{(\rm TPL)}_4 \chi)}\,,
\end{align}
with
\begin{center}
\begin{tabular}{lcl}
$p^{(\rm TPL)}_0 =0.562679,$ && $p^{(\rm TPL)}_1 = -0.087062,$\\
$p^{(\rm TPL)}_2 = 0.001743,$ && $p^{(\rm TPL)}_3 = 25.850378,$\\
$p^{(\rm TPL)}_4 = 25.819795.$ &&
\end{tabular}
\end{center}
The equal-mass-limit fit reads
\begin{align}
f_{\rm EQ}(\chi) &= p^{(\rm TPL)}_0 + (p^{(\rm TPL)}_1 + p^{(\rm TPL)}_2 \chi)\nonumber\\
&\times \log{(p^{(\rm EQ)}_3 - p^{(\rm EQ)}_4 \chi)}\,,
\end{align}
with
\begin{center}
\begin{tabular}{lcl}
$p^{(\rm EQ)}_3 =10.262073,$ && $p^{(\rm EQ)}_4 = 7.629922.$
\end{tabular}
\end{center}
The global fit reads
\be
 f(\nu,\chi)= p^{(\rm TPL)}_0 + (p^{(\rm TPL)}_1 + p^{(\rm TPL)}_2 \chi) \log{(A_3 - A_4 \chi)}\,,
 \ee
 with
 \begin{align}
A_3 &= p_3^{(\rm EQ)} + 4 (p_3^{(\rm EQ)} - p_3^{(\rm TPL)}) (\nu - 1/4)\,,\nonumber\\
A_4 &= p_4^{(\rm EQ)} + 4 (p_4^{(\rm EQ)} - p_4^{(\rm TPL)}) (\nu - 1/4)\,.
\end{align}

\subsection{Slope of GW frequency at the peak}
The test-particle-limit fit reads
\begin{align}
f_{\rm TPL}(\chi) &= p^{(\rm TPL)}_0
 + (p^{(\rm TPL)}_1 + p^{(\rm TPL)}_2 \chi)  \nonumber\\
 &\times \log{(p^{(\rm TPL)}_3 - p^{(\rm TPL)}_4 \chi)}\,,
\end{align}
with
\begin{center}
\begin{tabular}{lcl}
$p^{(\rm TPL)}_0 =-0.011210,$ && $p^{(\rm TPL)}_1 = 0.004087,$\\
$p^{(\rm TPL)}_2 = 0.000633,$ && $p^{(\rm TPL)}_3 = 68.474666,$\\
$p^{(\rm TPL)}_4 = 58.301488.$ &&
\end{tabular}
\end{center}
The equal-mass-limit fit reads
\be
f_{\rm EQ}(\chi) = \sum_{i=0}^1 p^{(\rm EQ)}_i \chi^i\,,
\ee
with
\begin{center}
\begin{tabular}{lcl}
$p^{(\rm EQ)}_0 =0.011282,$ && $p^{(\rm EQ)}_1 = 0.000287.$
\end{tabular}
\end{center}
The global fit reads
\be
 f(\nu,\chi)= \sum_{i=0}^2 A_i \nu^i\,,
 \ee
where $A_0$ and $A_2$ are fixed by requiring that the test-particle-limit and equal-mass-limit fits are recovered exactly when $\nu = 10^{-3}$ and $\nu = 1/4$, respectively, and $A_1 = \sum_{k=0}^1 e_k\chi^k$, with
\begin{center}
\begin{tabular}{lcl}
$e_0 =  0.015743,$ && $e_1 = 0.022442.$
\end{tabular}
\end{center}

\section{Phenomenological merger-ringdown model}
\label{app:newRD}

Here, we discuss the performance of our phenomenological
merger-ringdown model (see Sec.~\ref{subsec:RDmodel}), 
which builds on previous work~\cite{Baker:2008mj,Damour:2014yha,Nagar:2016iwa}, 
and we also compare it to the model of Ref.~\cite{Nagar:2016iwa}. In
order to more easily describe the comparison that we perform, we first
summarize some similarities and differences between the two models.

\begin{figure*}[tb]
  \includegraphics[width=.95\columnwidth]{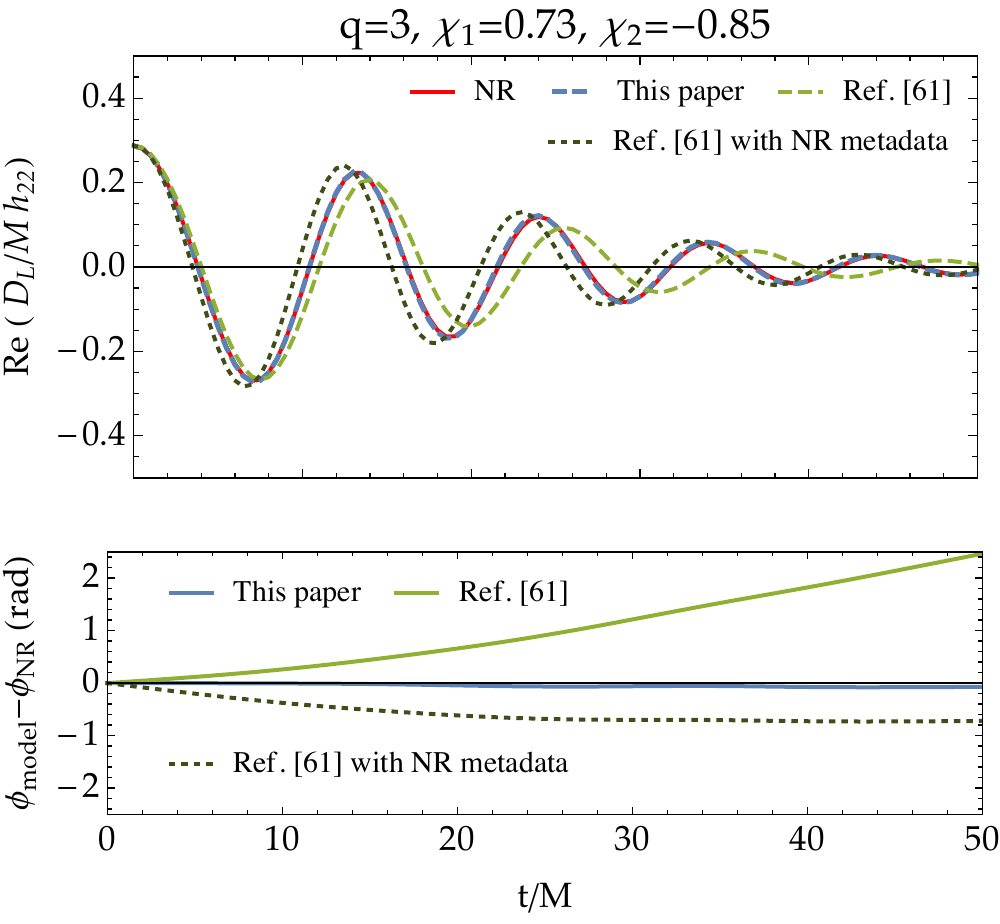}
  \hspace{1cm}
   \includegraphics[width=.95\columnwidth]{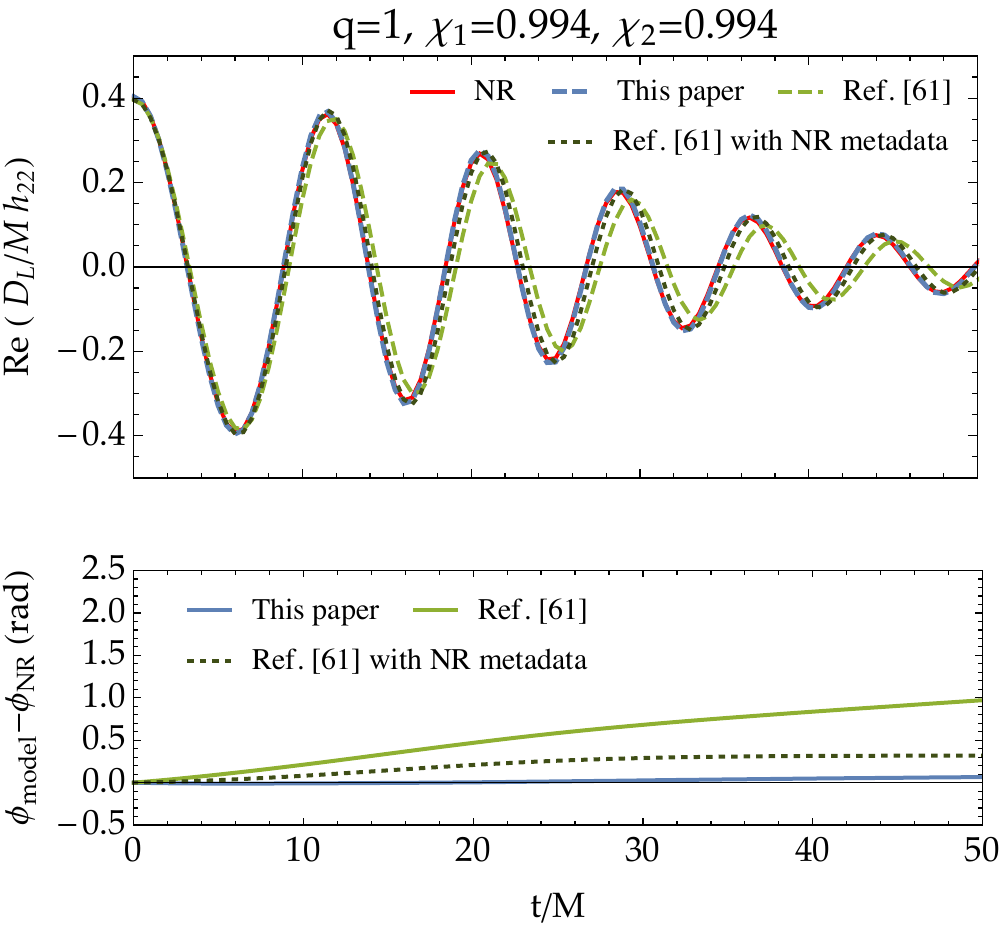}
   \caption{\label{fig:RDwfs}Waveform comparison in the time-domain between the merger-ringdown models of this paper and of Ref.~\cite{Nagar:2016iwa} and 
NR for two configurations in our NR catalog. The one in the left panel $(q,\chi_1,\chi_2)=(3,0.73,-0.85)$ was used in the calibration of the SEOBNRv4 model only. The one in the right panel, $(q,\chi_1,\chi_2)=(1,0.994,0.994)$ entered the calibration of both models. Phases are aligned at $t=0$ which corresponds to the amplitude peak. The top plots show the real part of each waveform while the bottom plots show the phase difference between the models and NR. The green line corresponds to the purely standalone model described in Ref. [61] whereas the dotted line labeled ``Ref.~[61]+NR metadata'' corresponds to the same model but with the QNM determined using the NR metadata for the remnant properties and the tables in Ref.~\cite{Berti:2005ys} instead of using the fits in Table I of Ref.~[61].}
\end{figure*}

The model of Ref.~\cite{Nagar:2016iwa} is a complete model of the post-merger phase in that it also specifies initial conditions at the time of merger: the value of the amplitude (its first derivative is 0
by construction) and of the frequency at that point are prescribed by
fits (functions of $\nu$ and the spin combination $a_0=(m_1
\chi_1+m_2 \chi_2)/M$ (see the last two rows of Table 1 in Ref.~\cite{Nagar:2016iwa}). By contrast, the model presented in
Sec.~\ref{subsec:RDmodel} attaches a ringdown portion to any
inspiral-plunge waveform by imposing a $C^1$ behaviour at the
attachment point (assumed to be the amplitude's peak). However, in
the context of the EOBNR model developed in this paper (\texttt{SEOBNRv4}),
the values of the amplitude and the frequency at the attachment point
are imposed by the explicit expressions (fits to NR) given in
Appendix~\ref{app:IV}. We can therefore obtain a full
model of the post-merger phase by simply combining the information in
Sec.~\ref{subsec:RDmodel} and in Appendix~\ref{app:IV}. 

Both models make use of a phenomenological ansatz where the dominant QNM is
factored out. In our EOBNR model, the value of the dominant (complex)
QNM frequency is obtained by first computing the mass and spin of the remnant object
from the fitting formulae in Refs.~\cite{Taracchini:2013rva,Hofmann:2016yih} (which were calibrated
using hundreds of publicly available NR waveforms) and then
interpolating results from Ref.~\cite{Berti:2005ys}. 
By contrast, Ref.~\cite{Nagar:2016iwa} directly provides fits allowing to reconstruct the real and imaginary part of the QNM as functions of $\nu$ and $a_0$.
  \footnote{
Note that the imaginary part is not directly fitted. Instead, fits for the
  frequency at merger and for a combination ($\Delta\omega$ in their notation) of the frequency at merger and of the real part of the QNM are given.
  }
The authors of Ref.~\cite{Nagar:2016iwa} however informed us \cite{NagarPrivate} that the comparisons to NR that are shown in their paper do not use those fits to reconstruct the QNM. Instead, for each NR waveform against which their model is compared, the mass and spin of the remnant BH is read from the NR metadata and used together with the tables in Ref.~\cite{Berti:2005ys} to determine the QNM. This leads to an improved behaviour as the real part of the QNM, a crucial ingredient of the model, is determined much more precisely than using the fits. However, such a procedure can only be applied at points in parameter space where an NR waveform exists and cannot be considered as part of a final standalone analytical model. In Fig.~\ref{fig:RDwfs}, we show the performance using both implementations. We expect that using the same tools as the ones in our model (very accurate fits existing in the literature for the remnant properties and interpolation of the tables in Ref.~\cite{Berti:2005ys}) to determine the QNM will lead to an intermediate performance, likely closer to using the NR metadata.

Given a set of physical parameters $(m_1,m_2,\chi_1,\chi_2)$ for the
binary components, our model provides a complete prescription
for the ringdown waveform (by which we mean the portion of waveform
starting at the amplitude peak of the full signal). For the model of Ref.~\cite{Nagar:2016iwa},
we additionally need the mass of the final BH. Since here we want
to compare the models to NR, we know the \emph{exact}
value for the final mass for each configuration and use it for the
model of Ref.~\cite{Nagar:2016iwa}. For our EOBNR model, we keep using the value provided by the fit in
Ref.~\cite{Taracchini:2013rva}. We can then compare directly to NR (without
using any further information from the NR waveform itself). All phases
are set to 0 at $t=0$.
Two examples are shown in Fig.~\ref{fig:RDwfs}, namely the $(q,\chi_1,\chi_2)=(3,0.73,-0.85)$ configuration (left panel),  
which was used to calibrate our model but not the model of Ref \cite{Nagar:2016iwa} and the $(q,\chi_1,\chi_2)=(1,0.994,0.994)$ configuration (right panel), which was used in both models. As we can see, whereas in our merger-ringdown model (i.e., \texttt{SEOBNRv4}), the dephasing remains of the order of $0.1$ rad throughout the merger-ringdown phase, in the model of Ref.~\cite{Nagar:2016iwa} 
it grows to more than 1 rad. Using the NR metadata to determine the QNM instead of the fits provided in Table I improves the behaviour of the model of Ref.~\cite{Nagar:2016iwa} by removing the linear drift of the phase, since the ringdown frequency is now exactly known. Even using this additional information coming from NR, the asymptotic dephasing reaches several tenths of radians.

 \begin{figure}[tb]
   \includegraphics[width=.9\columnwidth]{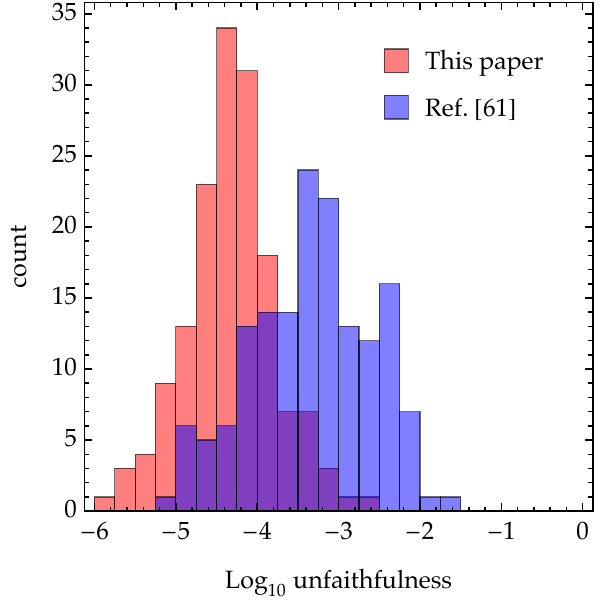}
    \caption{ \label{fig:rdcomparisonshist}Comparison of the performance of the merger-ringdown models of this paper and of Ref.~\cite{Nagar:2016iwa} over the catalog of 147 NR waveforms used in this paper. For each configuration, the faithfulness of each model to NR is computed at a total mass such that the merger frequency is at $50$Hz.}
 \end{figure}

Going beyond this time-domain comparison, we can also try to quantify
the performance of the models in terms of faithfulness. In order to avoid
computing matches between post-merger waveforms only (which would
require tapering the signal at $t=0$), we attach the post-merger
portion predicted by each model to the NR waveform itself cut at the
peak. However, there is an obvious problem in doing this. The
resulting hybrid waveforms would not even be continuous at the peak
since both the fits of Appendix~\ref{app:IV} and the last two rows of Table I of
Ref.~\cite{Nagar:2016iwa} do not \emph{exactly} reproduce the NR
values. For the purpose of this comparison only, we therefore replace
the values predicted by the fits in each model by the actual NR
values.\footnote{There is an additional subtlety in doing this in the
  model of Ref.~\cite{Nagar:2016iwa}. The QNM frequency is not directly provided as a
  fit. Using the NR value for $\omega_{22}$ but keeping the fit value
  for their $\Delta \omega$ leads to a slightly modified value of the
  dominant QNM frequency. To make sure that the QNM frequency
  predicted by the model in Ref.~\cite{Nagar:2016iwa} is preserved, we first compute it using
  $\omega_1=M_{\rm BH} \omega_{22}^{\rm mrg} - \Delta \omega$ where $\omega_{22}^{\rm mrg}$ and $ \Delta \omega$ are evaluated using the
  fits, and then use the NR value for $\omega_{22}^{\rm mrg}$
  elsewhere, redefining $ \Delta \omega = \omega - M_{\rm BH}
  \omega_{22}^{\rm NR}$ to ensure continuity at $t=0$.} We then
compute the faithfulness between the original NR waveform and the hybrid ones,
choosing a total mass $M$ such that the peak of
the waveform lies at 50\,Hz so that the merger-ringdown is in the most
sensitive spot of the detector's noise curve. The distribution of
faithfulness obtained across the catalog is shown in
Fig.~\ref{fig:rdcomparisonshist}. In most cases (all cases for \texttt{SEOBNRv4}),
both models lead to negligible unfaithfulness from the point of
view of data-analysis applications. The model of Ref.~\cite{Nagar:2016iwa} however features a
tail extending above the $1\%$ level, mainly composed of waveforms with very
antisymmetric spin configurations (such as the one shown in the left panel of Fig.~\ref{fig:RDwfs}) that were not included in the calibration of this model since
they were not publicly available at the time.
In summary, we find that calibrating our phenomenological expressions for the merger-ringdown part of the model to an extensive NR catalog such as the one used in this paper is crucial to obtain a highly-accurate model everywhere in parameter space. New, future NR waveforms 
will allow us to further test and extend the model.

\section{Existing numerical-relativity simulations used in this work}
\label{app:NRsims}

In addition to the new NR waveforms listed in Table
\ref{table:nrsims}, we have also used previously-produced waveforms,
described in Secs.~\ref{sec:original}--\ref{sec:bam}.  Some waveforms
were used only for validation of the model; in this case, they
follow the calibration waveforms and are separated by a horizontal line.  The
table columns are the same as in Table \ref{table:nrsims}.

\newcommand{\numoriginal}{38}
\newcommand{\numpublicnew}{9}
\newcommand{\numchu}{94}

\subsection{SXS waveforms from Ref.~\cite{Mroue:2013xna}}
\label{sec:original}
\begingroup
 \begin{longtable}{ccccccc}
  \doubleline
  ID & $q$ & $\chi_1$ & $\chi_2$ & $e$ & $M\omega_{22}$ & $N_\mathrm{orb}$\\
\hline
SXS:BBH:0004 & $1.0$ & $-0.50$ & $+0.00$ & $3.7\times 10^{-4}$ & $0.01151$ & $30.2$ \\
SXS:BBH:0005 & $1.0$ & $+0.50$ & $+0.00$ & $2.5\times 10^{-4}$ & $0.01227$ & $30.2$ \\
SXS:BBH:0007 & $1.5$ & $+0.00$ & $+0.00$ & $4.2\times 10^{-4}$ & $0.01229$ & $29.1$ \\
SXS:BBH:0013 & $1.5$ & $+0.50$ & $+0.00$ & $1.4\times 10^{-4}$ & $0.01444$ & $23.8$ \\
SXS:BBH:0016 & $1.5$ & $-0.50$ & $+0.00$ & $4.2\times 10^{-4}$ & $0.01149$ & $30.7$ \\
SXS:BBH:0019 & $1.5$ & $-0.50$ & $+0.50$ & $7.6\times 10^{-5}$ & $0.01460$ & $20.4$ \\
SXS:BBH:0025 & $1.5$ & $+0.50$ & $-0.50$ & $7.6\times 10^{-5}$ & $0.01456$ & $22.4$ \\
SXS:BBH:0030 & $3.0$ & $+0.00$ & $+0.00$ & $2.0\times 10^{-3}$ & $0.01775$ & $18.2$ \\
SXS:BBH:0036 & $3.0$ & $-0.50$ & $+0.00$ & $5.1\times 10^{-4}$ & $0.01226$ & $31.7$ \\
SXS:BBH:0045 & $3.0$ & $+0.50$ & $-0.50$ & $6.4\times 10^{-4}$ & $0.01748$ & $21.0$ \\
SXS:BBH:0046 & $3.0$ & $-0.50$ & $-0.50$ & $2.6\times 10^{-4}$ & $0.01771$ & $14.4$ \\
SXS:BBH:0047 & $3.0$ & $+0.50$ & $+0.50$ & $4.7\times 10^{-4}$ & $0.01743$ & $22.7$ \\
SXS:BBH:0056 & $5.0$ & $+0.00$ & $+0.00$ & $4.9\times 10^{-4}$ & $0.01589$ & $28.8$ \\
SXS:BBH:0060 & $5.0$ & $-0.50$ & $+0.00$ & $3.4\times 10^{-3}$ & $0.01608$ & $23.2$ \\
SXS:BBH:0061 & $5.0$ & $+0.50$ & $+0.00$ & $4.2\times 10^{-3}$ & $0.01578$ & $34.5$ \\
SXS:BBH:0063 & $8.0$ & $+0.00$ & $+0.00$ & $2.8\times 10^{-4}$ & $0.01938$ & $25.8$ \\
SXS:BBH:0064 & $8.0$ & $-0.50$ & $+0.00$ & $4.9\times 10^{-4}$ & $0.01968$ & $19.2$ \\
SXS:BBH:0065 & $8.0$ & $+0.50$ & $+0.00$ & $3.7\times 10^{-3}$ & $0.01887$ & $34.0$ \\
SXS:BBH:0148 & $1.0$ & $-0.44$ & $-0.44$ & $2.0\times 10^{-5}$ & $0.01634$ & $15.5$ \\
SXS:BBH:0149 & $1.0$ & $-0.20$ & $-0.20$ & $1.8\times 10^{-4}$ & $0.01614$ & $17.1$ \\
SXS:BBH:0150 & $1.0$ & $+0.20$ & $+0.20$ & $2.9\times 10^{-4}$ & $0.01591$ & $19.8$ \\
SXS:BBH:0151 & $1.0$ & $-0.60$ & $-0.60$ & $2.5\times 10^{-4}$ & $0.01575$ & $14.5$ \\
SXS:BBH:0152 & $1.0$ & $+0.60$ & $+0.60$ & $4.3\times 10^{-4}$ & $0.01553$ & $22.6$ \\
SXS:BBH:0153 & $1.0$ & $+0.85$ & $+0.85$ & $8.3\times 10^{-4}$ & $0.01539$ & $24.5$ \\
SXS:BBH:0154 & $1.0$ & $-0.80$ & $-0.80$ & $3.3\times 10^{-4}$ & $0.01605$ & $13.2$ \\
SXS:BBH:0155 & $1.0$ & $+0.80$ & $+0.80$ & $4.7\times 10^{-4}$ & $0.01543$ & $24.1$ \\
SXS:BBH:0156 & $1.0$ & $-0.95$ & $-0.95$ & $5.4\times 10^{-4}$ & $0.01643$ & $12.4$ \\
SXS:BBH:0157 & $1.0$ & $+0.95$ & $+0.95$ & $1.4\times 10^{-4}$ & $0.01535$ & $25.2$ \\
SXS:BBH:0158 & $1.0$ & $+0.97$ & $+0.97$ & $7.9\times 10^{-4}$ & $0.01565$ & $25.3$ \\
SXS:BBH:0159 & $1.0$ & $-0.90$ & $-0.90$ & $5.6\times 10^{-4}$ & $0.01588$ & $12.7$ \\
SXS:BBH:0160 & $1.0$ & $+0.90$ & $+0.90$ & $4.2\times 10^{-4}$ & $0.01538$ & $24.8$ \\
SXS:BBH:0166 & $6.0$ & $+0.00$ & $+0.00$ & $4.4\times 10^{-5}$ & $0.01940$ & $21.6$ \\
SXS:BBH:0167 & $4.0$ & $+0.00$ & $+0.00$ & $9.9\times 10^{-5}$ & $0.02054$ & $15.6$ \\
SXS:BBH:0169 & $2.0$ & $+0.00$ & $+0.00$ & $1.2\times 10^{-4}$ & $0.01799$ & $15.7$ \\
SXS:BBH:0170 & $1.0$ & $+0.44$ & $+0.44$ & $1.3\times 10^{-4}$ & $0.00842$ & $15.5$ \\
SXS:BBH:0172 & $1.0$ & $+0.98$ & $+0.98$ & $7.8\times 10^{-4}$ & $0.01540$ & $25.4$ \\
SXS:BBH:0174 & $3.0$ & $+0.50$ & $+0.00$ & $2.9\times 10^{-4}$ & $0.01337$ & $35.5$ \\
SXS:BBH:0180 & $1.0$ & $+0.00$ & $+0.00$ & $5.1\times 10^{-5}$ & $0.01227$ & $28.2$ \\

  \doubleline
\end{longtable}
\label{tab:original}
\endgroup

\subsection{SXS waveforms from Ref.~\cite{Kumar:2015tha,Lovelace:2010ne,Scheel:2014ina}}
\label{sec:publicnew}
\begingroup
\begin{longtable}{ccccccc}
  \doubleline
  ID & $q$ & $\chi_1$ & $\chi_2$ & $e$ & $M\omega_{22}$ & $N_\mathrm{orb}$\\
\hline
SXS:BBH:0177 & $1.0$ & $+0.99$ & $+0.99$ & $1.3\times 10^{-3}$ & $0.01543$ & $25.4$ \\
SXS:BBH:0178 & $1.0$ & $+0.99$ & $+0.99$ & $8.6\times 10^{-4}$ & $0.01570$ & $25.4$ \\
SXS:BBH:0202 & $7.0$ & $+0.60$ & $+0.00$ & $9.0\times 10^{-5}$ & $0.01324$ & $62.1$ \\
SXS:BBH:0203 & $7.0$ & $+0.40$ & $+0.00$ & $1.4\times 10^{-5}$ & $0.01322$ & $58.5$ \\
SXS:BBH:0204 & $7.0$ & $+0.40$ & $+0.00$ & $1.7\times 10^{-4}$ & $0.01044$ & $88.4$ \\
SXS:BBH:0205 & $7.0$ & $-0.40$ & $+0.00$ & $7.0\times 10^{-5}$ & $0.01325$ & $44.9$ \\
SXS:BBH:0206 & $7.0$ & $-0.40$ & $+0.00$ & $1.6\times 10^{-4}$ & $0.01037$ & $73.2$ \\
SXS:BBH:0207 & $7.0$ & $-0.60$ & $+0.00$ & $1.7\times 10^{-4}$ & $0.01423$ & $36.1$ \\
\hline
SXS:BBH:0306 & $1.3$ & $+0.96$ & $-0.90$ & $1.5\times 10^{-3}$ & $0.02098$ & $12.6$ \\

  \doubleline
\end{longtable}
\endgroup

\subsection{SXS waveforms from Ref.~\cite{Chu:2015kft}}
\label{sec:chu}
\begingroup
\begin{longtable}{ccccccc}
  \doubleline
  ID & $q$ & $\chi_1$ & $\chi_2$ & $e$ & $M\omega_{22}$ & $N_\mathrm{orb}$\\
\hline
\endfirsthead
ID & $q$ & $\chi_1$ & $\chi_2$ & $e$ & $M\omega_{22}$ & $N_\mathrm{orb}$\\
\hline
\endhead
SXS:BBH:0290 & $3.0$ & $+0.60$ & $+0.40$ & $9.0\times 10^{-5}$ & $0.01758$ & $24.2$ \\
SXS:BBH:0291 & $3.0$ & $+0.60$ & $+0.60$ & $5.0\times 10^{-5}$ & $0.01764$ & $24.5$ \\
SXS:BBH:0289 & $3.0$ & $+0.60$ & $+0.00$ & $2.3\times 10^{-4}$ & $0.01711$ & $23.8$ \\
SXS:BBH:0285 & $3.0$ & $+0.40$ & $+0.60$ & $1.6\times 10^{-4}$ & $0.01732$ & $23.8$ \\
SXS:BBH:0261 & $3.0$ & $-0.73$ & $+0.85$ & $1.0\times 10^{-4}$ & $0.01490$ & $21.5$ \\
SXS:BBH:0293 & $3.0$ & $+0.85$ & $+0.85$ & $9.0\times 10^{-5}$ & $0.01813$ & $25.6$ \\
SXS:BBH:0280 & $3.0$ & $+0.27$ & $+0.85$ & $9.7\times 10^{-5}$ & $0.01707$ & $23.6$ \\
SXS:BBH:0257 & $2.0$ & $+0.85$ & $+0.85$ & $1.1\times 10^{-4}$ & $0.01633$ & $24.8$ \\
SXS:BBH:0279 & $3.0$ & $+0.23$ & $-0.85$ & $6.0\times 10^{-5}$ & $0.01629$ & $22.6$ \\
SXS:BBH:0274 & $3.0$ & $-0.23$ & $+0.85$ & $1.6\times 10^{-4}$ & $0.01603$ & $22.4$ \\
SXS:BBH:0258 & $2.0$ & $+0.87$ & $-0.85$ & $1.8\times 10^{-4}$ & $0.01612$ & $22.8$ \\
SXS:BBH:0248 & $2.0$ & $+0.13$ & $+0.85$ & $7.0\times 10^{-5}$ & $0.01552$ & $23.2$ \\
SXS:BBH:0232 & $1.0$ & $+0.90$ & $+0.50$ & $2.8\times 10^{-4}$ & $0.01558$ & $23.9$ \\
SXS:BBH:0229 & $1.0$ & $+0.65$ & $+0.25$ & $3.1\times 10^{-4}$ & $0.01488$ & $23.1$ \\
SXS:BBH:0231 & $1.0$ & $+0.90$ & $+0.00$ & $1.0\times 10^{-4}$ & $0.01487$ & $23.1$ \\
SXS:BBH:0239 & $2.0$ & $-0.37$ & $+0.85$ & $9.1\times 10^{-5}$ & $0.01478$ & $22.2$ \\
SXS:BBH:0252 & $2.0$ & $+0.37$ & $-0.85$ & $3.8\times 10^{-4}$ & $0.01488$ & $22.5$ \\
SXS:BBH:0219 & $1.0$ & $-0.50$ & $+0.90$ & $3.3\times 10^{-4}$ & $0.01484$ & $22.4$ \\
SXS:BBH:0211 & $1.0$ & $-0.90$ & $+0.90$ & $2.6\times 10^{-4}$ & $0.01411$ & $22.3$ \\
SXS:BBH:0233 & $2.0$ & $-0.87$ & $+0.85$ & $6.0\times 10^{-5}$ & $0.01423$ & $22.0$ \\
SXS:BBH:0243 & $2.0$ & $-0.13$ & $-0.85$ & $1.8\times 10^{-4}$ & $0.01378$ & $23.3$ \\
SXS:BBH:0214 & $1.0$ & $-0.62$ & $-0.25$ & $1.9\times 10^{-4}$ & $0.01264$ & $24.4$ \\
SXS:BBH:0209 & $1.0$ & $-0.90$ & $-0.50$ & $1.7\times 10^{-4}$ & $0.01137$ & $27.0$ \\
SXS:BBH:0226 & $1.0$ & $+0.50$ & $-0.90$ & $2.4\times 10^{-4}$ & $0.01340$ & $22.9$ \\
SXS:BBH:0286 & $3.0$ & $+0.50$ & $+0.50$ & $8.0\times 10^{-5}$ & $0.01693$ & $24.1$ \\
SXS:BBH:0253 & $2.0$ & $+0.50$ & $+0.50$ & $6.7\times 10^{-5}$ & $0.01397$ & $28.8$ \\
SXS:BBH:0267 & $3.0$ & $-0.50$ & $-0.50$ & $5.6\times 10^{-5}$ & $0.01410$ & $23.4$ \\
SXS:BBH:0218 & $1.0$ & $-0.50$ & $+0.50$ & $7.8\times 10^{-5}$ & $0.01217$ & $29.1$ \\
SXS:BBH:0238 & $2.0$ & $-0.50$ & $-0.50$ & $6.9\times 10^{-5}$ & $0.01126$ & $32.0$ \\
SXS:BBH:0288 & $3.0$ & $+0.60$ & $-0.40$ & $1.9\times 10^{-4}$ & $0.01729$ & $23.5$ \\
SXS:BBH:0287 & $3.0$ & $+0.60$ & $-0.60$ & $7.0\times 10^{-5}$ & $0.01684$ & $23.5$ \\
SXS:BBH:0283 & $3.0$ & $+0.30$ & $+0.30$ & $7.6\times 10^{-5}$ & $0.01646$ & $23.5$ \\
SXS:BBH:0282 & $3.0$ & $+0.30$ & $+0.00$ & $7.5\times 10^{-5}$ & $0.01629$ & $23.3$ \\
SXS:BBH:0281 & $3.0$ & $+0.30$ & $-0.30$ & $6.7\times 10^{-5}$ & $0.01618$ & $23.2$ \\
SXS:BBH:0277 & $3.0$ & $+0.00$ & $+0.30$ & $7.0\times 10^{-5}$ & $0.01595$ & $22.9$ \\
SXS:BBH:0284 & $3.0$ & $+0.40$ & $-0.60$ & $1.5\times 10^{-4}$ & $0.01656$ & $22.8$ \\
SXS:BBH:0278 & $3.0$ & $+0.00$ & $+0.60$ & $2.1\times 10^{-4}$ & $0.01623$ & $22.8$ \\
SXS:BBH:0256 & $2.0$ & $+0.60$ & $+0.60$ & $7.8\times 10^{-5}$ & $0.01598$ & $23.9$ \\
SXS:BBH:0230 & $1.0$ & $+0.80$ & $+0.80$ & $1.3\times 10^{-4}$ & $0.01542$ & $24.2$ \\
SXS:BBH:0255 & $2.0$ & $+0.60$ & $+0.00$ & $4.0\times 10^{-5}$ & $0.01580$ & $23.3$ \\
SXS:BBH:0276 & $3.0$ & $+0.00$ & $-0.30$ & $6.7\times 10^{-5}$ & $0.01559$ & $23.0$ \\
SXS:BBH:0251 & $2.0$ & $+0.30$ & $+0.30$ & $7.5\times 10^{-5}$ & $0.01514$ & $23.5$ \\
SXS:BBH:0250 & $2.0$ & $+0.30$ & $+0.00$ & $7.5\times 10^{-5}$ & $0.01503$ & $23.2$ \\
SXS:BBH:0271 & $3.0$ & $-0.30$ & $+0.00$ & $6.3\times 10^{-5}$ & $0.01508$ & $22.5$ \\
SXS:BBH:0249 & $2.0$ & $+0.30$ & $-0.30$ & $7.2\times 10^{-5}$ & $0.01478$ & $23.2$ \\
SXS:BBH:0275 & $3.0$ & $+0.00$ & $-0.60$ & $1.2\times 10^{-4}$ & $0.01569$ & $22.6$ \\
SXS:BBH:0254 & $2.0$ & $+0.60$ & $-0.60$ & $6.0\times 10^{-5}$ & $0.01541$ & $22.9$ \\
SXS:BBH:0269 & $3.0$ & $-0.40$ & $+0.60$ & $1.2\times 10^{-4}$ & $0.01563$ & $22.3$ \\
SXS:BBH:0225 & $1.0$ & $+0.40$ & $+0.80$ & $3.5\times 10^{-4}$ & $0.01536$ & $23.5$ \\
SXS:BBH:0270 & $3.0$ & $-0.30$ & $-0.30$ & $6.2\times 10^{-5}$ & $0.01482$ & $22.8$ \\
SXS:BBH:0245 & $2.0$ & $+0.00$ & $-0.30$ & $6.8\times 10^{-5}$ & $0.01441$ & $23.0$ \\
SXS:BBH:0242 & $2.0$ & $-0.30$ & $+0.30$ & $6.7\times 10^{-5}$ & $0.01417$ & $23.1$ \\
SXS:BBH:0223 & $1.0$ & $+0.30$ & $+0.00$ & $6.7\times 10^{-5}$ & $0.01402$ & $23.3$ \\
SXS:BBH:0241 & $2.0$ & $-0.30$ & $+0.00$ & $6.6\times 10^{-5}$ & $0.01394$ & $23.1$ \\
SXS:BBH:0240 & $2.0$ & $-0.30$ & $-0.30$ & $6.4\times 10^{-5}$ & $0.01359$ & $23.5$ \\
SXS:BBH:0222 & $1.0$ & $-0.30$ & $+0.00$ & $7.4\times 10^{-5}$ & $0.01324$ & $23.6$ \\
SXS:BBH:0228 & $1.0$ & $+0.60$ & $+0.60$ & $3.2\times 10^{-4}$ & $0.01543$ & $23.5$ \\
SXS:BBH:0247 & $2.0$ & $+0.00$ & $+0.60$ & $1.0\times 10^{-4}$ & $0.01530$ & $22.6$ \\
SXS:BBH:0263 & $3.0$ & $-0.60$ & $+0.60$ & $1.9\times 10^{-4}$ & $0.01526$ & $22.0$ \\
SXS:BBH:0266 & $3.0$ & $-0.60$ & $+0.40$ & $1.8\times 10^{-4}$ & $0.01488$ & $22.0$ \\
SXS:BBH:0227 & $1.0$ & $+0.60$ & $+0.00$ & $3.1\times 10^{-4}$ & $0.01452$ & $23.1$ \\
SXS:BBH:0221 & $1.0$ & $-0.40$ & $+0.80$ & $2.7\times 10^{-4}$ & $0.01440$ & $22.7$ \\
SXS:BBH:0237 & $2.0$ & $-0.60$ & $+0.60$ & $6.1\times 10^{-5}$ & $0.01433$ & $22.6$ \\
SXS:BBH:0244 & $2.0$ & $+0.00$ & $-0.60$ & $7.5\times 10^{-5}$ & $0.01422$ & $23.2$ \\
SXS:BBH:0217 & $1.0$ & $-0.60$ & $+0.60$ & $1.5\times 10^{-4}$ & $0.01421$ & $22.7$ \\
SXS:BBH:0215 & $1.0$ & $-0.60$ & $-0.60$ & $1.8\times 10^{-4}$ & $0.01189$ & $25.8$ \\
SXS:BBH:0262 & $3.0$ & $-0.60$ & $+0.00$ & $2.0\times 10^{-4}$ & $0.01473$ & $22.5$ \\
SXS:BBH:0213 & $1.0$ & $-0.80$ & $+0.80$ & $1.4\times 10^{-4}$ & $0.01435$ & $22.3$ \\
SXS:BBH:0265 & $3.0$ & $-0.60$ & $-0.40$ & $9.0\times 10^{-5}$ & $0.01422$ & $23.4$ \\
SXS:BBH:0264 & $3.0$ & $-0.60$ & $-0.60$ & $2.8\times 10^{-4}$ & $0.01410$ & $23.4$ \\
SXS:BBH:0224 & $1.0$ & $+0.40$ & $-0.80$ & $2.5\times 10^{-4}$ & $0.01361$ & $22.9$ \\
SXS:BBH:0236 & $2.0$ & $-0.60$ & $+0.00$ & $1.2\times 10^{-4}$ & $0.01361$ & $23.4$ \\
SXS:BBH:0216 & $1.0$ & $-0.60$ & $+0.00$ & $2.6\times 10^{-4}$ & $0.01300$ & $23.6$ \\
SXS:BBH:0235 & $2.0$ & $-0.60$ & $-0.60$ & $6.1\times 10^{-5}$ & $0.01274$ & $25.1$ \\
SXS:BBH:0220 & $1.0$ & $-0.40$ & $-0.80$ & $1.0\times 10^{-4}$ & $0.01195$ & $25.7$ \\
SXS:BBH:0212 & $1.0$ & $-0.80$ & $-0.80$ & $2.4\times 10^{-4}$ & $0.01087$ & $28.6$ \\
SXS:BBH:0303 & $10.0$ & $+0.00$ & $+0.00$ & $5.1\times 10^{-5}$ & $0.02395$ & $19.3$ \\
SXS:BBH:0300 & $8.5$ & $+0.00$ & $+0.00$ & $5.7\times 10^{-5}$ & $0.02311$ & $18.7$ \\
SXS:BBH:0299 & $7.5$ & $+0.00$ & $+0.00$ & $5.9\times 10^{-5}$ & $0.02152$ & $20.1$ \\
SXS:BBH:0298 & $7.0$ & $+0.00$ & $+0.00$ & $6.1\times 10^{-5}$ & $0.02130$ & $19.7$ \\
SXS:BBH:0297 & $6.5$ & $+0.00$ & $+0.00$ & $6.4\times 10^{-5}$ & $0.02082$ & $19.7$ \\
SXS:BBH:0296 & $5.5$ & $+0.00$ & $+0.00$ & $5.2\times 10^{-5}$ & $0.01668$ & $27.9$ \\
SXS:BBH:0295 & $4.5$ & $+0.00$ & $+0.00$ & $5.2\times 10^{-5}$ & $0.01577$ & $27.8$ \\
SXS:BBH:0259 & $2.5$ & $+0.00$ & $+0.00$ & $5.9\times 10^{-5}$ & $0.01346$ & $28.6$ \\
\hline
SXS:BBH:0292 & $3.0$ & $+0.73$ & $-0.85$ & $1.8\times 10^{-4}$ & $0.01749$ & $23.9$ \\
SXS:BBH:0268 & $3.0$ & $-0.40$ & $-0.60$ & $1.7\times 10^{-4}$ & $0.01473$ & $22.9$ \\
SXS:BBH:0234 & $2.0$ & $-0.85$ & $-0.85$ & $1.4\times 10^{-4}$ & $0.01147$ & $27.8$ \\
SXS:BBH:0273 & $3.0$ & $-0.27$ & $-0.85$ & $2.0\times 10^{-4}$ & $0.01487$ & $22.9$ \\
SXS:BBH:0210 & $1.0$ & $-0.90$ & $+0.00$ & $1.8\times 10^{-4}$ & $0.01248$ & $24.3$ \\
SXS:BBH:0260 & $3.0$ & $-0.85$ & $-0.85$ & $3.5\times 10^{-4}$ & $0.01285$ & $25.8$ \\
SXS:BBH:0302 & $9.5$ & $+0.00$ & $+0.00$ & $6.0\times 10^{-5}$ & $0.02366$ & $19.1$ \\
SXS:BBH:0301 & $9.0$ & $+0.00$ & $+0.00$ & $5.5\times 10^{-5}$ & $0.02338$ & $18.9$ \\
SXS:BBH:0272 & $3.0$ & $-0.30$ & $+0.30$ & $6.4\times 10^{-5}$ & $0.01521$ & $22.7$ \\
SXS:BBH:0246 & $2.0$ & $+0.00$ & $+0.30$ & $7.2\times 10^{-5}$ & $0.01514$ & $22.9$ \\

  \doubleline
\end{longtable}
\endgroup

\subsection{BAM waveform from Ref.~\cite{Khan:2015jqa}}
\label{sec:bam}

\begingroup
\begin{tabular}{ccccccc}
  \doubleline
  ID & $q$ & $\chi_1$ & $\chi_2$ & $e$ & $M\omega_{22}$ & $N_\mathrm{orb}$\\
\hline
BAMq8s85s85 & $8.0$ & $+0.85$ & $+0.85$ & $9.1\times 10^{-3}$ & $0.05476$ & $7.9$ \\

  \doubleline
\end{tabular}
\endgroup

\section{Faithfulness of \texttt{SEOBNRv4} against \texttt{IMRPhenomD}}
\label{app:faith}
In this Appendix we present faithfulness comparisons between the EOBNR model of
this paper  (\texttt{SEOBNRv4}) and the phenomenological
inspiral-merger-ringdown model (\texttt{IMRPhenomD})~\cite{Khan:2015jqa} for
specific ranges of total masses and $\chi_{\textrm{A}}$, in order to gain more
insight into the plots of Fig.~\ref{fig:faithsimv2v4}, where instead all
possible values of these parameters are put together. All plots in this Appendix
use the O1 PSD and the same color coding of Fig.~\ref{fig:faithsimv2v4}, namely:
(i) points with faithfulness above 97\% are white, (ii) points with faithfulness
below 73\% are red, and (iii) all remaining points are colored according to the
legend on the right. We consider three sets of BBHs, each containing $10^6$
points. Spins are uniformly sampled in $-0.99 \leq \chi_{1,2} \leq 0.99$.
Component masses are uniformly sampled in: (i) $1\,M_{\odot} \leq M \leq
25\,M_{\odot}$, subject to the constraint $4\,M_{\odot} \leq M \leq 25\,
M_{\odot}$, for the first set; (ii) $1\,M_{\odot} \leq M \leq 100\,M_{\odot}$,
subject to the constraint $25\,M_{\odot} \leq M \leq 100\, M_{\odot}$, for the
second set; (iii) $1\,M_{\odot} \leq M \leq 200\,M_{\odot}$, subject to the
constraint $100\,M_{\odot} \leq M \leq 200\, M_{\odot}$, for the third set. For
each set of BBHs, we split the data into configurations with
$|\chi_{\textrm{A}}| \leq 0.1$ and those with $|\chi_{\textrm{A}}| \geq 0.5$.
Regions of the $(\nu,\chi_{\textrm{eff}})$ plane that are excluded by the
constraints in $M$ or $\chi_{\textrm{A}}$ are shaded in grey.

\begin{figure}[htb]
\hspace{-0.7cm}\includegraphics[width=0.8\columnwidth]{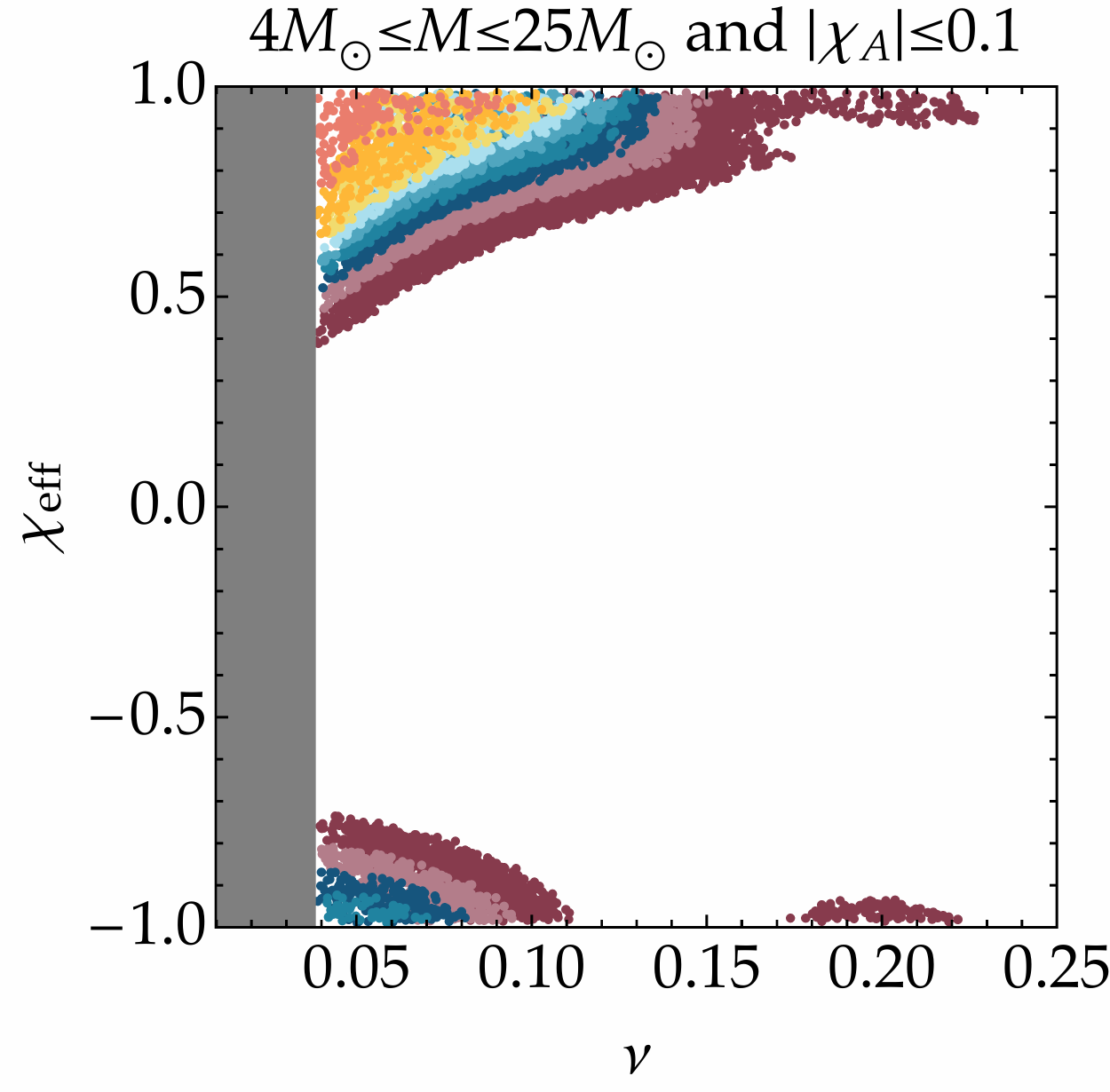} \raisebox{0.0\height}{\includegraphics[width=0.25\columnwidth]{figs/Faiv4PhDLeg.pdf}}

\hspace{-0.7cm}\includegraphics[width=0.8\columnwidth]{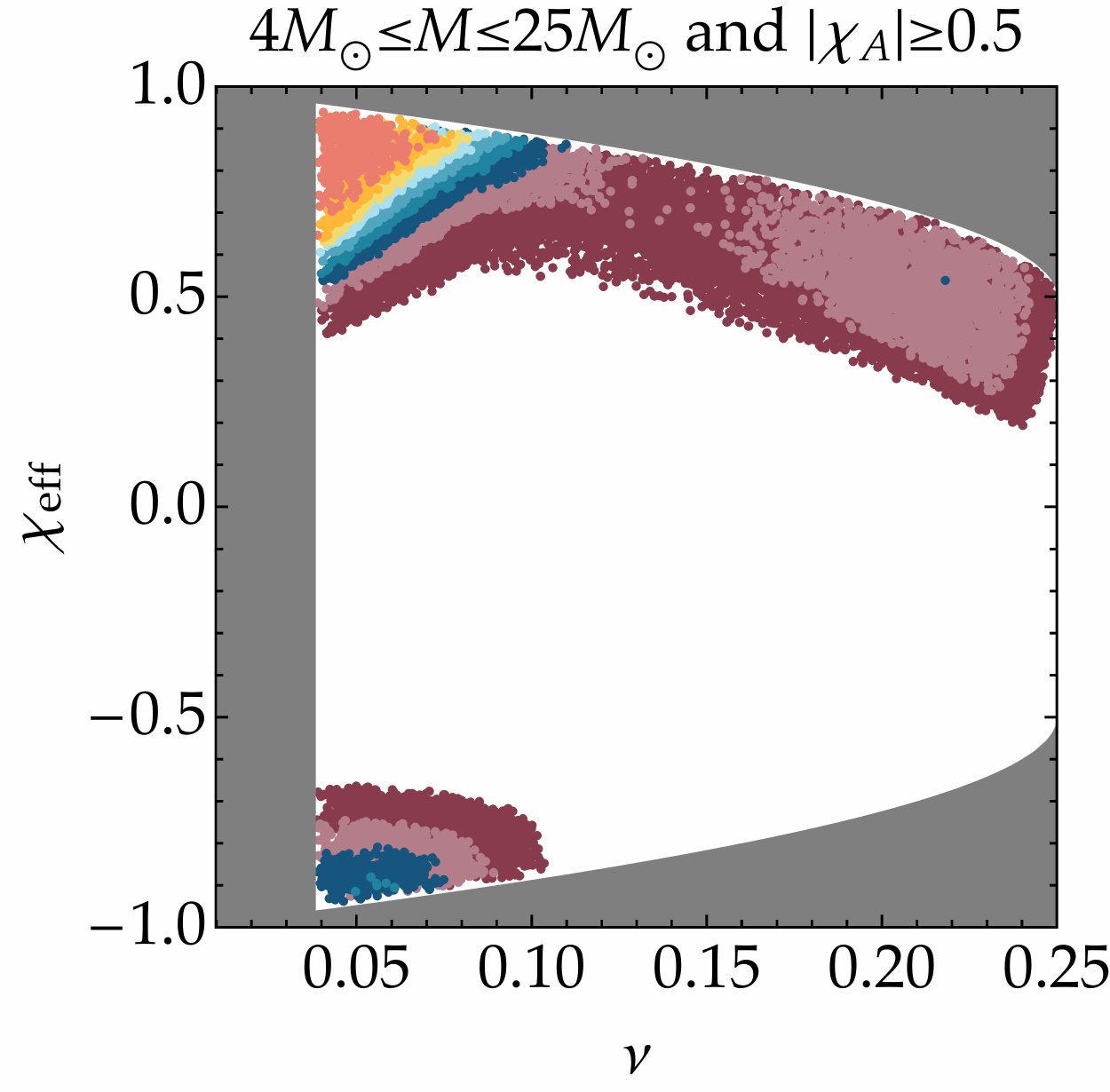} \raisebox{0.0\height}{\includegraphics[width=0.25\columnwidth]{figs/Faiv4PhDLeg.pdf}}
\caption{\label{fig:fsplitfaithM1} Faithfulness of the EOBNR model of this paper (\texttt{SEOBNRv4}) against the phenomenological inspiral-merger-ringdown model (\texttt{IMRPhenomD})~\cite{Khan:2015jqa} for $10^6$ random spinning, nonprecessing BBHs with $4\,M_{\odot} \leq M \leq 25\,M_{\odot}$ using the Advanced LIGO O1 noise PSD and a low-frequency cutoff of 25\,Hz. When plotting, we restrict the data in $\chi_{\textrm{A}}$ according the the values specified above each plot. Points with faithfulness above 97\% are not shown. Note that only 0.9\% (2.5\%) of points have faithfulness below 97\% when $|\chi_{\textrm{A}}|\leq 0.1$ ($|\chi_{\textrm{A}}|\geq 0.5$). Points with faithfulness $\leq 73\%$ are in red.}
\end{figure}

In Fig.~\ref{fig:fsplitfaithM1} we show results for the first set of BBHs, i.e.,
those with $4\,M_{\odot} \leq M \leq 25\, M_{\odot}$. This low-mass range
emphasizes the role of the inspiral in the computation of the matches. We notice
that points with faithfulness below 90\% are confined to mass ratios above $\sim
5$ and $|\chi_{\textrm{eff}}| \gtrsim 0.5$, irrespective of $\chi_{\textrm{A}}$,
that is in those regions where the signal spans many GW cycles and where
existing NR simulations do not provide strong constraints on models. On the
other hand, it is reassuring that at $q\lesssim 5$ and small values of
$|\chi_{\textrm{A}}|$ the differences in the low-frequency portion of the models
are always within 10\%. At large values of $|\chi_{\textrm{A}}|$, differences up
to 10\% can be found also in the comparable-mass regime.

\begin{figure}[htb]
\hspace{-0.7cm}\includegraphics[width=0.8\columnwidth]{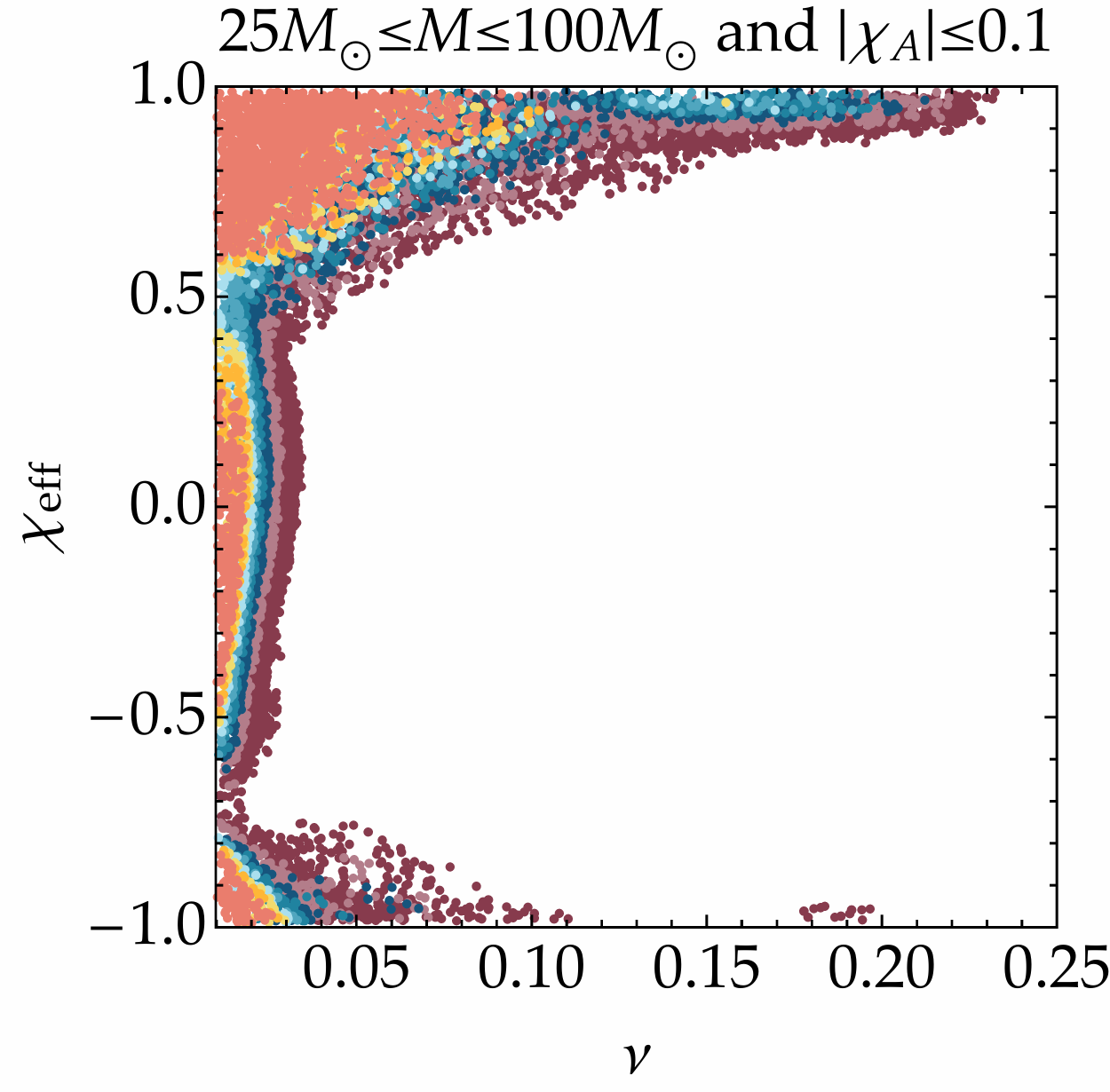} \raisebox{0.0\height}{\includegraphics[width=0.25\columnwidth]{figs/Faiv4PhDLeg.pdf}}

\hspace{-0.7cm}\includegraphics[width=0.8\columnwidth]{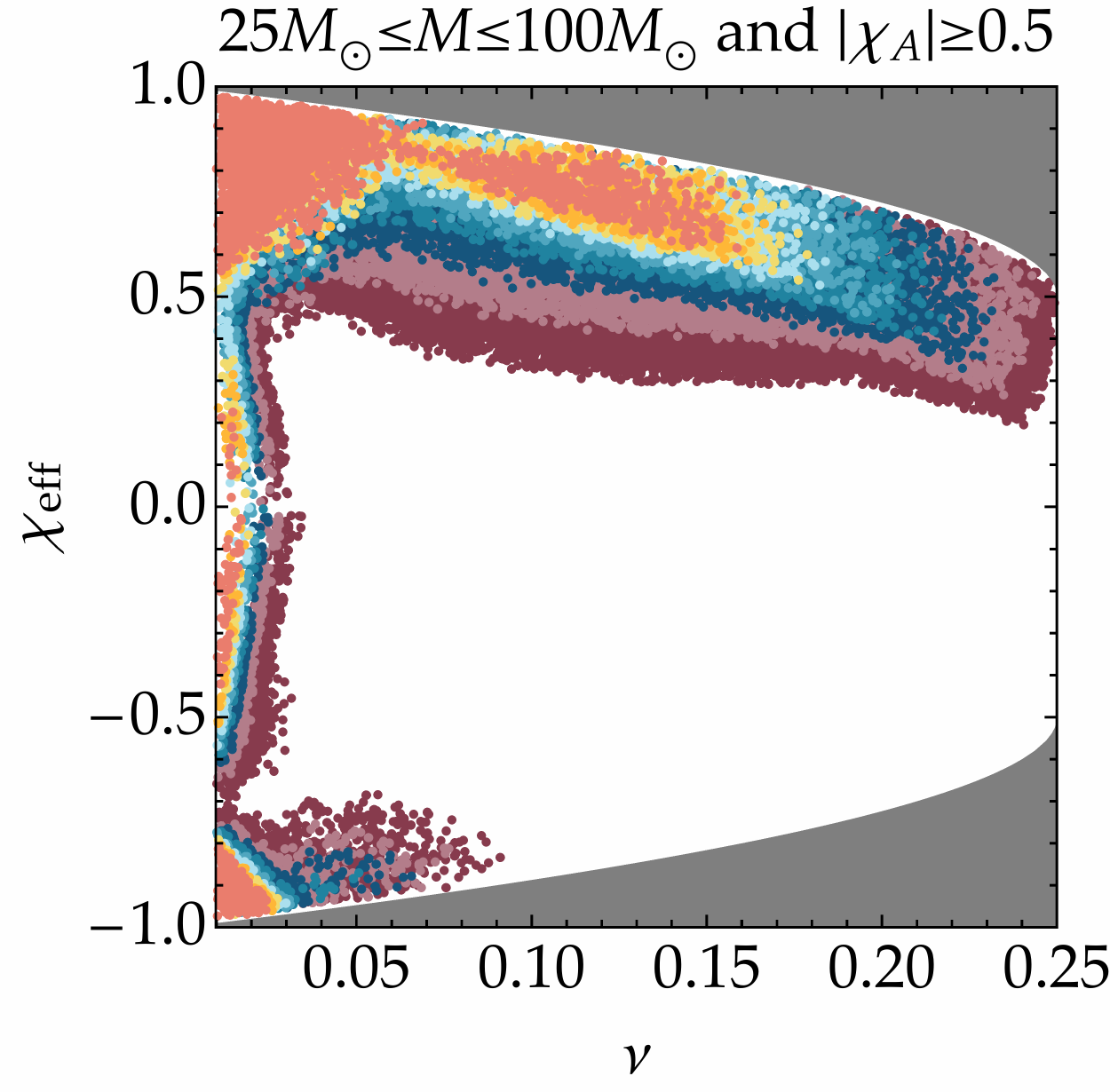} \raisebox{0.0\height}{\includegraphics[width=0.25\columnwidth]{figs/Faiv4PhDLeg.pdf}}
\caption{\label{fig:fsplitfaithM2} Same as Fig.~\ref{fig:fsplitfaithM1}, but now for BBHs with $25\,M_{\odot} \leq M \leq 100\,M_{\odot}$. Note that only 1.1\% (5.1\%) of points have faithfulness below 97\% when $|\chi_{\textrm{A}}|\leq 0.1$ ($|\chi_{\textrm{A}}|\geq 0.5$).}
\end{figure}

In Fig.~\ref{fig:fsplitfaithM2} we show results for the second set of BBHs,
i.e., those with $25\,M_{\odot} \leq M \leq 100\, M_{\odot}$. In this
intermediate-mass range both the inspiral and the merger-ringdown contribute to
the matches. At mass ratios above $\sim 8$, unsurprisingly, we find many points
with very poor faithfulness (below 80\%). For both approximants, this is the
region of extrapolation away from the respective domains of calibration. At
smaller mass ratios, whenever $|\chi_{\textrm{A}}|$ is small (top panel), the
faithfulness is good, as long as $\chi_{\textrm{eff}}$ does not exceed $\sim
0.9$. However, when $|\chi_{\textrm{A}}|$ is large (bottom panel), large
differences are found even for comparable masses and moderate values of
$\chi_{\textrm{eff}}$. This indicates that calibration to NR simulations is not
constraining even at these values of total mass. For large
$|\chi_{\textrm{A}}|$'s, even at small mass ratios, large unfaithfulness regions
are present. This is expected because of the poorer coverage of the
$\chi_{\textrm{A}}$ dimension with NR runs that entered the \texttt{IMRPhenomD}
calibration. 

\begin{figure}[htb]
\hspace{-0.7cm}\includegraphics[width=0.8\columnwidth]{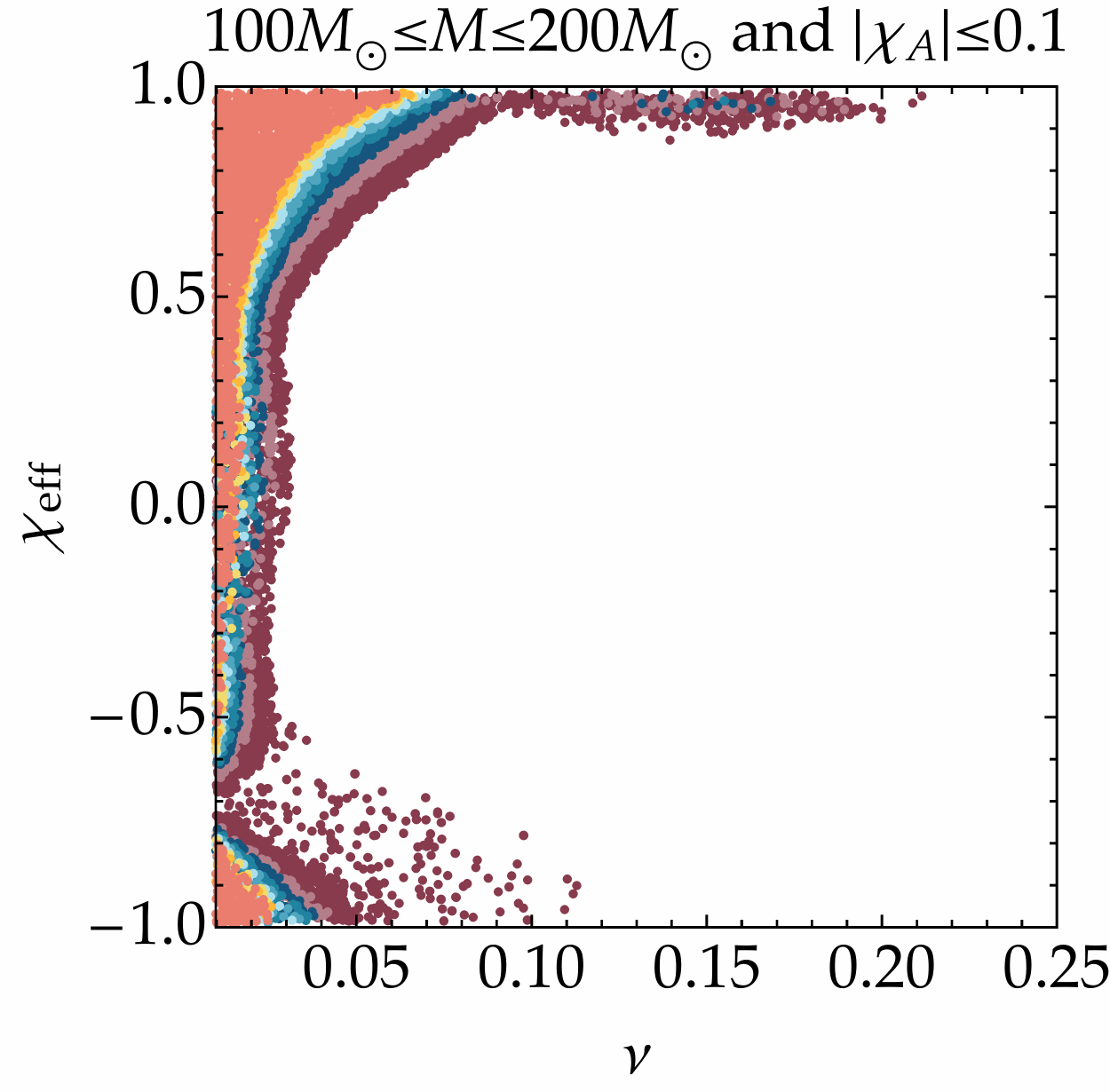} \raisebox{0.0\height}{\includegraphics[width=0.25\columnwidth]{figs/Faiv4PhDLeg.pdf}}

\hspace{-0.7cm}\includegraphics[width=0.8\columnwidth]{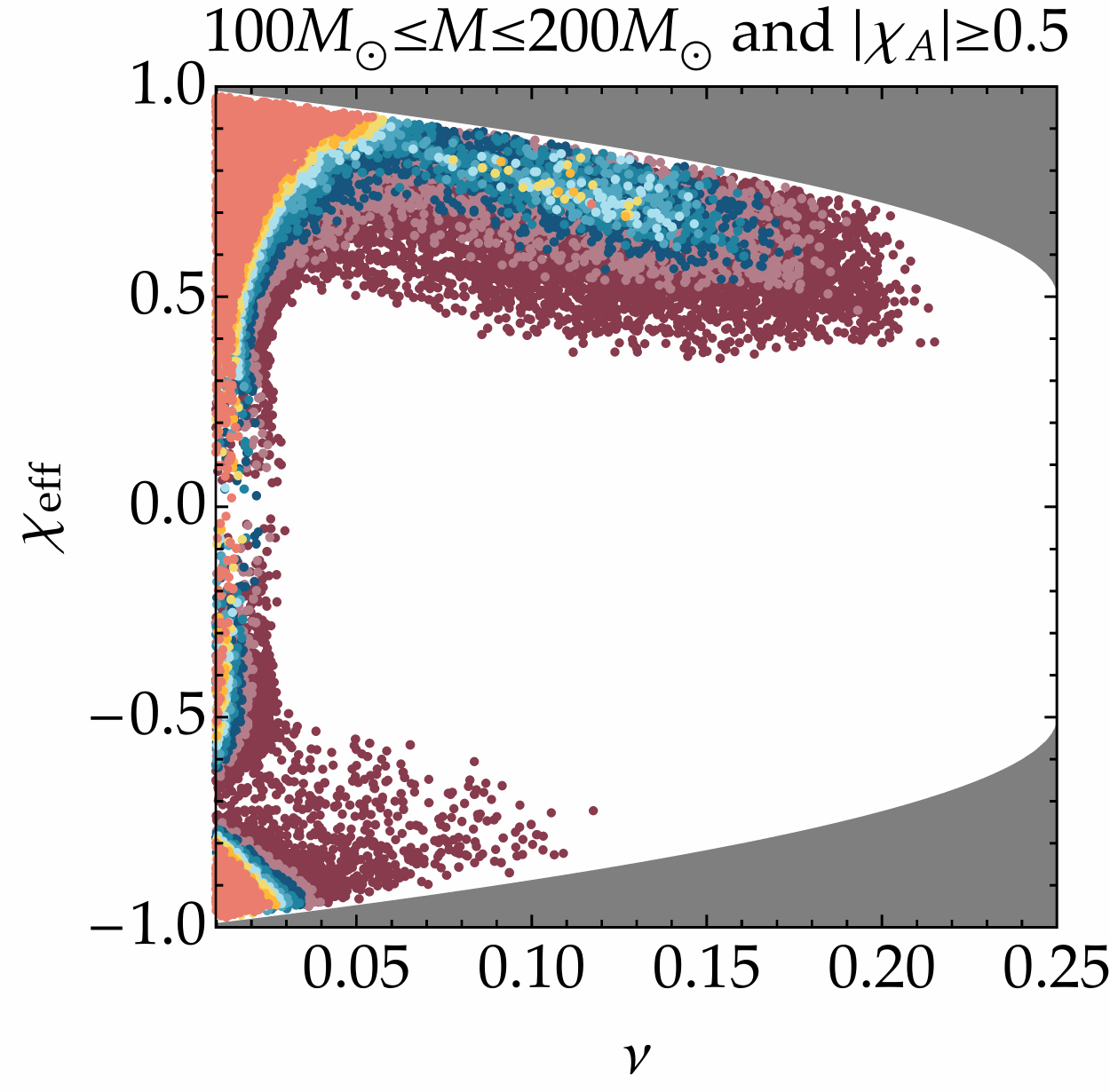} \raisebox{0.0\height}{\includegraphics[width=0.25\columnwidth]{figs/Faiv4PhDLeg.pdf}}
\caption{\label{fig:fsplitfaithM3} Same as Fig.~\ref{fig:fsplitfaithM1}, but now for BBHs with $100\,M_{\odot} \leq M \leq 200\,M_{\odot}$. Note that only 0.9\% (2.2\%) of points have faithfulness below 97\% when $|\chi_{\textrm{A}}|\leq 0.1$ ($|\chi_{\textrm{A}}|\geq 0.5$).}
\end{figure}

Finally, in Fig.~\ref{fig:fsplitfaithM3} we show results for the third set of
BBHs, i.e., those with $100\,M_{\odot} \leq M \leq 200\, M_{\odot}$. This
high-mass range emphasizes the role of the late inspiral and of the
merger-ringdown in the computation of the matches. Here we observe distinct
behaviors according to the range of $\chi_{\textrm{A}}$ that one considers. For
small values of $|\chi_{\textrm{A}}|$ (top panel), most of the
$(\nu,\chi_{\textrm{eff}})$ plane has faithfulness above 97\% thanks to the fact
that this is precisely the domain that is best constrained by existing NR
simulations. In particular, we observe that the white region amply encompasses
the location of the most extreme NR waveform that was included in the
calibration of \texttt{SEOBNRv4}, i.e., the \texttt{BAM} run at
$(q,\chi_1,\chi_2)=(8,0.85,0.85)$. At very large mass ratios, very large
differences between the models persist. For large values of $|\chi_{\textrm{A}}|$
(bottom panel), besides the difference at very large mass ratios, we observe
many points with faithfulness below 90\% for mass ratios as small as 3, for the
same reason mentioned above when discussing the intermediate-total-mass set with
large $|\chi_{\textrm{A}}|$'s. 

\FloatBarrier

\bibliography{inspire,etreferences}
\end{document}